\documentclass[opre,nonblindrev]{informs3} % current default for manuscript submission

\usepackage{float}
\usepackage{amsmath}
\usepackage{amssymb}
\usepackage{color}
\usepackage{rotating}
\usepackage{multirow}
\usepackage{latexsym}
\usepackage{secdot}
\usepackage{setspace}
\usepackage{booktabs}
\usepackage[us,12hr]{datetime}
\usepackage{bm}
\usepackage{rccol}
\usepackage{epsfig}
\usepackage[margin=1in,paper=letterpaper]{geometry}
\usepackage{dcolumn}
\usepackage{url}
\usepackage{amsfonts}
\usepackage{amssymb}
\usepackage{enumitem}
\usepackage{todonotes}
\usepackage{ulem}
\usepackage{amsmath}
\usepackage{wrapfig}
\usepackage{soul}
\usepackage{kantlipsum}
\usepackage{stackengine}
\allowdisplaybreaks

\newtheorem{thm}{Theorem}[section]

\newtheorem{lem}[thm]{Lemma}

\newtheorem{example}[thm]{Example}
\newtheorem{definition}[thm]{Definition}

\newtheorem{theorem}[thm]{Theorem}

{\theoremstyle{THkey}}

\OneAndAHalfSpacedXI % current default line spacing
\usepackage{natbib}
 \bibpunct[, ]{(}{)}{,}{a}{}{,}%
 \def\bibfont{\small}%
 \def\bibsep{\smallskipamount}%

\EquationsNumberedThrough    % Default: (1), (2), ...
\MANUSCRIPTNO{} 

\newcommand{\ts}{{\,}}

\def \ts {\,}

\def \that {{\widehat t}}

\def \Mhat {{\widehat M}}
\def \Nhat { {\widehat N}}
\def \Zhat {{\widehat Z}}
\def \Shat {{\widehat S}}

\def \Nbar {{\overline N}}

\def \Dcal {{\mathcal D}}

\def \Ocal {{\mathcal O}}

\def \uvec {\bm{u}}
\def \wvec {\bm{w}}

\def \gvec {\bm{g}}

\def \hvec {\bm{h}}

\def \tvec {\bm{t}}
\def \svec {\bm{s}}

\def \zhat {\widehat{z}}
\def \tvechat {\widehat{\bm{t}}}

\def \Stild {\widetilde{S}}
\def \tvectild {\widetilde{\bm{t}}}
\def \wvectild {\widetilde{\bm{w}}}

\def \ttild {\widetilde{t}}
\def \wtild {\widetilde{w}}

\def \Tree {{{\sf T}}}

\def \edge {{\sf E}}

\def \root {{\sf root}}

\def \children {{\sf Ch}}
\def \parent {{\sf Pa}}
\def \descendant {{\sf De}}

\def \Integer {{\mathbb Z}}
\def \path {{{\sf path}}}
\def \product {{\sf Prod}}
\def \nest {{\sf Nest}}

\def \grid {{\sf Grid}}
\def \bag {{\sf B}}

\def \idc {{\bf 1}}
\def \treewidth {{\sf tw}}
\newcommand{\Real}{\mathbb{R}}

\def \dom {{\sf Dom}}
\def \budget {{\sf Bud}}
\def \prefW {{\sf Pref}}
\def \action {{\mathcal A}}
\def \GNL {{GNL}}
\def \MNL {{\sf MNL}}
\def \MMNL {{\sf MMNL}}
\def \GRN {{\sf GRN}}
\def \reduceA {{\overline{\mathcal{A}}}}

\newcommand{\rubigg}[1]{\biggl \lceil #1 \biggr\rceil}
\newcommand{\ruBig}[1]{\Bigl \lceil #1 \Bigr\rceil}
\newcommand{\rubig}[1]{\bigl \lceil #1 \bigr\rceil}

\newcommand{\rd}[1]{\lfloor #1 \rfloor}

\newcommand{\rdbig}[1]{\bigl \lfloor #1 \bigr\rfloor}
\newcommand{\rdBig}[1]{\Bigl \lfloor #1 \Bigr\rfloor}

\newcommand{\pr}[1]{{\color{black}{#1}}}

\newcommand{\yw}[1]{{\color{black}{#1}}}

\newcommand{\opft}[1]{\textit{\texttt{#1}}}
\newcommand{\head}[1]{\opft{Head}\left(#1\right)}
\newcommand{\subtree}[1]{\opft{Subt}\left(#1\right)}
\newcommand{\tail}[1]{\opft{Tail}\left(#1\right)}
\newcommand{\vtcs}[1]{\opft{Vtcs}\left(#1\right)}

\makeatletter
\def\moverlay{\mathpalette\mov@rlay}
\def\mov@rlay#1#2{\leavevmode\vtop{%
   \baselineskip\z@skip \lineskiplimit-\maxdimen
   \ialign{\hfil$\m@th#1##$\hfil\cr#2\crcr}}}
\newcommand{\charfusion}[3][\mathord]{
    #1{\ifx#1\mathop\vphantom{#2}\fi
        \mathpalette\mov@rlay{#2\cr#3}
      }
    \ifx#1\mathop\expandafter\displaylimits\fi}
\makeatother
\newcommand{\cupdot}{\charfusion[\mathbin]{\cup}{\cdot}}
\newcommand{\bigcupdot}{\charfusion[\mathop]{\bigcup}{\cdot}}

\renewcommand{\qed}{\hfill \mbox{\raggedright \rule{0.1in}{0.1in}}}
\begin{document}
%%%%%%%%%%%%%%%%

\RUNAUTHOR{}
%\RUNAUTHOR{\normalfont Wang, Rusmevichientong, Topaloglu, \today ;}
\RUNTITLE{}
%\RUNTITLE{\normalfont{\hspace{-5pt}Tree Decomposition for Assortment Optimization}}

\TITLE{\large 
Tree Decomposition of Product and Nest Graphs for\\ 
Assortment Optimization under the Generalized Nested Logit Model
%\vspace{-10pt}
}

\ARTICLEAUTHORS{
\vspace{-10pt}
 \AUTHOR{\mbox{Yuheng Wang$^1$, Paat Rusmevichientong$^2$, Huseyin Topaloglu$^3$,}}
 \AFF{
 $^1$SKEMA Business School, Universit\'e C\^ote d'Azur, Lille, France
 $^2$Marshall School of Business, Los Angeles, CA, USA
 $^3$School of Operations Research and Information Engineering, Cornell Tech, New York, NY 10044, USA
 }
 \vspace{-7pt}
 \EMAIL{\footnotesize yuheng.wang1@skema.edu, rusmevic@marshall.usc.edu, topaloglu@orie.cornell.edu}
 \\
 \vspace{-1mm}
 {{\footnotesize November 10, 2026}}
 \vspace{-3mm}
}

\ABSTRACT{%
The generalized nested logit model extends the nested logit model by allowing each product to belong to multiple nests.  The overlapping nests in the generalized nested \pr{logit} model can capture complex substitution patterns, and the model has found applications in many areas.  We study the assortment optimization problem under the generalized nested logit model 
and prove that it is NP-hard when there are only two nests or when any two distinct nests have at most one product in common.  Motivated by these complexity results, we introduce novel concepts of product and nest graphs 
\yw{that capture which products are included in which nests. We}
use \pr{the tree decompositions} \yw{of the graphs} to characterize when the assortment optimization problem admits efficient solution methods. When the treewidth of a product graph is bounded, the assortment optimization problem can be solved  in polynomial time.  If the treewidth of the product graph is unbounded but the treewidth of the nest graph is bounded, then there exists a fully polynomial-time approximation scheme. These results represent the first polynomial-time algorithms for assortment optimization under the generalized nested logit model.  Our numerical experiments on synthetic and real datasets 
show that the generalized nested logit model provides a better fit to the data than many existing models in the literature.}

\KEYWORDS{\vspace{-2mm} Generalized nested logit model, assortment optimization, polynomial-time algorithms} 
%\HISTORY{This paper was first submitted on April 12, 1922 and has been with the authors for 83 years for 65 revisions.}

\maketitle

\vspace{-10mm}

\section{Introduction}
\label{sec:introduction}
\vspace{-1mm}

Discrete choice models have been widely used in operations management to capture how customers choose among alternatives and to represent the demand process. The multinomial logit model is one of the most commonly used discrete choice models; it assumes that a customer assigns independent Gumbel-distributed utilities to each product and chooses the product with the highest utility. The nested logit model generalizes the multinomial logit model by partitioning the set of products into disjoint nests and allows utilities of products within the same nest to be correlated, but the utilities of products in different nests are independent.  
The generalized nested logit model, also called the cross-nested logit model in the literature, extends the nested logit model by allowing overlapping nests, so that each product can belong to multiple nests, enabling it to capture complex correlation patterns among the product~utilities.

Researchers have recognized that the generalized nested logit model can capture more complex substitution patterns than the multinomial logit or nested logit models, and they have applied it to modeling customer choices in a variety of applications.  Despite the abundant literature on the applications of generalized nested logit models, \pr{there are hardly any papers on} operational decisions under this model. In this paper, we study assortment optimization under the generalized nested logit model. Under this model, %the outside option belongs to its own dedicated nest, and 
a customer first chooses among the nests or the outside option. If the customer chooses a nest, she will choose one of the products in that nest. Below is a summary of our contributions.
%and derive tractable algorithms for the problem with performance guarantees.  

%Discuss what is a generalized nested logit (\gnl) model in broad term. First, talk about the classical nested logit that generalizes the multinomial logit choice model by allowing for correlation among utilities of products within the same nest. However, it requires the nest to be disjoint.  This creates a limitation.
%Then, comes the generalized nested logit model (also known in the literature as cross nested logit model).  
%
%
%Lots of work focus on apply GNL to various marketing, transportation, and economics applications. However, there are very few work focus on assortment optimization under the GNL. 

%\pr{
%Things to note here
%\begin{itemize}
%\item Should we call our model ``Generalized Nested Logit'' for consistency with Kooperman's paper?
%\item Need to cite the paper by Le + Mei in POM 2024
%\item Need to note that GNL is GEV, so it is RUM. Also briefly discuss the correlation structure.
%\end{itemize}
%}

%Key Contributions
%\begin{itemize}
%\item Novel concepts of Product Graph and Nest Graph
%\item We are the first to characterize the complexity of GNL assortment optimization with the tree width of product graph, and treewideth of nest graph
%\item Contrast this with Le + Mei and 2 other papers.  This is the first paper to offer a polynomial-time algorithm for assortment optimization under the Generalized Nested Logit model.
%\item Also, our numerics show that GNL provide  a much better fit than standard nested logit model (with non-overlapping nests)
%
%\end{itemize}
%}
%

\paragraph{\underline{New complexity results}:} 
%We focus on the generalized nested logit model in which an outside option appears only in its own dedicated nest, and all other nests can contain overlapping products but no outside option. 
\pr{
\cite{davis2014assortment} consider the nested logit model, which is a special case of the generalized nested logit in which the nests are disjoint, and prove that the corresponding assortment optimization problem can be solved in polynomial time via linear programming.}
Prior to our work, it was unknown whether the assortment optimization problem under the generalized nested logit model with overlapping nests is tractable.
Our paper settles this question. In contrast to the nested logit model, Theorem~\ref{thm:hardness_1} shows that the assortment optimization problem under the generalized nested logit model is NP-hard.  The proof of Theorem~\ref{thm:hardness_1} considers the setting with two nests, each of which contains multiple overlapping products. Theorem~\ref{thm:hardness_2} further shows that the problem remains NP-hard if any two distinct nests share at most one product. We later show that both of these instances admit fully polynomial-time approximation schemes.  

%These two complexity results enrich our understanding of assortment optimization under the generalized nested logit model with fully-captured nests.  
%that the assortment optimization problem is NP-hard
%for partially captured nests, but the problem is polynomial-time solvable via a linear program for fully captured nests.  
%\cite{le2024constrained} consider the assortment optimization problem under the generalized nested logit model with {\it partially captured nests} in which each nest can contain an outside option and show that the corresponding assortment optimization is NP-hard.   
%The placement of the outside option greatly affects the complexity of the assortment optimization problem.   

Because of its complexity, a common heuristic for solving the assortment optimization \yw{problem} under the generalized nested logit model is to convert it into the problem of finding a fixed point of some nonincreasing function $F: \Real_+ \to \Real_+$.   The challenge here is that for each $z \in \Real_+$, evaluating the function $F(z)$ requires solving a high-dimensional combinatorial optimization problem \citep{zhang2025exact}. For the special case of the nested logit model,
computing $F(z)$ can be done efficiently via linear programs \citep{davis2014assortment}.  \pr{For the generalized nested logit model, however, computing $F(z)$ is much more difficult.}
In Theorem \ref{thm:subproblem_hardness}, we prove that 
it is NP-hard to approximate the value of $F(z)$ to within any constant factor. This worst-case complexity result is surprising and raises the question of how to identify tractable instances of generalized nested logit models.

\paragraph{\uline{Using tree decomposition of product and nest graphs for assortment optimization}:} In Section~\ref{sec:product-and-nest-graphs}, 
we introduce novel concepts of product and nest graphs.  A product graph is an undirected graph in which each vertex corresponds to a product, and there is an edge between two products if they share a common nest. Analogously,  a nest graph is an undirected graph in which each vertex corresponds to a nest, and there is an edge between two nests if the two nests share at least one product in common.  The edges of these graphs capture how often products appear together and how often nests share common products. 
Tree decompositions of product and nest graphs give rise to a partition of vertices (Lemma \ref{lem:sep_vertices}), and we can use the treewidth --- a fundamental concept in graph theory and graph algorithms that measures how far a graph is from a tree ---  to identify instances where the assortment optimization problem can be solved in polynomial time or approximated with arbitrary accuracy. 
\pr{The connection between the complexity of assortment optimization and the treewidths of product and nest graphs is surprising.} 
Although there is a large literature on using tree decompositions to design algorithms and on using treewidth to characterize the algorithms' running times,  
to our knowledge,  this is the first paper to show how tree decomposition can be applied to assortment optimization under the generalized nested logit model. 

%We believe that our approach can be applied to other assortment optimization problems as well.

\paragraph{\uline{A polynomial-time algorithm for product graphs with bounded treewidths}:}  

In Section \ref{sec:product-graph-bounded-tree-width}, we show that when the product \yw{graph admits a bounded treewidth that is} independent of the number of products or nests, then 
we can use the tree decomposition to design a polynomial-time \pr{dynamic programming} algorithm to compute an optimal assortment.   To design our algorithm, we develop several new insights into the generalized nested logit model. First, we study nest-specific products that belong to only one nest and show that optimizing over them can be done independently within each nest. Furthermore, within each nest, instead of having to consider all subsets of nest-specific products, it suffices to only consider revenue-ordered subsets  (Theorem \ref{thm:nest-specific-products}).  This result allows us to restrict our attention to overlapping products that belong to at least two nests.  Second, we establish important structural properties of a tree decomposition of a product graph. 
Lemma \ref{lem:prod_graph_domain} shows that \pr{for each subtree in the tree decomposition of the product graph, products that share common nests are all located within the subtree, enabling us to use nodes in the tree decomposition as state variables in our dynamic programming algorithm.}
Lemma \ref{lem:prod_graph_nest_sep} further shows that the nests of the products associated with a subtree at a particular node can be partitioned as a disjoint union of nests associated with each child node. This partition of nests allows us to efficiently design a polynomial-time dynamic programming algorithm to compute an optimal assortment (Theorem~\ref{thm:exact_algo}).  To our knowledge, this is the first polynomial-time algorithm for assortment optimization under the generalized nested logit model.
%We believe that our approach can be applied to other choice models as well.

\paragraph{\uline{A fully polynomial-time approximation scheme for nest graphs with bounded treewidths}:}
Section~\ref{sec:fptas} considers the case where the treewidth of the product graph is unbounded, but the nest graph has a bounded treewidth. Theorem~\ref{thm:FPTAS} shows that we can still obtain a fully polynomial-time approximation scheme (FPTAS) in this case.  Our algorithm exploits special properties of the tree decomposition of the nest graph. 
\pr{Analogous to the properties of the tree decomposition of the product graph, we establish structural properties of the tree decomposition of the nest graph.
Lemma \ref{lem:nest_graph_domain} shows that for each subtree in the tree decomposition of the nest graph, nests that share common products are all located within the subtree. 
 Lemma \ref{lem:nest_graph_prod_sep} shows that products belonging to nests associated with a subtree at a particular node can be partitioned as a disjoint union of products associated with each child node.}  
In Theorem \ref{thm:fptas_dp_correctness}, we combine these structural insights with methods for solving multidimensional knapsack problems to design a dynamic programming method that yields an FPTAS.
\pr{The treewidths of product graph and nest graph can be very different, and thus, both the exact algorithm in Section \ref{sec:product-graph-bounded-tree-width} and the FPTAS in Section 
\ref{sec:fptas} are useful.  
Existing methods in the literature give an FPTAS whose running time is exponential in the number of nests.
Our proposed method, however, has a running time that is exponential in the treewidth of the nest graph. The treewidth of the nest graph is always bounded above by the number of nests, but the treewidth can be substantially smaller in many problem instances. Obtaining an FPTAS whose running time scales with the treewidth is the first result of its kind for this choice model.}
%Existing literature does not offer any polynomial-time algorithm with strong performance guarantees. 

\paragraph{\uline{Empirical results}:} \pr{
Using real transaction data from retailers,
we compare the predictive accuracy of the generalized nested logit model with that of the multinomial logit, nested logit, and mixture of multinomial logit models. Across a wide range of product categories, we find that the generalized nested logit model predicts customer purchases more accurately. We also observe similar results on synthetic datasets. This result shows that the generalized nested logit model captures the substitution patterns in retail~transactions better than the benchmarks we consider.
}

\vspace{-10pt}
\subsection{Literature Review}
\vspace{-0pt}

\pr{Belonging to the family of generalized extreme value models \citep{mcfadden1978modelling}, the generalized nested logit model includes the nested logit models as special cases. 
%The multinomial logit corresponds to the case in which each nest consists of a single product, and 
The nested logit corresponds to the case in which the nests are nonoverlapping.}
% The generalized nested logit model
% is a special case of the generalized extreme value model \citep{mcfadden1978modelling}, which includes the multinomial logit model as a special case. The 
% generalized nested logit model
% generalizes the nested logit model by allowing for overlapping nests, so each product may belong to multiple nests.  
Some researchers use the name cross-nested logit model \citep{bierlaire2006theoretical} to refer to the generalized nested logit model. %In this manuscript, 
We adopt the terminology of the generalized nested logit model, which was also used by \cite{wen2001generalized}, to emphasize that the model is a generalization of the nested logit model.

The  generalized nested logit model includes other choice models as special cases, including the paired combinatorial logit model in which each nest consists of a pair of products \citep{chu1989paired,koppelman2000paired}, the ordered GEV model in which there is a linear ordering on the set of products and each nest consists of a contiguous set of products \citep{small1987discrete}, and the product differentiation model in which each product is described by a vector of pre-selected features and is assigned to a nest based on the features \citep{bresnahan1996market}.

By allowing for overlapping nests, the product utilities under the generalized nested logit model can exhibit complex correlation patterns, and there is a lot of research on understanding these patterns; see, for example, \cite{abbe2007normalization,marzano2008covariance,marzano2013practically}.  Researchers have applied the generalized nested logit model to uncover complex substitution patterns in people migration \citep{beine2025new}, air carrier choices \citep{drabas2013modelling}, travel mode and departure time choices in Maryland-Washington D.C. region \citep{ding2015cross}, residential location choices \citep{yang2013cross}, and many others.

Yet, there has been very limited work on solving operational problems under the generalized nested logit model.  \cite{le2024constrained} are the first to consider the problem of constrained assortment optimization under the generalized nested logit model with partially captured nests, in which the outside option may appear in every nest. After showing that the unconstrained assortment optimization problem is NP-hard, they develop a discretization heuristic and solve the approximate problem as a mixed-integer linear program.  \cite{zhang2025exact} consider the cardinality-constrained version, develop a branch-and-bound method to solve the problem exactly, and design heuristics that yield near-optimal solutions in numerical experiments. \cite{pham2025constrained} consider the joint assortment and pricing  problem, reformulate the problem into bilinear and exponential-cone convex programs, and develop a heuristic to generate strong cutting planes within the branch-and-cut framework.  None of these methods is a polynomial-time algorithm in the worst case, and \pr{none} offers theoretical performance guarantees.

In contrast, 
we consider the generalized nested logit model \pr{with fully captured nests, in which the outside option appears only in its own dedicated nest,
and prove that the assortment optimization is NP-hard.}  We are the first to introduce novel concepts of product and nest graphs, along with their tree decompositions.  We show that if the treewidth of the product graph is bounded, then the corresponding assortment optimization problem can be solved in polynomial time. If the treewidth of the product graph is unbounded, but the treewidth of the nest graph remains bounded, then the assortment optimization problem admits a fully polynomial-time approximation scheme. To our knowledge, these are the first polynomial-time algorithms with provable performance guarantees for assortment optimization problems under the generalized nested logit model.

\pr{We note that using the approach in the seminal work of \cite{desir2022capacitated}, who consider a more general optimization problem, we can obtain an FPTAS for the assortment optimization problem under the generalized nested logit model, with a running time that increases exponentially with the number of nests. In contrast, our FPTAS has a running time that scales more gracefully with the treewidth of the nest graph.}  
Our work is also related to the large literature on assortment optimization under the multinomial logit 
\citep{TalluriVanRyzin:2004,sumida2021revenue}, 
the nested logit 
\citep{davis2014assortment,Feldman1}, and 
the mixture of multinomial logit
\citep{MendezDiazEtal:2010,Desir1}, and many others; see \cite{Ma2024:AssortmentOverview} for an overview.

%
%\paragraph{Generalized Nested Logit Model:} There are a lot of work in analyzing the \gnl model, looking at correlations, etc. Give some history here about how the model comes about. See some of the papers.  15 papers here?
%
%This is a great place to discuss the terminology. Cross-nested versus Generalized Nested Logit model.
%
%\paragraph{Assortment Optimization:}  Lots of work on assortment optimization under the classical nested logit model, where each product belongs to exactly one nest.  Need to look up Huseyin's paper for citations here. [7-8 papers] However, there is very few work on assortment optimiztion under the generalized nested logit.  Only found 3 papers here (1 in EUJOR, 1 in POM, and 1 in ArXiv). Give a brief description here, but all of these papers do NOT provide polynomial-time algorithms.  

%\paragraph{Tree Decomposition and Treewidths:} Classic CS concepts. Provide some cites here. To our knowledge, we are the first to apply these concepts to assortment optimization problem under the \gnl model.

The treewidth is one of the most fundamental concepts in graph theory and graph algorithms.  The concept was introduced and rediscovered multiple times; see, for example,
\cite{halin1976s,robertson1984graph,robertson1986graph}. It is well known that many NP-hard optimization problems on graphs can be solved in polynomial-time when \yw{the} treewidth of the input graph is bounded; see, for example,
\cite{bodlaender1988dynamic,courcelle1990monadic,courcelle1992monadic}. To our knowledge, however, this is the first paper that establishes connections between the complexity of the assortment optimization \yw{problem} and the treewidths of the product and nest graphs. We believe the concept of treewidth and tree decomposition, on appropriately constructed graphs, can be applied to other  problems.

\vspace{-10pt}
\subsection{Organization}
\vspace{-10pt}

Section \ref{sec:model} describes the assortment optimization problem  and establishes its computational complexity. Section \ref{sec:product-and-nest-graphs} introduces the product and nest graphs, establishes properties associated with their tree decompositions, and outlines the solution framework.  Section \ref{sec:product-graph-bounded-tree-width} describes a \yw{fully} polynomial-time algorithm for the case where the product graph has a bounded treewidth. 
Section~\ref{sec:fptas} describes an FPTAS for the case where the nest graph has bounded treewidth.  Section \ref{sec:empirical} contains numerical experiments on  real datasets and Section~\ref{sec:conclusion} concludes with directions for future research.

%\section{Model Description} 
\section{Assortment Optimization under the Generalized Nested Logit Model} \label{sec:model}

Following the formulation in \cite{wen2001generalized} and \cite{abbe2007normalization}, the generalized nested logit model is defined as follows.
We have a universe of $n$ products indexed by \mbox{$N = \{1, 2, \ldots, n\}$}, and the outside option is denoted by product $0$, corresponding to the no-purchase option.  Under the generalized nested logit model, there is a collection of possibly overlapping subsets of products indexed by $K = \{1, 2, \dots, m\}$; we refer to these subsets as nests. For each nest $k \in K$, let $M_k \subseteq N$ denote the set of products belonging to nest $k$. 
Each product $i \in N$ can belong to multiple nests.  %We consider the fully captured nest setting, so 
\pr{The outside option belongs to its own nest,} which consists only of product $0$.

Letting $U_i$ denote the utility of product $i \in N \cup \{0\}$,  we assume that 
  $U_i = \log w_i + \epsilon_i$, where \mbox{$w_i > 0$} denotes the product-specific constants, with $w_0 = 1$.  The random vector $(\epsilon_0, \epsilon_1, \ldots, \epsilon_n)$ has a distribution function given by: 
\mbox{$\exp \big( - G ( e^{-y_0}, e^{-y_1}, \ldots, e^{-y_n} ) \big)$} 
for all $(y_0, y_1, \ldots, y_n) \in \Real^{n+1}$.  The generating function $G$ is given by:
\mbox{$
	G(y_0, y_1, \ldots, y_n) = y_0 + \sum_{k \in K} \Big( \sum_{i \in M_k} \big( \alpha_{ik} y_i \big)^{1/\gamma_k} \Big)^{\gamma_k}
$}.
For each nest $k \in K$,  $\gamma_k \in (0, 1]$ denotes the nest-specific parameter, and $\alpha_{ik} \in \Real_+$ captures the degree of membership of product $i$ in nest $k$.  \pr{By definition}, $\alpha_{ik} = 0$ for all $i \notin M_k$, and $\sum_{k \in K} \alpha_{ik} = 1$.

The utility specification above shows that the generalized nested logit model is an instance of the Generalized Extreme Value model \citep{mcfadden1978modelling}. \pr{If} we offer an assortment $S \subseteq N$, the probability $\phi_i(S)$ that a customer will choose product $i \in S \cup \{0\}$ is  given by:
\vspace{5pt}
$$
	\phi_i(S) = \begin{cases}
	\sum_{k \in K}  \displaystyle \frac{ \big( \sum_{j \in S}   (\alpha_{jk} w_j)^{1/\gamma_k}  \big)^{\gamma_k}}{ 1 + \sum_{\ell \in K}  \big( \sum_{j \in S}  (\alpha_{j \ell} w_j)^{1/\gamma_\ell}  \big)^{\gamma_\ell}}\times 
\frac{ ( \alpha_{ik} w_i )^{1/\gamma_k}  }{ \sum_{j \in S} ( \alpha_{jk} w_j )^{1/\gamma_k}  } & \quad \text{ if } i \in S \\[20pt]
	 \displaystyle \frac{1}{ 1 + \sum_{\ell \in K}  \big( \sum_{j \in  S}  (\alpha_{j \ell} w_j)^{1/\gamma_\ell}  \big)^{\gamma_\ell}} & \quad \text{ if } i = 0 \\
	\end{cases}
\vspace{5pt}
$$
For ease of exposition, for each $i \in N$ and $k \in K$, let $v_{ik} ~\equiv~ (\alpha_{ik} w_i)^{1/\gamma_k}$ denote the preference weight of product $i$ within nest $k$, with  $v_{ik} = 0$ if $i \notin M_k$. Given an assortment $S \subseteq N$, let $V_k(S) =  \sum_{i \in S} v_{ik}$  denote the total preference weight of the offered products in nest $k$.  The choice probability $\phi_i(S)$ arises from a two-stage decision-making process.  First, the customer decides to make a purchase from nest $k \in K$ with probability
$
Q_k(S) \;=\;  V_k(S)^{\gamma_k}/ (\,1 + \sum_{\ell \in K} V_\ell(S)^{\gamma_\ell} ).
$
Second, given that a customer decides to purchase in nest $k$, the probability that she selects product $i \in S \cap M_k$ is given by
$
P_{ik}(S) = v_{ik}/V_k(S).
$
Therefore, if we offer a subset of products $S$, then a customer chooses a product $i$ with probability $\phi_i(S) = \sum_{k \in K} Q_k(S) P_{ik}(S)$. 
We use $i$ and $j$ to index products and $k$ and $\ell$ to index nests.

%To formulate the assortment optimization problem, 
For each $j \in N$, let $f_j > 0$ denote the revenue of product $j$.  Consider an arbitrary assortment $S \subseteq N$. For each $k \in K$, 
conditional on the customer choosing nest $k$, the expected revenue $R_k(S)$ from the offered products in nest $k$ is given by
$
R_k(S) \;=\; \sum_{i \in S} f_i P_{ik}(S) \;=\; \frac{\sum_{i \in S} f_i v_{ik}  }{V_k(S)}.
$
Letting $\Pi(S)$  denote the expected revenue from offering assortment $S$, we  have that
$$
\Pi(S) ~=~ \sum_{i \in S} f_i \phi_i(S) ~=~ \sum_{k \in K } Q_k(S) R_k(S) = \sum_{k \in K} \frac{ V_k(S)^{\gamma_k} R_k(S) }{1 + \sum_{\ell \in K} V_\ell(S)^{\gamma_\ell}}.
$$
We want to find an assortment that maximizes the expected revenue obtained from a customer, corresponding to the following optimization problem:
\begin{align}\label{eqn:cnl_assort} \tag{{\sf GNL-Assort}}
    Z^* ~=~ \max_{S \subseteq N  } ~ \Pi(S).
\end{align}

%As noted in Section \ref{sec:introduction}, the generalized nested logit model includes several well-known discrete choice models as special cases, including the multinomial logit, the nested logit, and the paired combinatorial logit. 
%One way to interpret the generality of the \gnl model is through the degree of overlaps across different nests. 
%induced correlation structure on product utilities. It is well known that the CNL model introduces correlation among products within the same nest (\cite{train2009discrete}). 
We note that the 
%multinomial logit model is a special case of the generalized nested logit model in which each product forms its own nest, so $K = N$ and $M_k = \{k\}$ for all $k$. The 
nested logit model is a special case of the generalized nested logit model in which the nests are assumed to be disjoint; that is, $M_k \cap M_{k'} = \varnothing$ for $k \ne k'$. 
%Under the pair combinatorial logit model, there is a nest for every possible pair of products, and thus, each product belongs to $n-1$ nests simultaneously.   
\pr{\cite{davis2014assortment} proves that the assortment optimization under the nested logit model with partially captured nests (in which the outside option can appear in every nest) is NP-hard, but the problem can be solved in polynomial time in the setting with fully captured nests (in which the outside option appears only in its own dedicated nest).
\cite{le2024constrained} prove that the assortment optimization problem under the generalized nested logit model with partially captured nests is also NP-hard. However, the complexity of the problem with fully captured nests remains unknown. 
The following theorem settles this question by showing that the assortment optimization problem under the generalized nested logit model is
 NP-hard even for fully captured nests. The proof is in Appendix~\ref{sec:hardness_1}.}

\begin{theorem}[Two Fully Captured Overlapping Nests]\label{thm:hardness_1}
The \ref{eqn:cnl_assort} problem is \mbox{NP-hard} when there are two fully captured overlapping nests.
\end{theorem}
\vspace{-5pt}
Theorem \ref{thm:hardness_1} uses an instance 
in which the two nests share all but one product.  Complementing this result, we consider the opposite extreme, where any two nests share at most~one~product.

%The proof is in Appendix~\ref{sec:hardness_1}, uses a reduction from the partition problem, and considers an instance %of the generalized nested logit model 
%in which the two nests share all but one product.  To complement this result, we consider the opposite extreme, where any two nests share at most one~product.

\begin{theorem}[Fully Captured Nests Sharing At Most One Product]
\label{thm:hardness_2}
The \mbox{\ref{eqn:cnl_assort}} problem is NP-hard even when $|M_k \cap M_{k'}| \leq 1$ for all nests $k \ne k'$.
\end{theorem}
\vspace{-5pt}
The proof of Theorem \ref{thm:hardness_2} is in Appendix~\ref{sec:hardness_2}, uses a reduction from the partition problem, and considers an instance in which the number of nests is half the number of products.  We note that  \cite{zhang2020assortment} prove that under the paired combinatorial logit model, the assortment optimization problem is also NP-hard, but they use an instance in which the number of nests is at least the number of products raised to the power of $3/2$, so the number of nests is much larger than the number of products.  
Theorem  \ref{thm:hardness_2} shows that the problem remains NP-hard even when any two nests share at most one product, and the number of nests is only half the number of products. 

In the next section, we introduce product and nest graphs and show how their treewidths can be used to characterize the complexity of the assortment optimization problem. It will turn out that the instances in Theorems  \ref{thm:hardness_1} and \ref{thm:hardness_2} admit a fully polynomial-time approximation scheme.

\section{Product and Nest Graphs}
\label{sec:product-and-nest-graphs}
\vspace{-5pt}

To capture the degrees of overlap across nests, we introduce product and nest graphs. These graphs represent the overlaps among products and nests from complementary perspectives. The structural properties of these graphs will allow us to identify conditions under which the assortment optimization problem is tractable.

\vspace{5pt}
\noindent {\bf \underline{Product Graph:}}  
For each $k \in K$, define
\(
\Mhat_k = \{ i \in M_k : \exists \, k' \in K \setminus \{k\} \text{ such that } i \in M_{k'} \},
\)
which corresponds to the set of products in $M_k$ that also appear in at least one other nest; that is, \mbox{$\Mhat_k = \bigcup_{\ell \ne k} (M_\ell \cap M_k)$}.
Letting $\Nhat = \bigcup_{k \in K} \Mhat_k$,  products in $\Nhat$ are referred to as {\it overlapping products} because they belong to at least two nests, and products in $N \setminus \Nhat$ are {\it nest-specific} products because they belong to only one nest. The product graph is then defined as $G_\product = (\Nhat, E_\product)$, where an edge $\{i,j\} \in E_\product$ exists between $i,j \in \Nhat$ if and only if there exists a nest $k \in K$ such that 
$i \in \Mhat_k$ and $j \in \Mhat_k$. %Two products are adjacent in the product graph if there exists a nest that contains both products. 
Nest-specific products that belong to exactly one nest never appear in $G_\product$. 
We show later that these products can be handled separately in the assortment optimization problem.

\vspace{5pt}
\noindent {\bf \underline{Nest Graph:}}  
The second graph, denoted by $G_\nest = (K,E_\nest)$, captures how nests overlap with one another. In the nest graph, each vertex corresponds to a nest $k \in K$, and an edge $\{k,\ell\} \in E_\nest$ exists if and only if $M_k \cap M_\ell \neq~\varnothing$. Thus, two nests are adjacent in $G_\nest$ if they share at least one product. The nest graph therefore encodes the connectivity structure among nests that arises from products that belong to multiple nests.

The following examples illustrate the product and nest graphs.  As a convention, we use a circle to represent a vertex in a product graph and a square to represent a vertex in a nest graph.

\begin{figure}[ht]
  \centering
  \begin{minipage}{0.4\textwidth}
    \centering    \includegraphics[
        height=1in,
        trim = 10pt 10pt 10pt 10pt,
        clip
    ]{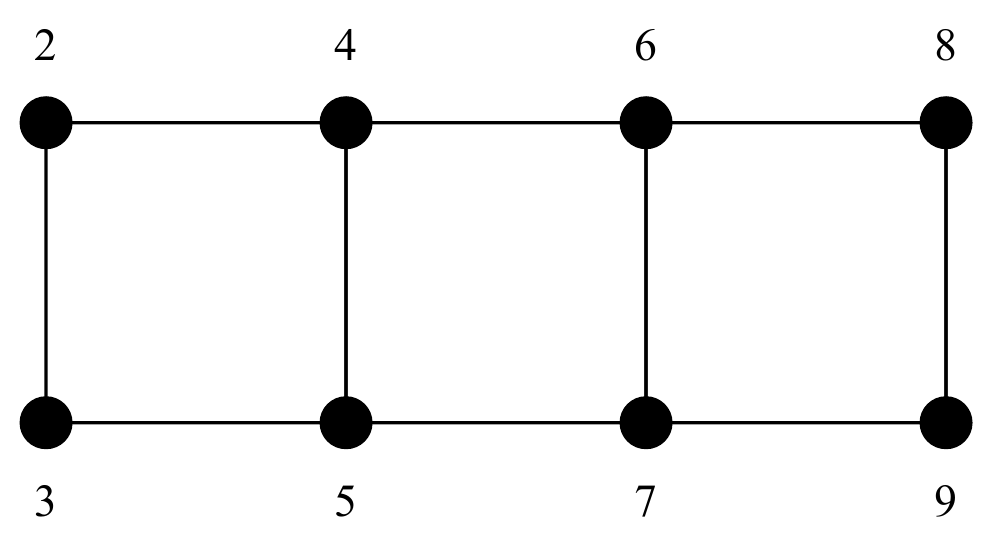}
  \end{minipage}~~~~~~~~~~~~~
  \begin{minipage}{0.4\textwidth}
    \centering    \includegraphics[
        height=1in,
		%width=0.5\linewidth,
        trim = 10pt 10pt 10pt 10pt,
        clip
    ]{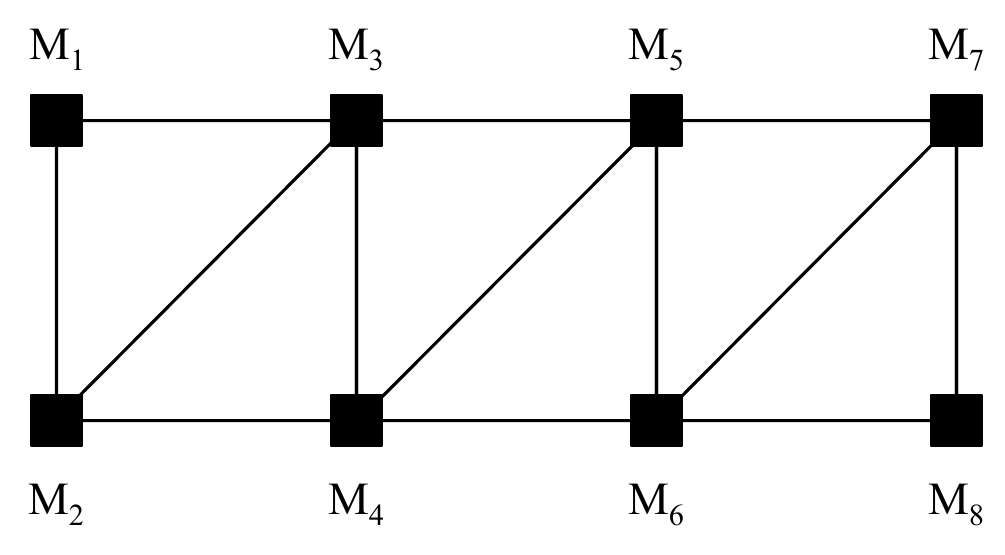}
  \end{minipage}
	\vspace{10pt}
  \caption{Left: The product graph of the ordered GEV model. Each circle corresponds to a product in $\Nhat$. Right: The associated nest graph. Each square corresponds to a nest in $K$.}
	\label{fig:ordered-GEV}
\end{figure}

\vspace{-22pt}
\begin{example}[Ordered GEV] \label{example:ordered-gev}
{\rm 
Suppose there are 10 products indexed by $\{1, 2, \ldots, 10\}$ and eight nests, where each nest is defined as \( M_i = \{i, i+1, i+2\} \) for \( i \in \{1,2,3,4,5,6,7,8\} \). This nest structure is known as the ordered GEV model \citep{small1987discrete}, which is motivated by settings in which alternatives are naturally ordered, for example by price or quality. In this example, we have \( \Nhat = \{2,3,4,5,6,7,8,9\} \). By construction of the ordered GEV model, there exists an edge between products \( i, j \in \Nhat \) if \( |i - j| \leq 2 \). The corresponding \pr{product and nest graphs} are in Figure \ref{fig:ordered-GEV}.}
\end{example}

\vspace{-0pt}
\begin{example}[Economy vs Business Seats in Parallel Flights]
%Basic vs Premium Versions] 
\label{example:basic-vs-premium}
{\rm 
We consider another example based on the product differentiation model \citep{bresnahan1996market}.  Consider a direct flight between two locations on a particular day. Suppose the airline offers $d$ different departure times during the day indexed by $t \in \{1, 2, \ldots, d\}$, and each customer can buy either a business (B) or economy (E) seat. Each product is thus an ordered pair $(t, c)$, where 
$t \in \{1, \ldots, d\}$ denotes the flight departure time and $c \in \{B, E\}$ denotes the seat type, so \mbox{$N = \{ (t, c) : t \in \{1, \ldots, d\}, c \in \{ B, E\} \}$}.  We construct the nests as follows. 
Let $M_{E} = \{ (t, E) \,:\, t = 1, \ldots, d\}$ be a nest consisting of all the economy-class seats and $M_{B} = \{ (t, B) \,:\, t = 1, \ldots, d\}$ a nest consisting of all the business-class seats. For $t = 1, \ldots, d$, let $M_t = \{ (t, B), (t, E)\}$ be a nest consisting of flights with a departure time $t$.   The nests $M_E$ and $M_B$ capture substitution among flight options for customers who are indifferent to departure times but care about service levels (either business- or economy-class). In contrast, each nest $M_t$ represents customers with a specific preference for flights departing at a particular time.
Figure \ref{fig:parallel-flights} shows the product and nest graphs when we have four different departure times ($d =4$).}
\vspace{-6mm}
\begin{figure}[ht]
  \centering
  \includegraphics
 % [
 %   trim={80pt 175pt 100pt 148pt},
 %   clip,
 %   width=0.4\textwidth
 % ]
 [scale = 0.4]{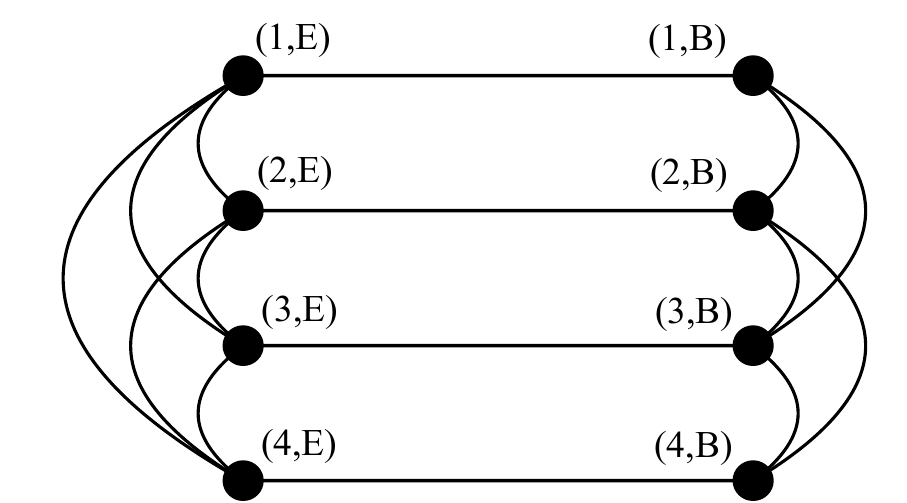}
  ~~~~~
  \includegraphics
  [scale = 0.4
%    trim={80pt 175pt 100pt 148pt},
%    clip,
%    width=0.4\textwidth
  ]{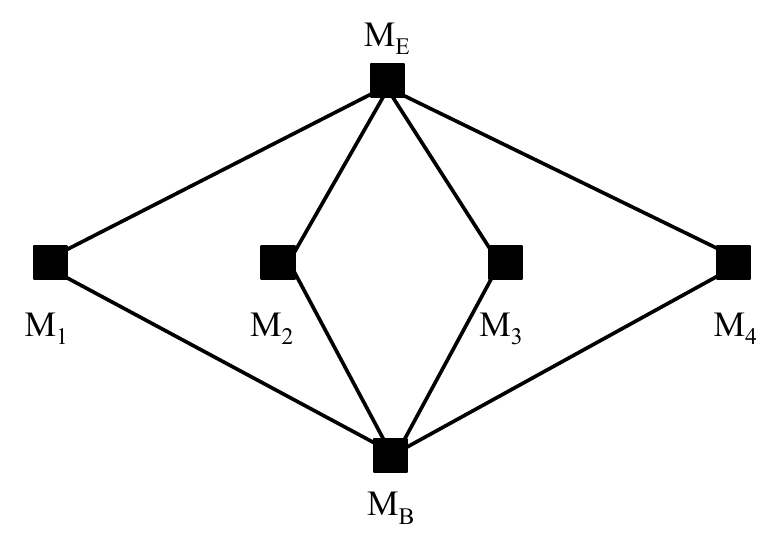}

  \vspace{-4pt}
  \caption{Left: The product graph of the model in Example \ref{example:basic-vs-premium} with $d=4$. On each side is a complete graph with 4 vertices. Right: The associated nest graph.}
  \label{fig:parallel-flights}
\end{figure}
\end{example}
%\prc{Give two more examples here. One for the GNL used in Theorem 2.1 and another one for the example used in Theorem 2.2}

\vspace{-30pt}
The product and nest graphs highlight different but complementary aspects of the generalized nested logit model. The product graph focuses on relationships among products that create non-trivial dependence across nests, while the nest graph summarizes the connectivity pattern among nests induced by products. These graph representations will %be crucial to our algorithmic results, which 
provide a unified framework for understanding when the assortment optimization problem is computationally tractable.

\vspace{-5pt}
\subsection{Tree Decomposition and Treewidth}

To analyze the complexity of the product and nest graphs, we make use of the {\it tree decomposition}. We begin by introducing its formal definition. 

\vspace{-5pt}
\begin{definition}[Tree decomposition]
Given an undirected graph $G = (V,E)$, a tree decomposition of $G$ is a connected tree $\Tree$ on a new set of nodes $\bag$, where each $b \in \bag$ corresponds to a subset $\vtcs{b}$ of vertices of $V$ and is referred to as a bag of vertices. The tree decomposition has the following three~properties: 
\begin{enumerate}
    \item $\bigcup_{b \in \bag} \vtcs{b} = V$.
    \item For every edge \pr{$(i,j) \in E$}, there exists $b \in \bag$ such that $\{i,j\} \subseteq \vtcs{b}$.
    \item For any $a,b,c \in \bag$, if $b$ lies on the path in $\Tree$ from $a$ to $c$, then $\vtcs{a} \cap \vtcs{c} \subseteq \vtcs{b}$.
\end{enumerate}
\end{definition}
The width of the tree decomposition $\Tree$ is equal to the size of the largest \pr{bag (set of vertices)} minus one; that is, $\max_{b \in \bag} |\vtcs{b}| - 1$.  A trivial tree decomposition is a tree with a single bag consisting of all vertices of the original graph.
The {\it treewidth } of a graph $G$, denoted by $\treewidth(G)$, is the minimum width among {all} tree decompositions of $G$. \yw{Throughout our paper, we also assume without loss of generality that each tree decomposition has at most $|V|$ nodes. This follows by removing ``redundant" bags; see \cite{kleinberg2006algorithm} Section 10.4 for details.}

\begin{figure}[h]
  \centering
  \begin{minipage}{0.48\textwidth}
    \centering    
    \includegraphics[
        height=1in,
        trim = 0em 0em 0em 0em,
        clip
    ]{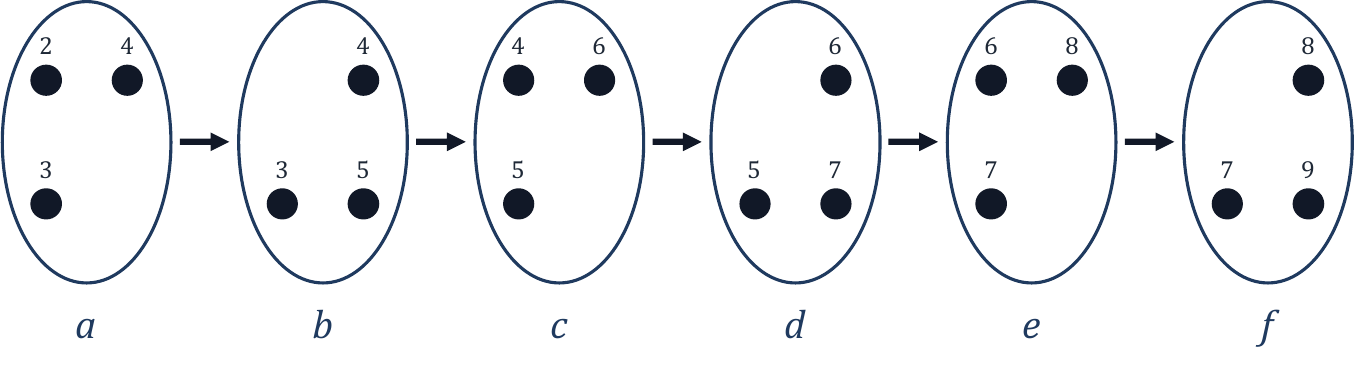}
  \end{minipage}~~~~~~~~~~~~~~\hspace{15pt}
  \begin{minipage}{0.48\textwidth}
    \centering    
    \includegraphics[
        height=1in,
		%width=0.6\linewidth,
        trim = 0em 0em 0em 0em,
        clip
    ]{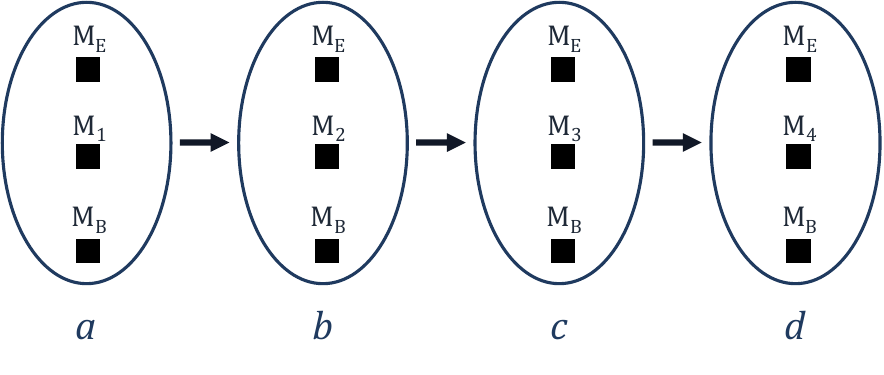}
  \end{minipage}
    \caption{Left: A tree decomposition of the product graph for Example \ref{example:ordered-gev}. Right: A tree decomposition of the nest graph for Example \ref{example:basic-vs-premium}.}
    \label{fig:tree_composition_exp}
	\vspace{-20pt}
\end{figure}

To illustrate a tree decomposition, we reuse the graph structures introduced in the previous section as an illustrative example. The left side of Figure~\ref{fig:tree_composition_exp} presents a valid tree decomposition of the product graph in Example \ref{example:ordered-gev}, and the right side shows a valid tree decomposition of the nest graph in Example \ref{example:basic-vs-premium}. The directed edges indicate the orientation of the tree, with node $a$ corresponding to the root of the tree. Each node in the tree decomposition corresponds to a subset of vertices from the original graph, either a subset of products for the product graph or a subset of nests for the nest graph. In this example, every node in the decomposition contains three vertices, implying that the tree decomposition has width of $2$. Under this example, the treewidth is also $2$.

%\begin{figure}[h]
%    \centering
%    \includegraphics[
%        width=0.6\linewidth,
%        trim = 0em 20em 0em 15em,
%        clip
%    ]{./pictures/product_graph_tree_decomposition.pdf}
%    \caption{A tree decomposition of the product graph for Example \ref{example:ordered-gev}. }
%    \label{fig:tree_composition_exp}
%\end{figure}

To streamline the discussion, we introduce notation to describe the  components of a tree decomposition. Given a node $b \in \bag$, we write $\parent(b)$ for the parent of $b$ in the tree, $\children(b)$ for the set of its children, and $\descendant(b)$ for the set of nodes that can be reached \pr{from} node $b$ in the directed tree, the descendants of $b$. By convention, we let $b \in \descendant(b)$. We then define the head of $b$ as $\head{b} = \vtcs{b} \cap \vtcs{\parent(b)}$ and $\tail{b} = \vtcs{b} \setminus \head{b}$.  Note that
$\head{b} \cap \tail{b} = \varnothing$ and $\vtcs{b} = \head{b} \cupdot \tail{b}$; we use $\cupdot$ to denote disjoint union. Finally, we define $\subtree{b} = \big(\; \bigcup_{a \in \descendant(b)} \vtcs{a} \; \big) \setminus \head{b}$, which collects all vertices appearing in bags of $b$ and its descendants, excluding those belonging to $\head{b}$. We refer to this as the subtree under node~$b$. \yw{We let $\root$ denote the root node of the tree and, by convention, we set $\head{\root}=\varnothing$.} To illustrate these definitions, we refer to the tree decompositions shown in Figure~\ref{fig:tree_composition_exp}. Consider node $c$ of the tree in the left figure.
 Its parent is node $b$, and its child is node $d$. Moreover, the set of descendants of $c$ is $\{c, d, e, f\}$. In this case, we have $\head{c} = \{4,5\}$, $\tail{c} = \{6\}$, and $\subtree{c} = \{6,7,8,9\}$.
For the right tree in Figure~\ref{fig:tree_composition_exp}, we have that $\head{c} = \{ M_E, M_B \}$, $\tail{c} = \{ M_3 \}$, and \mbox{$\subtree{c} = \{ M_3, M_4\}$}.
We note the difference in fonts: $\parent(b)$, $\children(b)$, and $\descendant(b)$ refer to nodes in the tree decomposition, while $\head{b}$, $\vtcs{b}$, $\tail{b}$, and $\subtree{b}$ refer to \pr{collections} of vertices in the original product graph or nest graph.

Intuitively, the set $\head{b}$ acts as a separator between $\subtree{b}$ and the vertices outside the subtree $V \setminus \subtree{b}$. In particular, there cannot exist an edge connecting a product in $\subtree{b}$ to a product in $V \setminus \subtree{b} \setminus \head{b}$, since such an edge would violate \pr{Property 2} of a tree decomposition and lead to a contradiction. %\prc{As stated, this is not correct. For example, in Figure \ref{fig:tree_composition_exp}, $6 \in \subtree{c}$ and $5 \in \vtcs(c)\setminus \subtree{c}$ and there is an edge connecting the two nodes. I think it should be $V \setminus \subtree{b} \setminus \head{b}$.}
This separation implies that the analysis of $\subtree{b}$ can be confined entirely to the subtree rooted at $b$ \pr{and $\head{b}$}. The next lemma focuses on the set of vertices in $\subtree{b}$ and shows that $\subtree{b}$ can be partitioned into disjoint subsets corresponding to the subtrees of its children together with $\tail{b}$.  \pr{We will apply this lemma to the tree decompositions of the product and nest graphs (Lemmas \ref{lem:prod_graph_nest_sep} and \ref{lem:nest_graph_prod_sep}) to establish the separations of the nests and products based on the tree decompositions.}

%We use $\bigcupdot$ to denote a disjoint~union.

\begin{lem}[Separation of Vertices] \label{lem:sep_vertices}
    For  $b \in \bag$,  $\subtree{b} =  \left( \bigcupdot_{a \in \children(b)} \subtree{a} \right) \bigcupdot \tail{b}  $. 
\end{lem}

\noindent {\it Proof:} Because $\subtree{b} = (\subtree{b} \setminus \tail{b}) \bigcupdot \tail{b}$, it  suffices to show that \mbox{$\subtree{b} \setminus \tail{b} = \bigcupdot_{a \in \children(b)} \subtree{a}$}. To see this, we first obtain the following chain of~equalities:
\begin{align*}
        \subtree{b}\setminus \tail{b}  
        & ~=~ 
         \bigg( \bigcup_{a \in \descendant(b)} \vtcs{a} \setminus \head{b} \bigg) \setminus \tail{b}   \overset{(i)}{=}  \bigg( \bigcup_{a \in \descendant(b) \setminus \{b\}} \vtcs{a} \bigg) \setminus \vtcs{b}
         \\
         &~\overset{(ii)}{=}~
         \bigg( \bigcup_{a \in \children(b)} \pr{\Big(\; \subtree{a} \cup \head{a} \;\Big)}\bigg) \setminus \vtcs{b} 
         \overset{(iii)}{=}    
         \bigcup_{a \in \children(b)} \subtree{a},
\end{align*}
where $(i)$ follows from $\head{b} \cupdot \tail{b} = \vtcs{b}$, and $(ii)$ follows from the definition of subtree. The last equality $(iii)$ follows because \mbox{$\subtree{a} \cap \vtcs{b} = \varnothing$} for $a \in \children(b)$. This holds because a path from a descendant of $a$ to $b$ must go through the node $a$. By Property~3 of the tree decomposition,  \pr{if $k \in \subtree{a} \cap \vtcs{b}$, then we must have \mbox{$k \in \vtcs{a}$}.} Because $b = \parent(a)$, it follows that $k \in \vtcs{a} \cap \vtcs{\parent(a)} = \head{a}$, which raises a contradiction since $\head{a}$ and $\subtree{a}$ are disjoint. Also, notice that $\head{a}\subseteq \vtcs{b}$ so we have \mbox{$(\subtree{a} \cup \head{a}) \setminus \vtcs{b} = \subtree{a}$}.
    
To complete the proof, it \pr{suffices} to show that for $a, a' \in \children(b)$ \yw{with $a \neq a'$ }, we have $\subtree{a} \cap \subtree{a'} = \varnothing$. We prove a stronger claim here, where we show $\subtree{a}  \cap (\subtree{a'} \cup \head{a'}) = \varnothing$ for $a, a' \in \children(b)$ \yw{with $a \neq a'$ }. By contradiction, we assume that $\subtree{a} \cap (\subtree{a'} \cup \head{a'}) \neq \varnothing$ so there exists a vertex $k$ such that $k \in \subtree{a} $ and $k \in \subtree{a'} \cup \head{a'} $. This indicates that there must exist a node in $\descendant(a)$ that contains $k$ and similarly, there must also exist a node in $ \descendant(a')$ that contains $k$ as well. Recall that both node $a$ and $a'$ are the children of $b$ so any path from a node in $\descendant(a)$ to another node in $\descendant(a')$ must pass through \pr{$a$, $a'$ and $b$}. By Property 3 of the tree decomposition, this indicates $k \in \vtcs{a}$ and $k \in \vtcs{b}$. Because $b = \parent(a)$, we have $k \in \head{a}$, but this is impossible because $\head{a} \cap \subtree{a} = \varnothing$! Therefore, it must be the case that $\subtree{a} \cap \subtree{a'} = \varnothing$.
\qed

%\prc{Should we add a discussion here about how to obtain a ``good'' tree decomposition?}

\yw{To close this section, we summarize two existing algorithms for constructing tree decompositions. When the treewidth of a graph is bounded by a constant, \cite{bodlaender1993linear} provides a linear-time algorithm that constructs a tree decomposition with the treewidth. For general graphs, \cite{feige2005improved} develop a polynomial-time algorithm that constructs a tree decomposition of width $O(w\sqrt{\log w})$, where $w$ is the treewidth of the graph.}

\subsection{Solution Framework}

Following the approach of \cite{davis2014assortment} and \cite{zhang2020assortment}, we reformulate the \ref{eqn:cnl_assort} problem into an equivalent formulation that involves finding a fixed point of a nonincreasing function.
To see this, note that for each \mbox{$S \subseteq N$},
$
\Pi(S) = \sum_{k \in K} \frac{ V_k(S)^{\gamma_k} R_k(S) }{1 + \sum_{\ell \in K} V_\ell(S)^{\gamma_\ell}} \geq z
$
if and only if 
$
\sum_{k \in K} V_k(S)^{\gamma_k} \big(R_k(S) - z\big) \geq z.
$
We define the function $F : \mathbb{R}_+ \to \mathbb{R}_+$ as follows: for each $z \in \Real_+$,
$
F(z) \;=\; \max_{S \subseteq N } \left\{ \sum_{k \in K} V_k(S)^{\gamma_k} \big(R_k(S) - z\big) \right\}.
$
The following lemma establishes the connection between this function and the optimal value $Z^*$ of the \ref{eqn:cnl_assort}~problem.

\begin{lem}\label{lem:cnl_opt}
The optimal value $Z^*$ of the \ref{eqn:cnl_assort} problem is the unique fixed point of the function~$F$; that is, $F(Z^*) = Z^*$.
Moreover, if $
    S^* = \argmax_{S \subseteq N } \left\{ \sum_{k \in K} V_k(S)^{\gamma_k} \big(R_k(S) - Z^* \big) \right\}$, then $S^*$ is an optimal solution to the \ref{eqn:cnl_assort} problem.
\end{lem}

The proof of Lemma~\ref{lem:cnl_opt} follows the same argument as in \cite{gallego2014constrained} and is therefore omitted. Consider the equation $z = F(z)$. Since the left-hand side is strictly increasing in $z$, while $F(z)$ is continuous and nonincreasing, the equation $ z = F(z)$ admits a unique solution. This solution can be computed via a standard binary search over an interval that contains the optimal objective value. At each iteration of the binary search, we evaluate $F(z)$ at the midpoint of the current interval and update the interval depending on whether $F(z)$ lies above or below $ z$. Such procedures are standard in the revenue management literature; see, for example, \cite{alfandari2021exact}. The above argument implies that if $F(z)$ can be evaluated exactly in polynomial time, then the optimal objective value $\hat{z}$ can be computed within an additive error $\epsilon$,  with a running time that is polynomial in the input size and $\log(1/\epsilon)$. Moreover, if $F(z)$ cannot be evaluated exactly but admits a fully polynomial-time approximation scheme, then a similar procedure yields a  fully polynomial-time approximation scheme for the \ref{eqn:cnl_assort} problem. For completeness, we state and prove this result in Appendix \ref{sec:using_fixed_point_proof}.

 %\prc{Use Yuheng's proof instead. It's much easier.}

\begin{thm}[Using $F(z)$ to Solve \ref{eqn:cnl_assort} Problem] 
\label{thm:using-fixed-point-to-solve-assortment-problem}
If we can compute $F(z)$ exactly in polynomial time for every $z \in \Real_+$, then we can solve the \ref{eqn:cnl_assort} problem within an {\it additive error}~$\epsilon$, with a running time that is polynomial in the input size and $\log(1/\epsilon)$.
%$n$, $m$, $\log \max_{i \in N, k \in K} v_{ik}$, and $\log(1/\epsilon)$.  
If there is a \pr{fully} polynomial-time approximation scheme 
to approximate $F(z)$  for every $z \in \Real_+$, then we also have a fully polynomial-time approximation scheme for the \ref{eqn:cnl_assort} problem with a {\it multiplicative error}~$\epsilon$, with a running time that is polynomial in the input size and $1/\epsilon$.
\end{thm}

By Theorem \ref{thm:using-fixed-point-to-solve-assortment-problem}, in the remainder of this manuscript, we focus on developing efficient methods for approximately evaluating $F(z)$ {\it at a fixed value of $z$}. Note that the objective function of $F(z)$ can be written as $\sum_{k \in K} V_k(S)^{\gamma_k} \big(R_k(S) - z\big) = \sum_{k \in K} \frac{\sum_{i \in S} v_{ik} (f_i - z) }{V_k(S)^{1 - \gamma_k}}$. To streamline our discussion, we let $r_i = f_i - z$ and we focus on solving the following optimization problem
\begin{equation} \label{eqn:sub_param_new} \tag{{\sf Aux-Assort}}
    F^*=\max_{S \subseteq N} ~ \sum_{k \in K} \frac{\sum_{i \in S} v_{ik} r_i }{V_k(S)^{1 - \gamma_k}}. 
\end{equation}
In the \ref{eqn:sub_param_new} problem above, the revenue $r_i$ may be negative. However, it is without loss of generality to assume $r_i \ge 0$ for all $i \in N$. \pr{To see this, for each $S \subseteq N$, let \mbox{$\eta_i(S) = \sum_{k \in K} v_{ik}/ \big( V_k(S) \big)^{1-\gamma_k}$}, and it follows that $F^* = \max_{S \subseteq N} \sum_{i \in S} r_i \eta_i(S)$.  For any assortment~$S$, let $B_S = \{i \in S \,:\, r_i < 0\}$ denote all the products in $S$ with negative revenue. Then, we have that $\sum_{i \in S \setminus B} r_i \eta_i(S\setminus B) \geq \sum_{i \in S} r_i \eta_i(S)$ because $\gamma_k \in (0,1]$ for all $k$, so for each $i \in S \setminus B$, \mbox{$\eta_i(S \setminus B) \geq \eta_i(S)$} by definition. Thus,  any assortment $S$ containing products with negative revenue  can be improved by removing those products.} We can therefore restrict attention to the case $r_i \ge 0$ for all $i \in N$. Our goal in the following sections is to develop an efficient way of computing or approximating $F^*$. To close this section, the following theorem shows that it is NP-hard to approximate the  \ref{eqn:sub_param_new} problem within any constant factor.

\begin{thm}[Hardness of the \ref{eqn:sub_param_new} Problem] \label{thm:subproblem_hardness}
    Letting $m = |K|$, the \ref{eqn:sub_param_new} problem is NP-hard to approximate within $O(1/m^{1-\delta} )$ for any $\delta > 0$. 
\end{thm}
The proof %of~this~theorem 
is deferred to  Appendix \ref{sec:subproblem_hardness}. It is based on a reduction from the maximum independent set problem, which is known to be NP-hard to approximate within any constant factor. We highlight two important remarks. First, the proof applies specifically to the \ref{eqn:sub_param_new} problem and does not directly imply that the overall assortment optimization problem under the generalized nested logit model inherits the same hardness of approximation. In fact, whether the \ref{eqn:cnl_assort} problem itself is NP-hard to approximate within $O(1/m^{1-\delta})$ for any $\delta > 0$ remains, to the best of our knowledge, an open question. Second, although the reduction shows that the \ref{eqn:sub_param_new} problem is NP-hard to approximate within any constant factor, the construction relies on instances with a highly intricate structure of the nests. As a result, this does not rule out the possibility that the \ref{eqn:sub_param_new} problem, and consequently the assortment optimization problem, may be tractable in settings with more structured nest configurations. We characterize such cases in the following sections.

\section{\mbox{A Polynomial-time Algorithm for Product Graphs with Bounded Treewidths}}
\label{sec:product-graph-bounded-tree-width}

%We develop an algorithm for the \ref{eqn:sub_param_new} problem based on the tree decomposition over the product graph $G_\product$. 

Letting the tree decomposition over the product graph be denoted by \mbox{$\Tree_\product = (\bag_\product, \edge_\product)$}, we will use $\Tree_\product$ to develop an algorithm for the \ref{eqn:sub_param_new} problem.
In this section, for any $p \in \bag_\product$, the set $\vtcs{p}$, $\head{p}$, $\subtree{p}$ and $\tail{p}$ correspond to sets of {\it products}. We first \yw{give} a few important structural properties of the tree decomposition over the product graph and then discuss a dynamic program to solve the \ref{eqn:sub_param_new} problem.

\subsection{Properties of Nest-Specific Products}
\label{sec:nest-specific-products-properties}

Recall from the previous section that the vertices of the product graph $G_\product$ consist of overlapping products that belong to at least two nests, corresponding to $\Nhat$. To analyze our algorithm, we first isolate the products specific to a single nest, corresponding to $N \setminus \Nhat$, which we refer to as the {\it nest-specific products}. For these products, we establish that the optimal assortment exhibits a revenue-ordered structure. To see this, we first rewrite the \ref{eqn:sub_param_new} problem  as follows:
$$
    F^* ~=~ \max_{S \subseteq N} \left\{ \sum_{k \in K} \frac{\sum_{i \in S} v_{ik} r_i }{V_k(S)^{1 - \gamma_k}} \right\} 
    ~\overset{(i)}{=} \max_{\Shat \subseteq \Nhat}~ \left\{~ \sum_{k \in K} ~\max_{T_k \subseteq M_k \setminus \Mhat_k}~ \frac{\sum_{i \in \Shat \cupdot T_k } v_{ik} r_i}{V_k(\Shat \cupdot T_k)^{1 - \gamma_k}} ~\right\},
$$
where $(i)$ follows from the fact that $M_k \setminus \Mhat_k$ consists of nest-specific products that only belong to nest $k$. Once we fix $\Shat \subseteq \Nhat$, the optimization over the remaining nest-specific products can be carried out independently within each nest.  %Therefore, to analyze the structure of the optimal solution for the nest-specific products for nest~$k$, it suffices to fix an arbitrary assignment for $\Shat \subseteq \Nhat$.
 The following theorem shows that once $\Shat$ is fixed, an optimal subset of nest-specific products is revenue-ordered.

\begin{thm}[Optimal Nest-Specific Products are Revenue-Ordered]
\label{thm:nest-specific-products}
For each $\Shat \subseteq \Nhat$ and $k \in K$, there exists a revenue-ordered subset that is an optimal solution to the optimization problem $\max_{T_k \subseteq M_k \setminus \Mhat_k}~ \big( \sum_{i \in \Shat \pr{\cupdot} T_k } v_{ik} r_i \big) / V_k(\Shat \pr{\cupdot} T_k)^{1 - \gamma_k}$; that is, there exists an optimal solution
 $T_k^*(\Shat)$ and  $\tau_k \in \pr{\Real}$ such that $T_k^*(\Shat) = \{ i \in M_k \setminus \Mhat_k \,:\, r_i \geq \tau_k \}$.
\end{thm}
\noindent{\it Proof:} 
Fix an arbitrary $\Shat \subseteq \Nhat$ and $k \in K$. Define $W: \Real_+^2 \to \Real_+$ as follows: for each \mbox{$(x,y) \in \Real^2_+$}, $W(x,y) = ( x + \sum_{i \in \Shat} v_{ik} r_i ) / ( y + \sum_{i \in \Shat} v_{ik})^{1-\gamma_k}$ . Because the numerator is affine in $x$ and the denominator is concave in $y$ since $0 < \gamma_k \leq 1$, $W$ is a quasiconvex function in $\Real_+^2$. Moreover,
$$
\max_{T_k \subseteq M_k \setminus \Mhat_k}~ \big( \sum_{i \in \Shat \pr{\cupdot} T_k } v_{ik} r_i \big) / V_k(\Shat \pr{\cupdot} T_k)^{1 - \gamma_k}
~=~ \max_{T_k \subseteq M_k \setminus \Mhat_k}  W\Big(  \sum_{i \in T_k}  v_{ik} r_i ~,~ \sum_{i \in T_k}  v_{ik} \Big)
$$
The result follows from Lemma~C.5 by \cite{SantiagoDesir:2023}, who considered the optimization problem $\max_{S \subseteq {\cal S}} W( \sum_{i \in S} a_i \,,\, \sum_{i \in S} b_i )$, where ${\cal S}$ denotes the collection of subsets, the vectors $\bm{a}, \bm{b}$ are nonnegative, and $W: \Real^2 \to \Real$ is a quasiconvex function. 
Letting $S^*$ denote an optimal assortment,  \cite{SantiagoDesir:2023} show that there exist $\lambda, \mu \in \Real$ such that $S^*$ is an optimal solution to \mbox{$\max_{S \in {\cal S}} \sum_{i \in S} (\lambda a_i + \mu b_i)$}.  Applying this result to our setting, ${\cal S} = 2^{M_k \setminus \Mhat_k }$, $a_i = v_{ik} r_i $ and $b_i = v_{ik}$ for all $i \in M_k \setminus \Mhat_k$. Thus, the optimal assortment $T_k^*(\Shat)$ can be found by solving the problem $\max_{T_k \subseteq M_k \setminus \Mhat_k} \sum_{i \in T_k} (\lambda v_{ik} r_i  + \mu v_{ik})$ for some \mbox{$\lambda, \mu \in \Real$}. Because the function $W$ is increasing in $x$ and decreasing in $y$, it follows that $\lambda > 0$. \pr{Because $\lambda > 0$,} factoring out $\lambda v_{ik}$ and writing $\mu/\lambda = -\tau_k$ for some \pr{$\tau_k \in \Real$}, it follows $T_k^*(\Shat)$ is an optimal solution to the problem  $\max_{T_k \subseteq M_k \setminus \Mhat_k} \sum_{i \in T_k} (r_i - \tau_k) v_{ik}$. Thus, $T_k^*(\Shat) = \{ i \in M_k \setminus \Mhat_k \,:\, r_i \geq \tau_k \}$, which is the desired result.
\qed

We note an immediate implication for the \ref{eqn:cnl_assort} problem.
Letting $S^*$ denote an optimal solution to the \ref{eqn:cnl_assort} problem, Theorem \ref{thm:nest-specific-products} implies that the set \mbox{$S^* \cap (M_k \setminus \Mhat_k)$} is revenue-ordered; that is, $S^* \cap  (M_k \setminus \Mhat_k) = \{\, i \in M_k \setminus \Mhat_k : r_i \geq \tau_k \,\}$ for some $\tau_k$.

\vspace{-10pt}
\subsection{Properties of Tree Decomposition over Product Graph}
\label{sec:product-graph-properties}

We discuss two important properties of a tree decomposition over the product graph. For each $i \in N$,  let $\nest(i) = \{ k \in K \,:\, i \in  M_k\}$ denote the set of nests that contain product $i$, and \mbox{$\nest(S) = \bigcup_{i \in S} \nest(i)$} denote the set of nests associated with products in $S$. For each $p \in \bag_\product$, the next lemma shows that all products belonging to the same nests as products in $\subtree{p}$ must be contained in $\subtree{p} \cupdot \head{p}$; that is, $\cup_{k \in \nest ( \subtree{p} )} \Mhat_k \subseteq \subtree{p} \cupdot \head{p}$. %Thus, the subtree $\subtree{p}$ contains all related products. 
\pr{This lemma ensures that computations involving products at node $p$ can be performed without having to consider products outside the subtree.}

\begin{lem}[Subtree Contains All Related Products] \label{lem:prod_graph_domain}
For $p \in \bag_\product$ and \mbox{$k \in \nest ( \subtree{p})$},  $\Mhat_k \subseteq \subtree{p} \cupdot \head{p} $.
\end{lem}
\noindent{\it Proof:}
The claim is trivial when $p = \root$ because $\subtree{\root} = \Nhat$, so assume that $p \ne \root$. Suppose, on the contrary, that there exists a nest $k \in \nest \big( \subtree{p} \big)$ and a product $i \in \Mhat_k$ such that $i \notin \subtree{p} \cup \head{p}$.  Because $k \in \nest \big( \subtree{p} \big)$, there exists a product $j \in \subtree{p}$ such that  $j \in \Mhat_k$. Let $a \in \descendant(p) $ be such that $j \in \vtcs{a}$. 
 
By definition of the product graph \pr{$G_\product$, $(i, j) \in E_\product$ }and by Property 2 of the tree decomposition, there exists a bag $c \in \bag_\product$ such that $\{i, j\} \subseteq \vtcs{c}$. Because \mbox{$i \notin \subtree{p} \cupdot \head{p}$}, we must have $c \notin \descendant(p)$. Therefore, node $c$ must not lie in the subtree under node $p$.  We consider the path from $c$ to $a$. Because node $c$ does not lie in the subtree under $p$ and node $a$ lies in the subtree under $p$, both $p$ and $\parent(p)$ are included in the path. By Property 3 of the tree decomposition, we must have $\vtcs{a} \cap \vtcs{c} \subseteq \vtcs{\parent(p)}$ and \mbox{$\vtcs{a} \cap \vtcs{c} \subseteq \vtcs{p}$}. Because $j \in \vtcs{a} \cap \vtcs{c}$, we have  $j \in \vtcs{\parent(p)}$ and \mbox{$j \in \vtcs{p}$}, which implies that  $j \in \head{p}$.  But we also have $ j \in \subtree{p}$. This is a contradiction because \mbox{$\subtree{p} \cap \head{p} = \varnothing$}! Thus, we must have  $\Mhat_k \subseteq \subtree{p} \cupdot \head{p}$. \qed

%We then consider a useful separation property over the nests induced by the tree decomposition over the product graph. 
Recall that Lemma~\ref{lem:sep_vertices} establishes a useful separation of the vertices based on the tree decomposition. Since we are working with the tree decomposition over the product graph, this result implies a separation of the products. In the next lemma, we show that a similar separation property holds for the nests as~well. \pr{The separation of nests across the children of node $p$ shown in this lemma allows us to perform computation at each child node separately and combine the results together, forming the basis for our dynamic programming method.}

\begin{lem}[Nest Separation in Product Graph] \label{lem:prod_graph_nest_sep}
For each $p \in \bag_\product$, 
    $$\nest(\subtree{p}) = \big\{k \in \nest(\tail{p}) : \Mhat_k \subseteq \vtcs{p} \big\} \bigcupdot \Big( \bigcupdot_{a \in \children(p)} \nest( \subtree{a}  )  \Big).$$
\end{lem}

\noindent{\it Proof:}
Consider an arbitrary $p \in \bag_\product$. Let  $\Dcal_p = \subtree{p} \setminus \vtcs{p} = \subtree{p} \setminus \tail{p}$, where the second equality follows because $\vtcs{p} = \head{p} \cupdot \tail{p}$ and $\subtree{p} \cap \head{p} = \varnothing$. 
%We first partition $\nest(  \subtree{p} )$ to two disjoint sets as follows:
Because
$
        \nest(  \subtree{p} ) = \Big( \nest( \subtree{p}) \setminus  \nest( \Dcal_p) \Big) \bigcupdot \nest( \Dcal_p)$, it suffices to prove the following~claims. 
  
\vspace{5pt}  
    {\bf Claim 1:} $ \nest( \subtree{p}) \setminus  \nest( \Dcal_p)  = \big\{k \in \nest(\tail{p}) : \Mhat_k \subseteq \vtcs{p} \big\}$.

    To prove Claim 1, let $K_p = \big\{k \in \nest(\tail{p}) : \Mhat_k \subseteq \vtcs{p} \big\}$. We will first show \mbox{$K_p \subseteq \nest\big( \subtree{p} \big) \setminus \nest \big(  \Dcal_p \big)$}. Consider an arbitrary $k \in K_p$. By definition, \mbox{$k \in \nest(\tail{p})$} and $\Mhat_k \subseteq \vtcs{p}$. \pr{Then,} $k \in \nest\big( \subtree{p} \big)$ because \mbox{$\tail{p} \subseteq \subtree{p}$}, and \mbox{$k \notin \nest \big(  \Dcal_p \big)$} because $\Mhat_k \subseteq \vtcs{p}$, so \mbox{$k \in \nest\big( \subtree{p} \big) \setminus \nest \big(  \Dcal_p \big)$}. This implies that \mbox{$K_p \subseteq \nest\big( \subtree{p} \big) \setminus \nest \big(  \Dcal_p \big)$}.  To establish the other direction, consider an arbitrary \mbox{$k \in \nest\big( \subtree{p} \big) \setminus \nest \big(  \Dcal_p \big)$}. By Lemma~\ref{lem:prod_graph_domain}, \mbox{$ \Mhat_k \subseteq \subtree{p} \cupdot \head{p} = \Dcal_p \cup \vtcs{p}$}. Because $k \notin \nest \big(  \Dcal_p \big)$  and $\Dcal_p = \subtree{p} \setminus \vtcs{p} = \subtree{p} \setminus \tail{p}$, it follows that $k \in \nest(\tail{p})$ and \mbox{$\Mhat_k \subseteq (\Dcal_p \cup \vtcs{p} ) \setminus \Dcal_p = \vtcs{p}$}, so $k \in K_p$. This implies $\nest\big( \subtree{p} \big) \setminus \nest \big(  \Dcal_p \big) \subseteq K_p$, and thus, \mbox{$K_p = \nest( \subtree{p}) \setminus  \nest( \Dcal_p)$}, proving Claim 1.
    
 \vspace{5pt}  
    {\bf Claim 2:} $\nest( \Dcal_p) = \bigcupdot_{a \in \children(p)} \nest(\subtree{a})$.

    To prove Claim 2, recall that Lemma \ref{lem:sep_vertices} gives $\Dcal_p =  \bigcupdot_{a \in \children(p)} \subtree{a} $. By definition, we must have $\nest(\bigcupdot_{a \in \children(p)} \subtree{a}) = \bigcup_{a \in \children(p)} \nest(\subtree{a})$.  To complete the claim, it thus suffices to show that $\nest(\subtree{a}) \cap \nest(\subtree{a'}) = \varnothing$ for $a, a' \in \children(\pr{p})$ such that $a \ne a'$.  Suppose, on the contrary, that there exists $k \in \nest(\subtree{a}) \cap \nest(\subtree{a'})$. Then, by Lemma~\ref{lem:prod_graph_domain}, we have $\Mhat_k \subseteq \subtree{a} \cup \head{a}$ and $\Mhat_k \subseteq \subtree{a'} \cup \head{a'}$. In Lemma \ref{lem:sep_vertices}, we show that \mbox{$\subtree{a}  \cap (\subtree{a'} \cup \head{a'}) = \varnothing$} so we must have \mbox{$\Mhat_k \subseteq (\subtree{a} \cup \head{a}) \cap (\subtree{a'} \cup \head{a'}) = \head{a} \cap \head{a'} $}. This indicates that $k \notin  \nest(\subtree{a}) $ and $k \notin \nest(\subtree{a'})$, but this is a contradiction! Therefore, Claim 2 holds. \qed

Lemmas~\ref{lem:prod_graph_domain} and \ref{lem:prod_graph_nest_sep} establish a strong separation property of the product graph under tree decomposition. Lemma~\ref{lem:prod_graph_domain} ensures that whenever a nest is associated with products in $\subtree{p}$, all of its overlapping products are fully contained in the subtree rooted at $p$. Building on this, Lemma~\ref{lem:prod_graph_nest_sep} shows how the nests associated with $\subtree{p}$ can be partitioned into disjoint components determined by $\vtcs{p}$ and the subtrees of its children. This decomposition highlights how both products and nests can be localized within subtrees, which is the key insight that allows us to design a dynamic programming algorithm when the product graph has a bounded treewidth.

\vspace{-10pt}
\subsection{A Dynamic Program to Solve the \ref{eqn:sub_param_new} Problem}
\label{sec:dynamic-program-bounded-bounded-product-treewidth}

We present a dynamic program to solve the \ref{eqn:sub_param_new} problem by exploiting the property of nest-specific products in Theorem \ref{thm:nest-specific-products} and the structure of the product graph through its tree decomposition as discussed in Lemmas \ref{lem:prod_graph_domain} and  \ref{lem:prod_graph_nest_sep}.    
Without loss of generality, we assume that the tree decomposition results in a single tree. If the decomposition yields a forest, we can process each tree independently and combine its value.  We can also assume that $\nest(\Nhat) = K$; that is, every nest contains at least one product that belongs to multiple nests. To see this, if there exists a nest in $K \setminus \nest(\Nhat)$ which contains only nest-specific products, we can introduce a dummy product with zero revenue that is shared among these nests, \pr{thus} ensuring that all nests are included in~$\nest(\Nhat)$. 

For each $k \in K$, let $\Ocal_k \subseteq 2^{M_k \setminus \Mhat_k}$ denote the collection of revenue-ordered subsets of $M_k \setminus \Mhat_k$; that is, $\Ocal_k = \Big\{ A \subseteq M_k \setminus \Mhat_k \,\big|\, A = \{ i \in M_k \setminus \Mhat_k \,:\, r_i \geq \tau \} \text{ for some } \tau \in \Real \Big\}$. Note that \pr{$|\Ocal_k| \leq n+1$} for all $k$.  Also, for each $S \subseteq N$, let $\psi_k(S) \;\equiv\; \big( \sum_{i \in S} v_{ik} r_i \big) / V_k(S)^{1-\gamma_k}$. \pr{By definition of \ref{eqn:sub_param_new},} %From the definition of the \ref{eqn:sub_param_new} problem, 
$$
	F^* ~=~ \max_{S \subseteq N} ~\sum_{k \in K} \psi_k(S) ~=~ \max_{S \subseteq \Nhat} ~\sum_{k \in K} ~\max_{A_k \in \Ocal_k} \psi_k( S \cupdot A_k)~,
$$
where the last equality follows because $v_{ik} = 0$ for all $i \notin M_k$, and thus, $\psi_k(S) = \psi_k(S \cap M_k)$ for all $S \subseteq N$. Moreover, by Theorem \ref{thm:nest-specific-products}, it suffices to only consider revenue-ordered subsets of $M_k \setminus \Mhat_k$.   Note that $M_k \setminus \Mhat_k$ consists only of products that belong only to nest $k$. Thus, it suffices to consider sets of the form $S \cupdot A_k$ where $S \subseteq \Nhat$ and $A_k \in \Ocal_k$.

We will use the tree decomposition $\Tree_{\product}$ of the product graph to develop a dynamic programming algorithm for computing $F^*$.  For each $p \in \bag_\product$, define a value function $J_p^\product: 2^{\head{p}} \to \Real$ as follows: for each $S \subseteq \head{p}$,
\begin{align*}
    J^{\product}_p \left( ~S~ \right) 
    & ~=~
    \max_{ T \subseteq \subtree{p} } \sum_{k \in \nest (\subtree{p} )} \max_{A_k \in \Ocal_k} \psi_k\big( S \cupdot T \cupdot A_k \big).
\end{align*}
\pr{Noting that $\head{p} \cap \subtree{p} = \varnothing$, the state variable $S$ captures the products in $\head{p}$ that will be offered in the assortment, and the value function $J^{\product}_p \left( S \right)$ captures the maximum  revenue that can be obtained from all products in the subtree of node $p$, given that $S$ is offered.
}
Because $\head{\root} = \varnothing$ and $\subtree{\root} = \Nhat$, $J^{\product}_\root \left( \varnothing \right) = \max_{T \subseteq \Nhat} \sum_{k \in K} \max_{A_k \in \Ocal_k} \psi_k( T \cupdot A_k)  = F^*$.

Our goal is to exploit the tree decomposition $\Tree_\product$  of the product graph to derive a dynamic programming equation for computing the value function. We give some intuition before stating the main result.
Fix a bag $p$ in the tree decomposition, and consider the set of products $\subtree{p}$ that are exclusively contained in the subtree rooted at $p$. Lemmas~\ref{lem:sep_vertices} and \ref{lem:prod_graph_nest_sep} imply two key separation properties. First, the products in $\subtree{p}$ can be partitioned into
$\bigcupdot_{a \in \children(p)} \subtree{a}$ and $\tail{p}$.
Second, $\nest(\subtree{p})$ can also be partitioned to
$
\big\{ k \in \nest(\tail{p}) : \Mhat_k \subseteq \vtcs{p} \big\}
$
and
$
\bigcupdot_{a \in \children(p)} \nest(\subtree{a})
$.

Together, these separations allow us to decompose the problem at node $p$ into independent subproblems over its children, plus a local term contributed by the products under $\vtcs{p}$. Finally, Lemma~\ref{lem:prod_graph_domain} shows that, to evaluate the contribution of $\nest(\subtree{p})$ to the objective, it suffices to know the offered products in $\subtree{p} \cup \head{p}$. Because the choice within $\subtree{p}$ is captured by the recursion, the only additional information we must carry is the assortment offered at $\head{p}$. This is the reason why we use the offered set in $\head{p}$ as our state variable.
In the dynamic program, the objective is to determine the optimal way to offer products in $\subtree{p}$, given fixed decisions made for products in $\head{p}$, in order to maximize the value contribution from the nests in $\nest(\subtree{p})$. The choice of $\nest(\subtree{p})$ is motivated by Lemma~\ref{lem:prod_graph_domain}, where $\Mhat_k \subseteq \subtree{p} \cupdot \head{p}$ for any $k \in \nest(\subtree{p})$ and this ensures that  the objective function $\sum_{k \in \nest (\subtree{p} )} \max_{A_k \in \Ocal_k} \psi_k( S \cupdot T \cupdot A_k)$ is well-defined given $S \subseteq\head{p}$ and $ T \subseteq \subtree{p} $.

The following theorem gives the dynamic programming recursion for the value function $J^{\product}_p$ and the running time for computing $F^*$.
\begin{thm}\label{thm:exact_algo}
The value function $J^\product_p$ satisfies the following dynamic programming equation: for each $p \in \bag_\product$ and $S \subseteq \head{p}$,
$$
J^\product_p(S) ~=~
\max_{T_p \subseteq \tail{p}} \left\{~ \sum_{\substack{ k \in \nest(\tail{p}) \\ s.t. ~ \Mhat_k \subseteq \vtcs{p} }} \max_{A_k \in \Ocal_k} \psi_k( S \cupdot T_p  \cupdot A_k ) 
    + 
    \sum_{a \in \children(p)} J^{\product}_a\Big( \big(S \cupdot T_p\big) \cap \head{a} \Big) ~\right\}~.
$$
Starting from the leaf nodes of $\Tree_\product$, the running time to compute $J^{\product}_\root \left( \varnothing \right)$ is $O\big( \, m \, n^2 \, 2^{\treewidth(G_\product)} \,\big)$.
\end{thm}
\noindent {\it Proof:}
We obtain the following chain of equalities
\begin{align*}
    & J^{\product}_p  \left( ~S~ \right)
    ~=~
    \max_{ T \subseteq \subtree{p} } \sum_{k \in \nest (\subtree{p} )} \max_{A_k \in \Ocal_k} \psi_k\big( S \cupdot T  \cupdot A_k \big) \\
    &  \overset{(i)}{=} \max_{ T \subseteq \subtree{p}}  \Bigg\{\sum_{\substack{ k \in \nest(\tail{p}) \\ s.t. ~ \Mhat_k \subseteq \vtcs{p} }} \max_{A_k \in \Ocal_k} \psi_k( S \cupdot T \cupdot A_k) ~+~ \sum_{a \in \children(p)} \sum_{\ell \in \nest( \subtree{a} ) } \max_{A_\ell \in \Ocal_\ell} \psi_\ell( S \cupdot T \cupdot A_\ell )\Bigg\} \\
    & \overset{(ii)}{=} 
    \max_{T_p \subseteq \tail{p} } \Bigg\{ \sum_{\substack{ k \in \nest(\tail{p}) \\ s.t. ~ \Mhat_k \subseteq \vtcs{p} }} \max_{A_k \in \Ocal_k} \psi_k( S \cupdot T_p \cupdot A_k ) \\
    &  ~~~~\qquad \qquad \qquad \qquad \qquad \qquad 
    + 
    \sum_{a \in \children(p)} \max_{T_a \subseteq \subtree{a}} \Bigg\{\sum_{\ell \in \nest( \subtree{a} ) } \max_{A_\ell \in \Ocal_\ell}  \psi_\ell( S \cupdot T_p \cupdot T_a \cupdot A_\ell) \Bigg\} \Bigg\} \\
    & \overset{(iii)}{=} \max_{T_p \subseteq \tail{p}} \Bigg\{ \sum_{\substack{ k \in \nest(\tail{p}) \\ s.t. ~ \Mhat_k \subseteq \vtcs{p} }}  \max_{A_k \in \Ocal_k} \psi_k( S \cupdot T_p  \cupdot A_k) 
    + 
    \sum_{a \in \children(p)} J^{\product}_a\Big( \big(S \cupdot T_p\big) \cap \head{a} \Big) \Bigg\} ,
\end{align*}
where $(i)$  is from 
\mbox{$\nest(\subtree{p}) = \big\{k \in \nest(\tail{p}) : \Mhat_k \subseteq \vtcs{p} \big\} \bigcupdot \Big( \bigcupdot_{a \in \children(p)} \nest( \subtree{a}  )  \Big)$}, by Lemma \ref{lem:prod_graph_nest_sep}.
For equality $(ii)$, note that \mbox{$\subtree{p} = \tail{p} \bigcupdot \big( \bigcupdot_{a \in \children(p)} \subtree{a} \big)$}
by Lemma \ref{lem:sep_vertices}. This means that each $T \subseteq \subtree{p}$ can be written as $T = T_p  \cupdot \big( \cupdot_{a \in \children(p)} T_a \big)$, where $T_p \subseteq \tail{p}$ and $T_a \subseteq \subtree{a}$ for all $a \in \children(p)$.  Because $\psi_k(X) = \psi_k(X \cap M_k)$ for all $X \subseteq N$, it follows that for every $k \in \nest(\tail{p})$ such that $\Mhat_k \subseteq \vtcs{p}$, 
$$
\max_{A_k \in \Ocal_k} \psi_k( S \cupdot T \cupdot A_k) 
=  \max_{A_k \in \Ocal_k} \psi_k( S \cupdot T_p \cupdot A_k) 
$$
because $T \cap \Mhat_k = T_p \cap \Mhat_k$.
Moreover, for each $a \in \children(p)$ and $\ell \in \nest(\subtree{a})$, it follows from Lemma~\ref{lem:prod_graph_domain} that $\Mhat_\ell \subseteq \subtree{a} \cupdot \head{a}$.  Because $\psi_\ell$ only depends on subsets of $M_\ell$ and $T \cap \Mhat_\ell = (T_p \cupdot T_a) \cap \Mhat_\ell$ since $a \in \children(p)$, it follows that 
$$
	\max_{A_\ell \in \Ocal_\ell} \psi_\ell( S \cupdot T \cupdot A_\ell )
	~=~ \max_{A_\ell \in \Ocal_\ell} \psi_\ell( S \cupdot T_p \cupdot T_a \cupdot A_\ell )~.
$$
The equality $(ii)$ then follows by optimizing separately over the products that belong to the subtree under the children of $p$ and  products that are specific to node $p$. 
The final equality $(iii)$ holds because for $\ell \in \nest(\subtree{a})$, $\psi_\ell$ only depends on subsets of $M_\ell$ and $\Mhat_\ell \subseteq \subtree{a} \cupdot \head{a}$ by Lemma~\ref{lem:prod_graph_domain}.  Thus, for each $T_a \subseteq \subtree{a}$, we have 
$$
\max_{A_\ell \in \Ocal_\ell}  \psi_\ell( S \cupdot T_p \cupdot T_a \cupdot A_\ell) ~=~   \max_{A_\ell \in \Ocal_\ell}  \psi_\ell \Big( \; \big( \,( S \cupdot T_p )  \cap  \head{a} \,\big) \; \cupdot T_a \cupdot A_\ell \Big)~,
$$
and the equality $(iii)$ then follows from the definition of the value function $J^\product_a$.

We will now analyze the running time. First, for every nest $k \in K$, we sort the revenue-ordered subsets in $\Ocal_k$ from smallest to largest, which takes $O(m n \log n)$ operations because there are $m$ nests and for each nest~$k$, $|\Ocal_k| \leq n$.
For each $p \in \bag_\product$, let $H_p : 2^{\head{p}} \times 2^{\tail{p}} \to \Real_+$ be defined by: for each $S \subseteq \head{p}$ and $T \subseteq \tail{p}$,
$$
	H_p(S, T) = \sum_{k \in \nest(\tail{p}) \,:\, \Mhat_k \subseteq \vtcs{p} } ~~ \max_{A_k \in \Ocal_k} \psi_k( S \cupdot T  \cupdot A_k).
$$
For each $(S,T) \in 2^{\head{p}} \times 2^{\tail{p}}$, 
computing the function $H_p(S,T)$ takes $O(m n)$ operations because for each nest $k$, computing $\max_{A_k \in \Ocal_k} \psi_k( S \cupdot T  \cupdot A_k)$
takes $O(n)$ operations \yw{as} there are $O(n)$ \mbox{revenue-ordered} subsets in $\Ocal_k$. \yw{Since} these revenue-ordered subsets have been sorted, we can evaluate the function $\psi_k(S \cupdot T \cupdot A_k)$ in $O(1)$ operations. To see this, recall that $\psi_k(X) = (\sum_{i \in X} v_{ik} r_i ) /  (\sum_{i \in X} v_{ik} )^{1-\gamma_k}$ for all $X$. To evaluate $\psi_k(S \cupdot T \cupdot A_k)$ for every \mbox{$A_k \in \Ocal_k$}, we maintain the sums in the numerator and denominator, and because the sets in $\Ocal_k$ are sorted, each one is 
obtained from the previous one by adding just one element; the sums can be updated in $O(1)$ operations.  Because the number of possible $(S,T) \in 2^{\head{p}} \times 2^{\tail{p}}$ is \mbox{$2^{|\vtcs{p}|} \leq 2^{\treewidth(G_\product) + 1}$} since $\treewidth(G_\product) = \max_{a \in \bag_\product} |\vtcs{a}| - 1$, it follows that computing the function $H_p$ takes $O\big( m \, n \, 2^{\treewidth(G_\product)} \big)$ operations. Because the number of bags in a \pr{non-redundant} tree decomposition of a product graph is $O(n)$,  computing the function $H_p$ for every $p \in \bag_\product$ takes $O( m \, n^2 \, 2^{\treewidth(G_\product)} )$~operations.

The dynamic programming equation can be written as follows: for each $S \subseteq \head{p}$,
\vspace{-3pt}
$$
J^\product_p(S) ~=~
\max_{T \subseteq \tail{p}} \Big\{~ H_p ( S, T) 
    + 
    \sum_{a \in \children(p)} J^{\product}_a\big( (S \cupdot T ) \cap \head{a} \big) ~\Big\}~.
$$
\vspace{-3pt}
Having computed $H_p$ for all $p \in \bag_\product$, we can recursively compute the value functions starting from the leaf nodes. The time  it takes to compute the value function $J^\product_p$ is \mbox{$O \Big( 2^{\treewidth(G_\product)} ( 1 + | \children(p) |) \Big)$}. Using the fact that $\sum_{p \in \bag_\product} |\children(p)|  = O(n)$, the total time to compute all the value functions is $O \big( n \, 2^{\treewidth(G_\product)}  \big)$. Putting everything together, the total running time is $O\big( \, m \, n^2 \, 2^{\treewidth(G_\product)} \,\big)$. \qed

In Appendix \ref{section:example-dp-product-graph}, we give a concrete example of how the dynamic programming algorithm in Theorem \ref{thm:exact_algo} can be applied to compute an optimal assortment for a generalized nested logit model with $10$ products and $7$ nests. For this example, the product graph has a treewidth of $2$.

\section{An FPTAS for Nest Graphs with Bounded Treewidths}
\label{sec:fptas}

A limitation of using the product graph is that the treewidth is lower bounded by the maximum number of products in $\Mhat_k$; that is,
\pr{$\treewidth(G_\product) \geq \max_{k \in K} |\Mhat_k| - 1$}.  
%As indicated by the hardness result, the assortment optimization under the GNL model can become NP-hard when there are two nests whose products completely overlap (Theorem \ref{thm:hardness_1}), or when multiple nests exhibit more complex overlapping patterns (Theorem \ref{thm:hardness_2}). 
To address this issue, we develop a fully polynomial-time approximation scheme (FPTAS) to solve the \ref{eqn:sub_param_new} problem by considering the tree decomposition over the nest graph, \pr{which we denote by}  $\Tree_\nest = (\bag_\nest, \edge_\nest)$.
%denote the tree decomposition over the nest graph. 
For any $q \in \bag_\nest$, {\it the sets $\vtcs{q}$, $\subtree{q}$, $\head{q}$, and $\tail{q}$ now correspond to sets of nests}. We first develop some structural properties of the tree decomposition over the nest graph and then introduce a dynamic program. To streamline the discussion, 
for each $Q \subseteq K$, 
let \mbox{$\product(Q) = \bigcup_{k \in Q} M_k$} 
denote all products that belong to at least one nest in $Q$.
%for $Q \subseteq K$.

\subsection{Properties of Tree Decomposition over Nest Graph}\label{sec:fptas_properties}
\vspace{-5pt}
We present two structural properties of the tree decomposition over the nest graph. 
The two lemmas are analogous to Lemmas \ref{lem:prod_graph_domain} and \ref{lem:prod_graph_nest_sep}, and we defer proofs to Appendix \ref{sec:fptas_additional}. The first lemma shows that for each $q \in \bag_\nest$ and $i \in \product(\subtree{q} )$, 
the set of nests that product $i$ belongs to, denoted by $\nest(i)$, must be entirely contained within the subtree rooted at node $q$. This ensures that local computations at node $q$ can be performed without considering nests outside its subtree. 
\vspace{-5pt}
\begin{lem}[Subtree and Head Contain All the Relevant Nests]  \label{lem:nest_graph_domain}
    For each $q \in \bag_\nest$ and \mbox{$i \in \product(\subtree{q})$}, we have $\nest(i) \subseteq \subtree{q} \cupdot \head{q}$.
\end{lem}
\vspace{-5pt}

The next lemma shows a separation property for products in $\product(\subtree{q} )$. A key element in the claim is the set of products defined by $ \big\{i \in \product(\tail{q}) : \nest(i) \subseteq \vtcs{q} \big\}$, which consists of products that belong to the nests in $\tail{q}$ with the additional requirement that $\nest(i)$ is fully contained within $\vtcs{q}$. In other words, it includes only those products whose associated nests do not extend beyond $\vtcs{q}$. 
We refer to these as the {\it local products of node $q$}. In the next lemma, we show that  $\product(\subtree{q} )$ can be partitioned into the local products of node $q$, and the products that belong to the children nodes of $q$. The proof is similar to the proof of Lemma \ref{lem:prod_graph_nest_sep}.
\vspace{-5pt}
\begin{lem}[Separation of Products in a Nest Graph] \label{lem:nest_graph_prod_sep}
    For each $q \in \bag_\nest$, 
    $$\product(  \subtree{q}  ) =  \big\{i \in \product(\tail{q}) : \nest(i) \subseteq \vtcs{q} \big\} ~ \bigcupdot \Big(\bigcupdot_{a \in \children(q)} \product(\subtree{a}) \Big). $$
\end{lem}
\vspace{-5pt}
Taken together, Lemmas~\ref{lem:nest_graph_domain} and~\ref{lem:nest_graph_prod_sep} establish a strong separation property of the nest graph under tree decomposition. We will use these properties to derive an FPTAS for the \ref{eqn:sub_param_new} problem.

\vspace{-6pt}
\subsection{Outline of the FPTAS}\label{fptas_outline}
\vspace{-7pt}
This section is devoted to providing an overview of the FPTAS based on the tree decomposition of the nest graph. The main result is stated in the following theorem. 
\vspace{-5pt}
\begin{theorem}[FPTAS] \label{thm:FPTAS}
    For any $\epsilon > 0$, there exists an algorithm that provides a \pr{$(1-\epsilon)$}-approximate solution to the \ref{eqn:sub_param_new} problem with running time $O\Big( ~m^2 \cdot ( \pr{2n/\epsilon})^{O(\treewidth(G_\nest))} ~ \Big)$.
\end{theorem}
\vspace{-5pt}
%\prc{Comment here on the case where $\treewidth(G_\nest) = 0$, which corresponds to the case of the classical nested logti setting.}
%
%\prc{Add a discussion here about finite number of nests. And the case where $\treewidth(G_\nest) = 0$.}

\pr{We make two observations. First, 
Theorems \ref{thm:FPTAS} and \ref{thm:using-fixed-point-to-solve-assortment-problem} yield an FPTAS for the \ref{eqn:cnl_assort} problem, and $\treewidth(G_\nest) \leq m$, so the running time of our method is competitive with existing FPTAS in the literature.
Second,}
for the instances of the generalized nested logit model 
used in Theorems~\ref{thm:hardness_1} and~\ref{thm:hardness_2}, we can easily check that the nest graphs have bounded treewidths. \pr{Consequently, we have FPTAS available for the instances used in Theorems \ref{thm:hardness_1} and~\ref{thm:hardness_2}.}
Our FPTAS consists of two main steps: 1) an {\it approximate} version of the \ref{eqn:sub_param_new} problem by restricting to a finite grid, and 2) a dynamic program that exploits the tree decomposition to solve the approximate problem efficiently.  We now describe these two steps in more details.

%We provide an overview of our algorithm, which proceeds in two steps. 
%In the first step, we construct an approximate version of Problem~\eqref{eqn:sub_param_new}. This approximation follows the same high-level idea as in \cite{desir2022capacitated}, and solving the approximate problem yields a $(1-2\epsilon)$-approximate solution to Problem~\eqref{eqn:sub_param_new}. However, directly solving the approximate version of the problem is computationally expensive and the dynamic program in \cite{desir2022capacitated} has a running time exponential in the total number of nests. To tackle this challenge, in the second step, we exploit additional structure through a tree decomposition of the nest graph. By designing a dynamic program that operates on this tree decomposition, we show that the running time can be improved so that it grows exponentially only in the treewidth of the nest graph, rather than in the total number of nests. Thus, the tree decomposition plays a central role in reducing the computational complexity of the algorithm.

To describe an approximate version of the \ref{eqn:sub_param_new} problem, for each $i \in N$, define a function $\theta_i: \Real_+^m \to \Real_+$ as follows:
for each $\svec \in \Real_+^m$, $\theta_i(\svec) = \sum_{k \in \nest(i)} v_{ik}r_i / s_k^{1-\gamma_k}$. 
The value $\theta_i(\svec)$ corresponds to the contribution of product~$i$ assuming we offer an assortment $S$ such that $V_k(S) = s_k$ for $k \in \nest(i)$.
It is easy to verify that the \ref{eqn:sub_param_new} problem is equivalent to the following:

\noindent \underline{An Equivalent Formulation of the \ref{eqn:sub_param_new} Problem:}
\vspace{-2pt}
$$ 
F^* ~=~ \max_{ \substack{S \subseteq N, ~ \tvec \in \mathbb{R}^{m}_+ }} \left\{  \sum_{i \in S } \theta_i(\tvec)~:~ V_k(S) \, \leq \, t_k  ~~ \forall k \in K \right\} ~.
$$
\vspace{-2pt}
Under the above formulation, we introduce a new set of decision variables that capture the value of $V_k(S)$. The benefit of this approach is that we can linearize the objective function and turn the problem into a knapsack-like problem. However, it also introduces challenges since the vector $\tvec$ is continuous. To tackle this, we will restrict $\tvec$ to a finite grid defined as follows. For each accuracy parameter $0 < \epsilon < 1$ and $k \in K$, define
$$
	\grid_k(\epsilon) = \{0\} \cup \bigcup_{j \in \yw{M_k}} \left\{ (1+\epsilon)^\ell ~:~ \ell = \bigg\lfloor \frac{\log v_{jk}}{\log(1+\epsilon)} \bigg\rfloor, \dots, \rubigg{\frac{\log(n v_{jk})  }{\log(1+\epsilon)}} \right\}~,
$$
and define $\grid(\epsilon) =  \grid_1(\epsilon) \times \dots \times \grid_m(\epsilon)$.  By our construction, if $S^*$ is an optimal solution to the \ref{eqn:sub_param_new} problem, we can always find $\tvec \in \grid$ such that $\frac{t_k}{1+\epsilon} \leq V_k( S^* ) \leq t_k$ for each $k \in K$. This is because $V_k(S^*)$ is between $v_{jk}$ and $n v_{jk}$ for some $j \in \yw{M_k}$, and our construction of $\grid_k(\epsilon)$ ensures that we discretize the interval $[v_{jk}, n v_{jk}]$ in an exponential grid of increment $(1+\epsilon)$. Note that the number of values in $\grid_k$ is $O\bigl( n\log(n)/ \epsilon \bigr)$.

%To tackle this, we develop an efficient rounding procedure for $\tvec$ that maintains sufficient accuracy. Let $S^*$ be the optimal solution for the \ref{eqn:sub_param_new} problem. Our objective is to construct, for each nest $k$, a collection of candidate values in $\grid_k$ such that there exists some $t_k \in \grid_k$ satisfying $\frac{t_k}{1+\epsilon} \leq V_k( S^* ) \leq t_k$. To achieve this, we let $B_{\text{min},k} = \max_{i \in S^* } v_{ik}$ and $B_{\text{max}, k} = n \cdot \max_{i \in S^* } v_{ik}$. Although $B_{\text{min},k}$ and $B_{\text{max},k}$ are not known in advance, they can be enumerated in $O(n)$ iterations. We construct a geometric grid over the interval $[B_{\text{min},k}, B_{\text{max},k}]$ as follows
%$$\big\{\,0\, \big\} \cup \left\{ (1+\epsilon)^\ell ~:~ \ell = \bigg\lfloor \frac{\log B_{\text{min},k}}{\log(1+\epsilon)} \bigg\rfloor, \dots, \rubigg{\frac{\log(B_{\text{max},k})}{\log(1+\epsilon)}} \right\},$$
%which contains $O\bigl(\log(n)/\epsilon\bigr)$ candidate values. Since we need to construct the geometric grid for each guess of $B_{\text{min},k}$ and $B_{\text{max}, k} $, in total $\grid_k$ contains $O\bigl(n\log(n)/\epsilon\bigr)$ number of candidate values. 

In addition to restricting  $\bm{t} \in \grid(\epsilon)$, \yw{we further round $v_{ik}$ using a rounding strategy similar to the one used in \cite{desir2022capacitated}.} We round $v_{ik}$ to $\rdbig{ \frac{4nv_{ik}}{\epsilon t_k}} $. Under this rounding strategy, if $\sum_{i \in S} v_{ik}  \leq t_k$, then $\sum_{i \in S} \rdbig{ \frac{4nv_{ik}}{\epsilon t_k}} \leq \rubig{\frac{4n}{\epsilon}}$.
% Let $\grid =  \grid_1 \times \dots \times \grid_m $, 
Our final reformulation of the \ref{eqn:sub_param_new} problem becomes
\begin{align}\label{eqn:nest_reformulate} \tag{{\sf Discretized-Aux-Assort}}
    \max_{ \substack{ S \subseteq N , ~ \tvec \in \grid(\epsilon)}} \left\{  \sum_{i \in S } \theta_i(\tvec) ~:~ \sum_{i \in S } \Big \lfloor \frac{4nv_{ik}}{\epsilon \, t_k} \Big \rfloor \leq  \ruBig{\frac{4n}{\epsilon}} ~~ \forall k \in K \right\}.
\end{align}
The following lemma shows that solving the \ref{eqn:nest_reformulate} problem gives a $(1-2\epsilon)$-approximate solution to the \ref{eqn:sub_param_new} problem. %The result is summarized in the following lemma. 
\vspace{-2mm}
\begin{lem} \label{lem:nest_approx}
If $\Shat$ is an optimal solution to the \ref{eqn:nest_reformulate} problem,  then $\Shat$ is a \mbox{$(1-2\epsilon)$-approximate} solution to the \ref{eqn:sub_param_new} problem.
\end{lem}
\vspace{-2mm}
The proof of Lemma~\ref{lem:nest_approx} follows arguments similar to those in \cite{desir2022capacitated} and is  in Appendix~\ref{sec:nest_approx}.
Note that
for each fixed $\tvec$, the \ref{eqn:nest_reformulate} problem 
is an $m$-dimensional knapsack problem with integer capacity consumption \citep{freville2004multidimensional}, so it
can then be solved exactly using a dynamic program.  This dynamic program maintains a state variable for each constraint, so the total number of states scales on the order of $O\!\left((n/\epsilon)^m\right)$, where $m$ is the number of nests.  Enumerating all possible values of~$\tvec$ requires a running time of $O\!\big( ( \, (n / \epsilon) \log n \, )^m \big)$.  Thus, this method has  a running time that is exponential in the number of nests. In the next section, we show that by exploiting the tree decomposition of the nest graph, this exponential dependence can be reduced so that the running time is exponential only in the treewidth of the nest graph.

%The \ref{eqn:nest_reformulate} is a multidimensional knapsack problem with integer capacity consumptions, so it can be solved exactly using dynamic programming with the number of nests $|K|$ is bounded \citep{kellerer2004multidimensional}. Note that if the  number of nests is bounded, so is the treewidth of the nest graph. We are interested in the setting in which the number of nests increases with the problem size, while the treewidth remains bounded. 
%
% 
%
%Before describing our algorithm for solving the \ref{eqn:nest_reformulate} problem, we briefly discuss an existing approach whose running time is exponential in the number of nests. Under this approach, we first enumerate all feasible values of the vector~$\tvec \in \grid(\epsilon)$. For each fixed~$\tvec \in \grid(\epsilon)$, the \ref{eqn:nest_reformulate} problem 
%is a $m$-dimensional knapsack problem \citep{freville2004multidimensional}, and 
%can then be solved exactly using a dynamic program. This dynamic program maintains a state variable for each constraint, so the total number of states scales on the order of $O\!\left((n/\epsilon)^m\right)$, where $m$ is the number of nests. Moreover, enumerating all possible values of~$\tvec$ requires running time
%$O\!\big( ( \, (n / \epsilon) \log n \, )^m \big)$. Consequently, this algorithm leads to a running time that is exponential in the number of nests. In the next section, we show that by exploiting the tree decomposition of the nest graph, this exponential dependence can be reduced so that the running time is exponential only in the treewidth of the nest graph.
\vspace{-5pt}
\subsection{A Dynamic Program to Solve the \ref{eqn:nest_reformulate} Problem}\label{sec:fptas_dp}

We develop a dynamic program based on the tree decomposition of the nest graph to solve the \ref{eqn:nest_reformulate} problem. 
\pr{Motivated by the separation of products in the tree decomposition of the nest graph,}
we define the following two notations: for each $q \in \bag_{\nest}$, 
$$
N_q = \product(\subtree{q}) \qquad \text{and} \qquad \Nbar_q = \big\{i \in \product(\tail{q}) : \nest(i) \subseteq \vtcs{q} \big\}.$$
Note that $\Nbar_q \subseteq N_q$ \pr{by Lemma \ref{lem:nest_graph_prod_sep}.}
We begin by providing some intuition for the dynamic program. Let $q \in \bag_{\nest}$ be a node in the tree decomposition of the nest graph, and let $(\, \Shat, \tvechat\,\,)$ denote an optimal solution to the \mbox{\ref{eqn:nest_reformulate}} problem. Our dynamic program proceeds by optimizing the offered products within $N_q$. We make two key observations. First, for any $i \in \Shat \cap N_q$, Lemma~\ref{lem:nest_graph_domain} implies that \mbox{$\nest(i) \subseteq \head{q} \cupdot \subtree{q}$}. Hence, to evaluate the contribution of product $i$ to the objective, it suffices to consider only the nests in \mbox{$\head{q} \cupdot \subtree{q}$}. 

Second, we distinguish between two types of nests: nests in $\subtree{q}$ and nests in $\head{q}$. On one hand, if $k \in \subtree{q}$, then $M_k \subseteq N_q$, so \mbox{$(\Shat \setminus N_q) \cap M_k = \varnothing$}, which implies that $v_{ik} = 0$ for all $i \in \Shat \setminus N_q$ and $\sum_{i \in \Shat \setminus N_q } \Big \lfloor \cfrac{4nv_{ik}}{\epsilon \, \that_k} \Big \rfloor = 0$.  The total preference weight for nest $k$ thus depends only on the products within~$N_q$. On the other hand, if $k \in \head{q}$, then $M_k \not\subseteq N_q$ and it is possible that $\sum_{i \in \Shat \setminus N_q } \Big \lfloor \cfrac{4nv_{ik}}{\epsilon \, \that_k} \Big \rfloor > 0$, reflecting products that are offered outside of $N_q$. Thus, when we optimize locally over $N_q$, the preference weight of nests in $\head{q}$ also depends on decisions made elsewhere in the tree. 
These two observations above indicate that, if our goal is to optimize the products offered within $N_q$, we must carefully track the information associated with nests in $\head{q}$.  

For $q \in \bag_\nest$, let $J^\nest_q$ denote the value function at node $q$ of $\Tree_\nest$. The domain of the function $J^\nest_q$ is $\budget_q \times \prefW_q$ and the action set is denoted by $\action_q$, where 
$\budget_q$, $\prefW_q$ and $\action_q$ are defined by:
\begin{align*}
\budget_q &~=~ \Big\{ \hvec \in \Integer_+^m ~\big|~ h_k = 0 ~\forall~k \notin \head{q} ~\text{and}~ 0 \leq h_k \leq \ruBig{\frac{4n}{\epsilon}} ~\forall~k \in \head{q}  \Big\}\\
\prefW_q &~=~ \big\{ \gvec \in \grid(\epsilon) ~\big|~ g_k = 0 ~\forall~k \notin \head{q}   \big\}\\
%\dom_q &~=~ \big\{ (\hvec,\gvec) \in \Integer_+^m \times \grid(\epsilon) ~\big|~ h_k = g_k = 0 ~\forall~k \notin \head{q} ~\text{and}~ 0 \leq h_k \leq \ruBig{\frac{4n}{\epsilon}} ~\forall~k \in \head{q}  \big\}\\
\action_q &~=~  \{ (S, \bm{t} ) ~\big|~ S \subseteq N_q, \bm{t} \in \grid(\epsilon) ~\text{and}~ t_k \,=\, 0 ~\forall~k \notin  \subtree{q} \}
\end{align*}
For each $(\hvec,\gvec) \in \budget_q \times \prefW_q$, we define
$$
    J^{\nest}_q(\hvec,\gvec)
    =
    \max_{(S,\bm{t}) \in \action_q}
    \Bigg\{
    \sum_{i \in S} \theta_i(\gvec + \tvec)
    \,:\,
    \sum_{i \in S}
    \Big \lfloor \frac{4nv_{ik}}{\epsilon\,g_k}     \Big \rfloor
    \leq
    h_k ~\forall k \in \head{q},~~
    \sum_{i \in S}
        \Big \lfloor \frac{4nv_{ik}}{\epsilon\,t_k}     \Big \rfloor
    \leq
        \Big \lceil \frac{4n}{\epsilon} \Big \rceil ~\forall k \in \subtree{q}
    \Bigg\}.
$$

We first give interpretations of the two state variables and the value functions. Recall  that  $(\, \Shat, \tvechat\,\,)$ denotes an optimal solution to the \mbox{\ref{eqn:nest_reformulate}} problem.  The state vector $\gvec \in \prefW_q$ keeps track of the discretized {\it preference weights} $\big( \that_k \,:\, k \in \head{q} \big)$ for all nests in $\head{q}$, and by definition, we set $g_k = 0$ for all $k \notin \head{q}$. Given the discretized preference weight vector $\gvec \in \prefW_q$, the vector $\hvec \in \budget_q$ keeps track of the {\it budget} allocated for each nest in $\head{q}$; that is, for each $k \in \head{q}$, 
$h_k$ keeps track of $\big \lceil \frac{4n}{\epsilon} \big \rceil  - \sum_{i \in \Shat \setminus N_q}  \big \lfloor \frac{4nv_{ik}}{\epsilon\, \that_k}  \big \rfloor$.  As noted earlier, for $k \in \head{q}$, the expression $\sum_{i \in \Shat \setminus N_q}  \big \lfloor \frac{4nv_{ik}}{\epsilon\, \that_k}  \big \rfloor$ can be positive. It is thus important to keep track of the remaining budget to ensure that the constraints for nests in $\head{q}$ are satisfied.

Given the state variables $(\hvec,\gvec)$, 
the value function $J^{\nest}_q(\hvec,\gvec)$ represents the maximum expected contribution by products in $N_q$. To see this, note that for each product $i \in N_q$, Lemma~\ref{lem:nest_graph_domain} ensures that $\nest(i) \subseteq \head{q} \cupdot \subtree{q}$, and thus, the contribution of product $i$ to the objective is
$$
\sum_{k \in \head{q}} \frac{v_{ik} r_i}{g_k^{1-\gamma_k}}
+
\sum_{k \in \subtree{q}} \frac{v_{ik} r_i}{t_k^{1-\gamma_k}}
=
\theta_i(\gvec + \tvec),
$$
where the equality follows because, by definition of \pr{$\prefW_q$} and $\action_q$,
$g_k + t_k = g_k$ for all $k \in \head{q}$, $g_k + t_k = t_k$ for all $k \in \subtree{q}$, and $g_k + t_k = 0$ for all $k \notin \head{q} \cupdot \subtree{q}$.

Finally, because $N_\root = N$, \mbox{$\head{\root} = \varnothing$}, and $\subtree{\root} = K$, it follows that \mbox{$\budget_\root = \prefW_\root = \{\bm{0}\}$}, \mbox{$\action_\root = \{ (S, \tvec) \,\big|\, S \subseteq N, \tvec \in \grid(\epsilon) \}$}, and $J^\nest_\root(\bm{0},\bm{0})$ is the optimal objective value of the  \mbox{\ref{eqn:nest_reformulate}} problem.  
 Our goal is to develop an efficient algorithm to compute $J^\nest_\root(\bm{0},\bm{0})$ based on the tree decomposition $\Tree_\nest$.

The challenge in computing $J^\nest_q(\hvec, \gvec)$ from the definition is twofold: the size of the action set $\action_q$ and the number of constraints. Recall that $\action_q ~=~  \{ (S, \bm{t} ) ~\big|~ S~\subseteq~N_q, \bm{t}~\in~\grid(\epsilon) ~\text{and}~ {t_k \,=\, 0} ~\forall~k \notin  \subtree{q} \}$. We thus have to consider all subsets of  $N_q = \product(\subtree{q})$ and search over all vectors $(t_k \in \grid_k(\epsilon) : k \in \subtree{q})$. So, the size of the action set is exponential in the size of \yw{$N_q$}, which can be as large as $n$.  Moreover, the number of constraints increases with the size of $\subtree{q}$.

Before addressing these computational challenges, we make an important observation. Note that, in the definition of $J^\nest_q$, we only consider constraints associated with nests \mbox{$k \in \head{q} \cupdot \subtree{q}$}. Also, by Lemma~5.1, nests outside $\head{q}\cup\subtree{q}$ do not affect the contribution of any product in $N_q$. The observation indicates that we can extend the definition of $J_q^\nest(\hvec,\gvec)$ to all $\gvec\in\grid(\epsilon)$ satisfying $g_k=0$ for $k\in\subtree{q}$. We summarize this result in the following lemma. The result follows directly from the preceding observation, so we omit the proof.
\vspace{-8pt}
\begin{lem}[Domain Extension]\label{lem:domain_extension}
For $q\in\bag_\nest$, $(\hvec,\gvec) \in \budget_q \times \prefW_q$, and
$\gvec'\in\grid(\epsilon)$ satisfying $g'_k=0$ for all
$k\in\head{q}\cup\subtree{q}$, we have
$
J_q^\nest(\hvec,\gvec+\gvec')
=
J_q^\nest(\hvec,\gvec).
$
\end{lem}

 The main result is then stated in the following theorem, which gives a dynamic programming equation for the value function $J^\nest_q$, allowing us to recursively compute the value function starting from the leaf nodes of $\Tree_\nest$, and eliminating the dependence on $\subtree{q}$.  For each $q \in \bag_\nest$, we define the reduced action set $
\reduceA_q
 = \{ (S, \bm{t} ) ~\big|~ S \subseteq \Nbar_q, \bm{t} \in \grid(\epsilon) ~\text{and}~ t_k \,=\, 0 ~\forall~k \notin  \tail{q} \}$. It is easy to check that $\reduceA_q \subseteq \action_q$ for all $q$.
%
%The goal in the rest of the section is to derive a recursive formulation of the value function $J^{\nest}_q( \, \hvec,\,\gvec\,)$. Given a fixed value of $\hvec$ and $\gvec$ such that $h_k = g_k = 0$ for all $k \notin \head{q}$, we consider the following optimization problem.
%\begin{align}\label{eqn:nest_graph_dp}
%\!\!\max_{\substack{
%    S_q \subseteq \Nbar_q \vspace{0.15em} \\
%    \wvec_a \in \mathbb{Z}^m_+,~ a \in \children(q)\vspace{0.15em} \\
%    \tvec \in \grid, ~s.t~ t_k = 0 ~\forall k \notin \tail{q} 
%}}~~~
%& \sum_{i \in S_q} \theta_i(\gvec + \tvec)
%+
%\sum_{a \in \children(q)} J^\nest_a\big( \wvec_a ,\, \gvec + \tvec \big)
%\\
%&
%\sum_{i \in S_q} \rdbig{\frac{4nv_{ik}}{\epsilon\,g_k}} + \sum_{a \in \children(q)} w_{ak} \leq h_k
%&&\!\! \forall k \in \head{q}, \notag
%\\
%&
%\sum_{i \in S_q} \rdbig{\frac{4nv_{ik}}{\epsilon\,t_k}} + \sum_{a \in \children(q)} w_{ak} \leq \ruBig{\frac{4n}{\epsilon}}
%&&\!\! \forall k \in \tail{q}. \notag
%\end{align}
%We show that Problem~\eqref{eqn:nest_graph_dp} provides an alternative characterization of $J^{\nest}_q(\hvec,\gvec)$ in the following Lemma.

%\prc{In the DP, we have $J^{\nest}_a(\wvec_a, \gvec + \tvec)$, but $\gvec + \tvec$ may NOT be in the $\prefW_q$. Am I missing something here?}
\vspace{-10pt}
\begin{thm}[Dynamic Programming Equation]
\label{thm:fptas_dp_correctness}
For $q \in \bag_\nest$ and $(\hvec, \gvec) \in \budget_q \times \prefW_q$,
\vspace{-5pt}
\begin{align*}
J^\nest_q(\hvec, \gvec) ~=~ \displaystyle
\max_{ (S_q,\tvec),\,  ( \wvec_a \,:\,  a \in \children(q) ) }
&~~~\sum_{i \in S_q} \theta_i(\gvec + \tvec)
+
\sum_{a \in \children(q)} J^\nest_a\big( \wvec_a ,\, \gvec + \tvec \big)
\\
{\rm s.t.} ~~~~~~~&
~~~\sum_{i \in S_q} \rdBig{\frac{4nv_{ik}}{\epsilon\,g_k}} ~+~ \sum_{a \in \children(q)} w_{ak} ~\leq~ h_k
&&\!\! \forall k \in \head{q}, \notag
\\
&
~~~\sum_{i \in S_q} \rdBig{\frac{4nv_{ik}}{\epsilon\,t_k}} ~+~ \sum_{a \in \children(q)} w_{ak} ~\leq~ \ruBig{\frac{4n}{\epsilon}}
&&\!\! \forall k \in \tail{q}, \notag\\
&~~~~(S_q, \tvec) \in \reduceA_q,~~~  \wvec_a \in \budget_a~~\forall~a \in \children(q)
\end{align*}
\end{thm}
\vspace{-5pt}
The proof of the theorem is given in Appendix \ref{sec:proof-dp-equation-nest-graph}. We provide some intuition here. By Lemma~\ref{lem:nest_graph_prod_sep}, we can partition $N_q$ into $\Nbar_q$ and $ \bigcupdot_{a \in \children(q)} N_a$. Because $\Nbar_q = \{ i \in \product(\tail{q}): \nest(i) \subseteq \vtcs{q} \}$, we can evaluate the function $\theta_i(\svec)$ using only $s_k$ for $k \in \vtcs{q}$ and we consider these products as the ``local" products of node $q$. The set of products $N_a$ corresponds to the product universe of the subproblem defined over node $a$. For each child $a \in \children(q)$ and nest $k \in \vtcs{q}$, we introduce an auxiliary decision variable $w_{ak}$ to represent %the rounded total preference weight contributed by products in $S \cap N_a$ to nest $k$. In other words, $w_{ak}$ encodes
how much of the budget for nest $k$ is allocated to subtree $a$. The first set of constraints then ensures that, for nests in $\head{q}$, the rounded preference weight from the local products in $S_q \subseteq \Nbar_q$ plus the allocations to the children do not exceed the budget $h_k$. The second set of constraints plays an analogous role for nests in $\tail{q}$. Taken together, the objective combines the contribution of the local products, $\sum_{i \in S_q} \theta_i(\gvec + \tvec)$, with the optimal contributions from each child subtree, captured by the value functions $J^\nest_a(\wvec_a,\gvec + \tvec)$. This yields a recursive computation of $J^{\nest}_q(\hvec,\gvec)$. \yw{Finally, for $a\in\children(q)$, we note that $\gvec+\tvec$ may not belong to $\prefW_a$. But its nonzero components outside $\head{a}$ correspond to nests outside $\head{a}\cup\subtree{a}$. Thus, by Lemma \ref{lem:domain_extension}, these components can be ignored when evaluating $J^\nest_a(\wvec_a,\gvec+\tvec)$.}

Given \pr{the result of} Theorem \ref{thm:fptas_dp_correctness},  
we can now give a proof of Theorem~\ref{thm:FPTAS}.

\noindent{\bf \underline{Proof of Theorem \ref{thm:FPTAS}:}} Theorem~\ref{thm:fptas_dp_correctness} gives a dynamic programming equation for the value function $J^{\nest}_q$.  We initialize the value functions at the leaf nodes and recursively compute the value function at 
each node in the tree decomposition $\Tree_\nest$, in a bottom-up fashion until we reach the \root \;node. As noted earlier, $J^\nest_\root(\mathbf{0}, \mathbf{0})$ yields the optimal solution to the  \ref{eqn:nest_reformulate} problem, which gives a $(1-2\epsilon)$-approximate solution for the \ref{eqn:sub_param_new} problem by Lemma~\ref{lem:nest_approx}.

%We now analyze the running time. 
%For each node $q \in \bag_\nest$, the domain of the value function~$J^\nest_q$ is $\budget_q \times \prefW_q$. By definition,  $|\budget_q| = O \big( (4n /\epsilon)^{|\head{q}|} \big) = O \Big(  (n/\epsilon)^{ O ( \treewidth( G_\nest) )} \Big)$, and
%$|\prefW_q| = O \big( ((n \log n) /\epsilon)^{|\head{q}|} \big) = O \Big(  (n/\epsilon)^{ O ( \treewidth( G_\nest) )} \Big)$.  The size of the state space is thus $ O \Big(  (n/\epsilon)^{ O ( \treewidth( G_\nest) )} \Big)$.   
\pr{To analyze the running time, the following claim shows the time to compute $J^\nest_q(\hvec,\gvec)$.}

%Moreover, the number of nodes in $\bag_\nest$ is at most $m = |K|$. To establish the desired result, it thus suffices to prove the following claim:

\noindent {\bf Claim:} For $q \in \bag_\nest$ and $(\hvec,\gvec) \in \budget_q \times \prefW_q$, we can compute $J^\nest_q(\hvec,\gvec)$ in $O \Big( m (n/\epsilon)^{ O ( \treewidth( G_\nest) )} \Big)$.

To establish the claim, fix an arbitrary $q \in \bag_\nest$ and $(\hvec,\gvec) \in \budget_q \times \prefW_q$.  We index the children of~$q$ as $\children(q)=\{1,2,\dots,C\}$. We first enumerate all possible values of $t_k \in \grid_k(\epsilon)$ for $k \in \tail{q}$. Since $|\tail{q}| \le \treewidth(G_\nest)+1$ and the set $\grid_k(\epsilon)$ has size $O( (n\log n)/\epsilon)$, the total number of choices of~$\tvec$ is bounded by $O\!\left((n/\epsilon)^{O(\treewidth(G_\nest))}\right)$. 

Let $\pr{\dom_q} = \{ \uvec \in \mathbb{Z}^m_+ \,\big|\, u_k = 0 ~\forall~k \notin \vtcs{q},  0 \leq u_k \leq \big \lceil \frac{4n}{\epsilon}\big\rceil ~\forall~k \in \vtcs{q} \}$. For each of the enumerated~$\tvec$, we consider the following collection of optimization problems. 
For $\ell \in \{0,1,\dots,C\}$ define the value function $Q^{\tvec}_\ell$ as follows: for each $\uvec \in \yw{\dom_q}$,
\begin{align*}
    Q^{\tvec}_\ell(\uvec) 
    =
    \max_{\substack{
        S_q \subseteq \Nbar_q \vspace{0.15em} \\
        \wvec_a \in \budget_a \\
		\forall a \in \{1,\dots,\ell\}
    }}
    \bigg\{
    \sum_{i \in S_q} \theta_i(\gvec + \tvec)
    +
    \sum_{a=1}^\ell J^\nest_a\big(\wvec_a,\gvec + \tvec\big)
    :
    \sum_{i \in S_q} \rdBig{\frac{4nv_{ik}}{\epsilon\,(g_k+t_k)}}
    +
    \sum_{a=1}^\ell w_{ak}
    \le u_k,\ \forall k \in \vtcs{q}
    \bigg\},
\end{align*}
where we use the fact that $g_k + t_k = g_k$ for all $k \in \head{q}$, $g_k + t_k = t_k$ for all $k \in \tail{q}$, and \mbox{$\vtcs{q} = \head{q} \cupdot \tail{q}$}.
%Intuitively, $Q_\ell(\uvec)$ represents the maximum total revenue that can be obtained by optimally selecting the local products $S_q$ and incorporating the first $\ell$ children's subproblems, subject to the knapsack constraints summarized by the capacity vector~$\uvec$. 
The base case $Q^{\tvec}_0(\uvec)$ corresponds to the situation where we ignore the contribution of children of~$q$, corresponding to
$$
    Q^{\tvec}_0(\uvec)
    =
    \max_{S_q \subseteq \Nbar_q}
    \bigg\{
    \sum_{i \in S_q} \theta_i(\gvec + \tvec)
    ~:~
    \sum_{i \in S_q} \rdBig{\frac{4nv_{ik}}{\epsilon\,(g_k+t_k)}}
    \le u_k,\ \forall k \in \vtcs{q}
    \bigg\},
$$
which is a multidimensional knapsack problem with integer capacities. Because $u_k \le \lceil 4n/\epsilon \rceil$ for each~$k$ by our definition of \yw{$\dom_q$} and $|\vtcs{q}| \leq \treewidth(G_\nest)+1$, the value $Q^{\tvec}_0(\uvec)$ can be computed via a standard algorithm for the multidimensional knapsack problem \citep{desir2022capacitated} with running time $O\big((n/\epsilon)^{\treewidth(G_\nest)}\big)$. Noting that $|\yw{\dom_q}| = O\big((n/\epsilon)^{\treewidth(G_\nest)}\big)$, it follows that we can compute the entire function $Q^{\tvec}_0$  in $O\big((n/\epsilon)^{O ( \, \treewidth(G_\nest) \, )}\big)$. Given $Q^{\tvec}_{\ell-1}$, we can then compute $Q^{\tvec}_{\ell}$ through the following recursion: for each \yw{$\uvec \in \dom_q$},
\begin{align*}
    Q^{\tvec}_\ell(\uvec)
    &=
    \max_{\wvec_\ell}
    \Big\{
        Q^{\tvec}_{\ell-1}(\uvec - \wvec_\ell)
        +
        J^\nest_\ell\big(\wvec_\ell,\gvec + \tvec\big)
	\;\big|\; \wvec_\ell \in \budget_\ell,~w_{\ell k} \leq u_k ~\forall~k \in \head{\ell}
    \Big\}\\
    &=
    \max_{\wvec_\ell \in \Integer_+^m }
    \Big\{
        Q^{\tvec}_{\ell-1}(\uvec - \wvec_\ell)
        +
        J^\nest_\ell\big(\wvec_\ell,\gvec + \tvec\big)
	\;\big|\;  0 \leq w_{\ell k} \leq u_k ~\forall~k \in \head{\ell},\, w_{\ell k} = 0 \,\,\forall\, k \notin \head{\ell}
    \Big\}~,
\end{align*}
where the last equality follows from the definition of $\budget_\ell$.
In the expression above, $\wvec_\ell$ specifies how much weight contributed by child~$\ell$ is allocated to nest \mbox{$k \in \head{\ell}$}. We do not assign weight to nests outside $\head{\ell}$, since those nests do not appear in the child's subproblem and therefore have no effect on its value function. For a fixed state $\uvec$, $w_{\ell k}$ can take any value in $\{0,\dots,u_k\}$ for $k \in \head{\ell}$, and therefore the number of feasible choices of $\wvec_\ell$ is at most $O\big((n/\epsilon)^{|\head{\ell}|}\big)$, which is bounded above by $O\big((n/\epsilon)^{\treewidth(G_\nest)}\big)$. Consequently, for each \yw{$\uvec \in \dom_q$}, the maximization step in the recurrence for $Q^{\tvec}_\ell(\uvec)$ can be evaluated in time $O\big((n/\epsilon)^{\treewidth(G_\nest)}\big)$, and the total number of states is also $O\big((n/\epsilon)^{\treewidth(G_\nest)}\big)$.  Computing $Q^{\tvec}_\ell$ thus takes  $O\big((n/\epsilon)^{O( \treewidth(G_\nest))}\big)$.  Since the number of children of $q$ is at most $m$, the total running time needed to compute $Q^{\tvec}_C$ is $O\!\big(
    m\big(n/\epsilon\big)^{O(\treewidth(G_\nest))}
\big)$.

%To recover the value $J^\nest_q(\hvec, \gvec)$, 
Define $\bar{\bm{u}} \in \dom_q$ as follows: \mbox{$\bar{u}_k = h_k \idc\{ k \in \head{q}\} + \rubig{\tfrac{4n}{\epsilon}} \idc\{ k \in \tail{q} \}$} for all $k \in K$.  It is easy to verify that $J^\nest_q(\hvec, \gvec) = \displaystyle \max_{\tvec} Q^{\tvec}_C(\bar{\bm{u}})$.
We can thus determine $J^\nest_q(\hvec, \gvec)$ with running time 
$O\Big(m \big(n/\epsilon\big)^{O(\treewidth(G_\nest))} \Big)$, which proves the claim.

%\pr{For a non-redundant tree decomposition, t}
The number of nodes in $\bag_\nest$ is at most $m = |K|$. 
For each node $q \in \bag_\nest$, the domain of the value function~$J^\nest_q$ is $\budget_q \times \prefW_q$, with $|\budget_q| = O \big( (4n /\epsilon)^{|\head{q}|} \big) = O \Big(  (n/\epsilon)^{ O ( \treewidth( G_\nest) )} \Big)$, and
$|\prefW_q| = O \big( ((n \log n) /\epsilon)^{|\head{q}|} \big) = O \Big(  (n/\epsilon)^{ O ( \treewidth( G_\nest) )} \Big)$.   \pr{Thus, the running time to compute
$J^\nest_q(\hvec, \gvec)$ for all $q$ and $(\hvec, \gvec)$ is $O\Big(m^2 \big(n/\epsilon\big)^{O(\treewidth(G_\nest))} \Big)$. Applying $\epsilon/2$ in place of $\epsilon$ yields a $(1-\epsilon)$-approximation for the \ref{eqn:sub_param_new} problem with running time $O\Big(m^2 \big(2 n/\epsilon\big)^{O(\treewidth(G_\nest))} \Big)$, which is the desired result.}
\qed

\vspace{-5pt}
\section{Empirical Study}\label{sec:empirical}

%In this section, 
We compare the predictive accuracy of the generalized nested logit (\GNL) model with that of the multinomial logit (MNL), nested logit (NL), and mixture of multinomial logit (MMNL) models, using the IRI Academic Dataset \citep{bronnenberg2008database}, which contains retail purchase transactions from grocery and drug stores across a variety of product categories. Our results show that the GNL model outperforms other benchmark models.  We also test the performance of GNL model on synthetic data; the results are similar and are shown in Appendix \ref{sec:numerics}.

\vspace{-8pt}
\subsection{Data Preprocessing } \label{sec:data}

The IRI Academic Dataset contains real-world transaction records across thirty product categories. Each purchased item is represented by its universal product code (UPC). Since many UPCs have very few sales records, we aggregate all items with the same vendor code (digits 4 through 8 of the UPC) into a unique ``product" to deal with data sparsity. The vendor here corresponds to the brand of the item but not the store where it is being sold. For each product category, the IRI dataset contains sales data from 50 U.S. regions, where each region contains transactions from stores operated by various retailers. We analyze each region separately. For each product category, we construct 50 separate datasets, with each dataset corresponding to the transaction data from one region. We retain only regions with at least 2,000 transactions for our analysis.

%Next, we discuss how we clean our transaction data 
For each category-region combination, we select the top 12 purchased products with the largest number of transactions to be the product universe. We combine the remaining transactions into a no-purchase option. We focus on 22 product categories with the highest number of transactions. This preprocessing approach closely follows the one in \cite{jagabathula2019limit}. We then convert the sales transactions for each category-region combination into a dataset represented by a set of tuples $\{(S_t, i_t): t = 1, 2, \dots, \tau \}$, where $S_t$ is the offered assortment and $i_t$ is the selected product in transaction $t$. We infer the offered assortment $S_t$ by assuming that all products for which a sale was observed during the week at the store are available in the offered assortment. We use all transactions in the first six weeks of 2007 as our dataset. 

\vspace{-5pt}
\subsection{Fitting the GNL Model}

In this section, we outline our procedure to estimate the GNL model. We consider a variant of the GNL model. Given an assortment $S_t$, the customer first selects a nest $k \in K$ with probability $Q_k(S_t) = (1 + V_k(S_t))^{\gamma_k}/\sum_{\ell \in K}(1 + V_\ell(S_t))^{\gamma_\ell} $. In the second stage, given that a customer decides to make a purchase in the nest $k$, she selects a product from $S_t \cap M_k$ with probability $P_{ik}(S_t) = v_{ik}/(1 + V_k(S_t))$, and decides not to purchase anything with probability $P_{0k}(S) = 1/(1 + V_k(S_t))$. Under this variant of the GNL model, we allow the customer to select the no-purchase option after the customer specifies the nests. The probability of observing a transaction $(S_t, i_t)$ under this GNL model is $\phi^{\GNL}_{i_t}(S_t) = \sum_{k \in K} Q_k(S_t) \cdot P_{i_t, k}(S_t)$.

We estimate the GNL model parameters by maximizing the likelihood function. For simplicity, we let $M_k = N$ for all $k \in K$ so every product can belong to every nest. The parameters are $\{v_{ik}: i \in N, k \in K\}$ and $\{\gamma_k: k \in K\}$. We follow the standard parametrization of the preference weights by letting $v_{ik} = e^{\mu_{ik}}$. Given the historical purchase data $\{(S_t,i_t): t=1,\ldots,\tau\}$, the log-likelihood function when customers follow the GNL model is
$
\mathcal{L}^{\GNL}
=
\sum_{t=1}^\tau
\log \phi_{i_t}^{\GNL}(S_t),
$
where $\phi_{i_t}^{\GNL}(S_t)$ is defined by the purchase probability above. We estimate the parameters by maximizing this log-likelihood over $\{\mu_{ik}: i \in N, k \in K\}$ and $\{\gamma_k: k \in K\}$ subject to $0 \leq \gamma_k \leq 1$ for all $k \in K$. The maximization problem is a \pr{nonconvex} continuous optimization problem. We use the nonconvex optimization solver \textsf{fmincon} in \textsf{MATLAB} to numerically solve this problem. 

We have discussed how to estimate the parameters of the GNL model for a fixed number of nests. We now discuss how we choose the appropriate number of nests. Recall that our dataset contains six weeks of transaction data. We use the first two weeks of data to estimate the model and the third week as a validation set to select the number of nests. We consider candidate numbers of nests in $\{1,2,3,4,5,6,7,8\}$. For each candidate, we estimate the GNL model using the training data and evaluate the log-likelihood of the fitted model on the validation set. We then select the number of nests that achieves the largest validation log-likelihood.

\vspace{-5pt}
\subsection{Benchmark Models}

We consider three additional benchmark models. The first is the standard MNL model, and the second is the NL model. When estimating the NL model, we adopt a setup similar to that of our GNL model, where the customer can select the no-purchase option after selecting a nest. The key difference between the NL model and GNL model is that, under the NL model, the nests are disjoint, whereas the GNL model allows a product to belong to multiple nests. Therefore, one key challenge is that the NL model also requires specifying what products each nest contains. We determine this using a greedy heuristic, which we describe in Appendix~\ref{sec:fitting}. Lastly, we estimate the MMNL model. We note that the MMNL model provides a strong benchmark within the RUM family because the MMNL model can approximate arbitrarily closely any discrete choice model consistent with the random utility maximization principle \citep{mcfadden2000mixed}. More details on the estimation procedure for the MMNL model are provided in Appendix~\ref{sec:fitting}.

To estimate the NL or MMNL models, we need to specify the number of nests or mixture components. We determine this number using the same validation procedure as for the GNL model, where we consider candidate numbers of nests or mixture components in $\{1,2,3,4,5,6,7,8\}$. For each candidate number of nests or mixture components, we estimate the NL or MMNL models using the first two weeks of data and evaluate its log-likelihood on the third week. We then select the number of nests or mixture components that achieves the highest validation log-likelihood. %We provide a more detailed discussion of the estimation procedure in Appendix~\ref{sec:fitting}.

\vspace{-5pt}
\subsection{Predicting Customer Purchases}
In this section, we evaluate the prediction performance of the fitted choice models. Recall that we use the last three weeks as our hold-out set. We evaluate the prediction performance by comparing the choice probabilities predicted by each model with the empirical choice frequencies observed in the hold-out set. To streamline our discussion, we let $\mathcal{S}$ denote the collection of unique assortments observed in the hold-out data, let $n_S$ denote the number of observed transactions under assortment $S$, and let $n_{i,S}$ denote the number of times product $i$ is selected under assortment $S$. We define $\bar{p}_{i,S}$ as the empirical choice frequency calculated from the sales transaction data, i.e., $
\bar{p}_{i,S} = \frac{n_{i,S}}{n_S},
$ and let $\hat{p}_{i,S}$ denote the choice probability predicted by a given fitted choice model. We then measure the prediction error using the root mean square error (RMSE), defined as
$
100 \times
\left(
\frac{
\sum_{S \in \mathcal{S}}
\sum_{i \in S \cup \{0\}}
\left(
\hat{p}_{i,S}-\bar{p}_{i,S}
\right)^2
}{
\sum_{S \in \mathcal{S}} |S \cup \{0\}|
}
\right)^{1/2}.
$
A smaller RMSE indicates that the fitted model more accurately captures the choice frequencies. 

In Table~\ref{tab:fit}, we report the out-of-sample predictive performance of the MNL, NL, MMNL, and GNL models across the 22 product categories. The first column lists the product category, and the second column reports the number of regions included in the analysis after restricting to regions with at least 2,000 transactions. The next four columns report the average RMSE of the MNL, NL, MMNL, and GNL models across the included regions. 
%The final column reports the number of regions in which the GNL model achieves a lower RMSE than the MMNL model.
The results in Table~\ref{tab:fit} demonstrate that the GNL model generally improves upon the MNL, NL, and MMNL benchmarks. Averaged across the 22 product categories, the prediction error of the GNL model is 2.94, compared with 3.32 for the MNL model, 3.03 for the NL model and 2.98 for the MMNL model. This corresponds to reductions of approximately 11\%, 3.1\% and 1.1\% respectively. At the product category level, the GNL model achieves a lower average prediction error than the MMNL model in 19 of the 22 categories. For the three categories in which the GNL model performs worse, the performance gap is very small. %The improvement is also consistent at the region level. Across the 1,082 category-region combination in our analysis, the GNL model outperforms the MMNL model in 605 cases, which corresponds to approximately 56\% of the datasets. 

We further test whether the improvement of the GNL model over the MMNL model is statistically significant. We conduct a one-sided paired $t$-test on the RMSE values across regions within each product category. Across the 22 product categories, we find that the GNL model significantly outperforms the MMNL model in 11 categories. Moreover, when conducting the same one-sided paired $t$-test across all 1,082 category-region combinations, we find that the GNL model significantly outperforms the MMNL model overall. These results indicate that the GNL model provides a consistent improvement over the MMNL benchmark, while substantially outperforming the standard MNL and NL models.

\begin{table}[htbp]
\footnotesize
\centering
\begin{tabular}{lccccc}
\toprule
Category & \# Reg. & MNL & NL & MMNL & GNL \\
\midrule
beer     & 47 & 4.17 & 3.81 & 3.55 & 3.49$^{***}$ \\
carbbev  & 50 & 2.33 & 2.06 & 1.99 & 1.98 \\
cigets   & 50 & 3.82 & 3.56 & 3.56 & 3.44$^{**}$ \\
coffee   & 50 & 4.06 & 3.51 & 3.40 & 3.38$^{*}$ \\
coldcer  & 50 & 2.31 & 2.12 & 2.10 & 2.06$^{*}$ \\
deod     & 50 & 3.80 & 3.77 & 3.78 & 3.75 \\
fzdinent & 50 & 2.77 & 2.37 & 2.38 & 2.36 \\
fzpizza  & 50 & 3.17 & 2.87 & 2.80 & 2.78 \\
hhclean  & 50 & 2.94 & 2.72 & 2.67 & 2.68 \\
hotdog   & 44 & 3.63 & 3.21 & 3.04 & 3.05 \\
laundet  & 50 & 2.88 & 2.56 & 2.51 & 2.46$^{***}$ \\
\midrule
\textbf{Avg} & 49.18 & 3.26 & 2.96 & 2.89 & 2.86 \\
\bottomrule
\end{tabular}
~~~
\begin{tabular}{lccccc}
\toprule
Category & \# Reg. & MNL & NL & MMNL & GNL \\
\midrule
margbutr & 50 & 2.30 & 2.09 & 2.08 & 2.03$^{*}$ \\
mustketc & 49 & 3.17 & 3.07 & 2.92 & 2.91 \\
peanbutr & 46 & 3.25 & 2.81 & 2.80 & 2.74$^{**}$ \\
saltsnck & 50 & 2.70 & 2.43 & 2.43 & 2.40$^{***}$ \\
shamp    & 50 & 4.04 & 3.99 & 3.91 & 3.92 \\
soup     & 50 & 2.95 & 2.75 & 2.71 & 2.69$^{*}$ \\
spagsauc & 50 & 3.36 & 3.02 & 2.95 & 2.94 \\
toitisu  & 47 & 3.51 & 3.02 & 2.99 & 2.94$^{**}$ \\
toothbr  & 49 & 4.94 & 4.74 & 4.74 & 4.64$^{**}$ \\
toothpa  & 50 & 2.92 & 2.85 & 2.84 & 2.84 \\
yogurt   & 50 & 3.93 & 3.43 & 3.29 & 3.26 \\
\midrule
\textbf{Avg} & 49.18 & 3.37 & 3.11 & 3.06 & 3.03 \\
\bottomrule
\end{tabular}
\vspace{0.5em}

\caption{Prediction performance across product categories. Stars indicate the statistical significance of the difference between the GNL and MMNL models: \textsuperscript{*}$p<0.10$, \textsuperscript{**}$p<0.05$, and \textsuperscript{***}$p<0.01$.}
\label{tab:fit}
\end{table}
\vspace{-15pt}
We highlight that by construction, the MMNL and GNL models have exactly the same number of model parameters. Moreover, both models are sufficiently flexible to capture general substitution patterns under the RUM framework. Therefore, the consistent improvement of the GNL model over the MMNL model cannot simply be attributed to a larger number of parameters or greater model flexibility. Rather, the results suggest that the particular structure imposed by the GNL model may be better suited for capturing the substitution patterns observed in our retail datasets.

\vspace{-2pt}
\section{Conclusion}
\label{sec:conclusion}

We introduced product and nest graphs associated with  the generalized nested logit model, showed how the treewidths of these graphs can be used to characterize the complexity of the assortment optimization problem under this choice model, and developed exact and approximate algorithms for the assortment problem.
 We also evaluated the generalized nested logit model on both real and synthetic datasets and showed that it fits the data better than many existing models in the literature. There are many exciting directions for future research. We believe  the concept of treewidths on appropriately defined graphs can be applied to other assortment optimization problems as well.  It is also interesting to explore whether we can extend our algorithms to handle cardinality or capacity constraints.

\noindent{{\bf AI Use Declaration:}} There has been no AI use in any stage of writing this paper.

\vspace{-1mm}
\bibliographystyle{ormsv080}

\renewcommand*{\bibfont}{\footnotesize}
\renewcommand*{\bibfont}{\normalfont\footnotesize\linespread{1}\selectfont}
\setlength{\bibsep}{1.0pt}
\bibliography{references}

\newpage

\ECSwitch

\begin{center}
\vspace{-9mm}
{\large \bf \uline{Online Appendix}\\}
{\large \bf Tree Decomposition of Product and Nest Graphs for Assortment Optimization under the Generalized Nested Logit  Model} 
\end{center}

\begin{APPENDICES}

%\section{Computational Complexity}

\section{Proof of Results in Section \ref{sec:model}}

To characterize the computational complexity of the  \ref{eqn:cnl_assort} problem,  we consider the feasibility version of this problem. In the feasibility version, given an expected revenue threshold $L$, we ask whether there exists $S \subseteq N$ such that $\Pi(S) \geq L$, which is equivalent to 
\begin{align}\label{eqn:fractional}
    \sum_{k \in K}  ~\frac{\sum_{i \in S }v_{ik}(f_i - L)}{V_k(S)^{1- \gamma_k}} \geq L.
\end{align}
To show that the feasibility version of the \pr{assortment optimization problem under the} generalized nested logit model is NP-complete, we use a reduction from the partition problem, which is a well known NP-complete problem; see \cite{GaJo79}. The partition problem is defined as follows. 
We are given a set of items $N = \{1,\ldots,n\}$. The size of item $i$ is $c_i \in \mathbb Z_+$. Assuming that the total size of the items is even, set $\sum_{i \in N} c_i = 2 \ts T$ for some $T \in \mathbb Z_+$. In the partition problem, we ask whether there exists $S \subseteq N$ such that $\sum_{i \in S} c_i = T$. 

\subsection{Proof of Theorem \ref{thm:hardness_1}}\label{sec:hardness_1}

 Before proceeding with the proof, we first analyze a property of the function $g(x) = \frac{6 + x}{\sqrt{1 + x}} + \sqrt{x}$. Let $\alpha = \frac{13 - 5\sqrt{5}}{22}$. We observe that $\alpha$ is the maximizer of the function $g(x)$ for $x \in [0, 2\alpha]$.
    
Consider an instance $\{ c_1, c_2, \dots, c_n \}$ of the Partition problem. We construct an instance of the \mbox{\ref{eqn:cnl_assort}} problem with $n+1$ products and for $i \in \{ 1, 2, \dots, n \}$, we let
$$f_i = 1 + g(\alpha), \qquad v_{i1} = v_{i2} = \frac{\alpha}{T} c_i.$$
For the product $n+1$, we let:
$$f_{n+1} = 6+ g(\alpha), \qquad v_{n+1, 1} = 1, \qquad v_{n+1, 2} = 0. $$ 
We also let $\gamma_1 = \gamma_2 = 1/2$. Notice that product $n+1$ has the largest revenue so including the product strictly increases the expected revenue. Therefore, it is sufficient to consider the assortment with structure $ S\cup \{n+1\}$ for $ S\subseteq \{1, 2, \dots, n\}$. In the rest of the proof, we show that there exists $S\subseteq \{1, 2, \dots, n\}$ with $\Pi(S \cup \{n+1\}) \geq g(\alpha)$ if and only if there exists $S\subseteq \{1, 2, \dots, n\}$ such that $\sum_{i \in S} c_i = T$. To see this, it follows from \eqref{eqn:fractional} that the inequality $\Pi(S \cup \{n+1\}) \geq g(\alpha)$ holds if and only if 
\begin{align}\label{eqn:simplify}
    \frac{\sum_{i \in S\cup \{n+1\}} v_{i1}(f_i - g(\alpha)) }{V_1(S\cup \{n+1\})^{1-\gamma_1}} + \frac{\sum_{i \in S\cup \{n+1\}} v_{i2}(f_i - g(\alpha)) }{V_2(S\cup \{n+1\})^{1-\gamma_2}} \geq g(\alpha).
\end{align} 
By how we define our instance of the \ref{eqn:cnl_assort} problem, we have:
$$V_1(S \cup \{n+1\}) = 1 + \frac{\alpha}{T} \sum_{i \in S} c_i, \quad V_2(S \cup \{n+1\}) = \frac{\alpha}{T} \sum_{i \in S} c_i.$$
We then have \eqref{eqn:simplify} equivalent with:
$$\frac{6 + \frac{\alpha}{T} \sum_{i \in S} c_i }{\sqrt{1 + \frac{\alpha}{T} \sum_{i \in S} c_i}} + \sqrt{\frac{\alpha}{T} \sum_{i \in S} c_i} \geq g(\alpha).$$
Notice that $\sum_{i \in S} c_i \leq 2T$ so $\frac{\alpha}{T} \sum_{i \in S} c_i \in [0, 2\alpha] $. Since function $g(x) = \frac{6+x}{\sqrt{1+x}} + \sqrt{x}$ attains its maximum at $x = \alpha$ , we must have $\frac{6 + \frac{\alpha}{T} \sum_{i \in S} c_i }{\sqrt{1 + \frac{\alpha}{T} \sum_{i \in S} c_i}} + \sqrt{\frac{\alpha}{T} \sum_{i \in S} c_i} \geq g(\alpha)$ if and only if $\frac{\alpha}{T} \sum_{i \in S} c_i = \alpha$. The last equality simplifies to $\sum_{i \in S} c_i = T$. Therefore, we show that there exists $S\subseteq \{1, 2, \dots, n\}$ with $\Pi(S \cup \{n+1\}) \geq g(\alpha)$ if and only if $\sum_{i \in S} c_i = T$ and the \ref{eqn:cnl_assort} problem is NP-hard. 

\subsection{Proof of Theorem \ref{thm:hardness_2}}\label{sec:hardness_2}

The proof outline follows similar construction as proof of Theorem \ref{thm:hardness_1}. We first analyze the property of the function $g(x) = \sqrt{x} - \frac{x}{2}$. We observe that the maximizer of the function $g(x)$ is at $ x = 1 $ with $g(1) = 1/2$.

Consider an instance $\{ c_1, c_2, \dots, c_n \}$ of the Partition problem. Notice that we can scale all the values $c_i$ by the same factor without changing the optimal solution, so without loss of generality we scale $c_i$ in a way that $\sum_{i \in N} c_i = 2$ and $T = 1$. We construct an instance of the \ref{eqn:cnl_assort} problem with $2n$ products and $n+1$ nests. The nests structure is specified as follows. Let $M_j = \{j, j + n\}$ for $j \leq n$ and $M_{n+1} = \{1, 2, \dots, n\}$. We let $\gamma_j = 1/2$ for all nests. We then specify the preference weight and revenues. In the following discussion, we denote $f = \frac{3}{2}\sqrt{2} + 2$. For product $i \in \{1, 2, \dots, n\}$, notice that we have $i \in M_i$ and $i \in M_{n+1}$. We let:
$$v_{ii} = c_i^2, \quad v_{i, n+1} = c_i, \quad v_{ij} = 0~\forall j \notin \{i, n+1\}, \quad f_i = 2f + 3/2.$$
For product $i \in \{n+1, n+2, \dots, 2n\}$, we must have $i$ contained in nest $ M_{i-n}$ so we let:
$$v_{i, i-n} = c_{i-n}^2,\quad v_{ij} = 0~\forall j \neq i-n, \quad f_i = 3f + 1/2.$$
Let $N = \{1,2,\ldots,n\}$ and $N' = \{n+1,n+2,\ldots,2n\}$. \yw{Since every product in $N'$ generates higher revenue than any product in $N$, adding any product from $N'$ to an assortment strictly increases its expected revenue.} Therefore, it is sufficient to consider the assortment with the structure $S \cup N'$ for $S \subseteq N$.  In the rest of the proof, we show that there exists $S\subseteq \{1, 2, \dots, n\}$ with $\Pi(S \cup N' ) \geq 2f + 1/2$ if and only if there exists $S\subseteq \{1, 2, \dots, n\}$ such that $\sum_{i \in S} c_i = 1$. 

We proceed by using \eqref{eqn:fractional}. Notice that if we offer an assortment $S$, we have $(S \cup N') \cap M_j = \{j, j+n\}$ if $j \in S$ and $(S \cup N') \cap M_j = \{j + n\}$ if $j \in N \setminus S $. Also, we have $(S \cup N') \cap M_{n+1} = S $. Considering all the parameters we set, we have for a nest $j$
\begin{align*}
    \frac{\sum_{i \in S \cup N' }v_{ij}(f_i - 2f - 1/2)}{V_j(S \cup N')^{1- \gamma_j}} = \begin{cases}
        \frac{c_j + c_j f }{\sqrt{2}} & \text{ if } j \in S \\
        c_j f & \text{ if } j \in N \setminus S \\
        \sqrt{\sum_{i \in S} c_i} &  \text{ if } j = n+1.
    \end{cases}
\end{align*}
Therefore, we have:
\begin{align*}
    \sum_{j =1}^{n+1} ~\frac{\sum_{i \in S \cup N' }v_{ij}(f_i - 2f - 1/2)}{V_j(S \cup N')^{1- \gamma_j}} & =  \sum_{j \in S} \frac{c_j + c_j f }{\sqrt{2}} +  \sum_{j \in N \setminus S} c_j f + \sqrt{\sum_{i \in S} c_i} \\
    & = \sum_{j \in S} \left( \frac{c_j}{\sqrt{2}} + (\frac{1}{\sqrt{2}} - 1)c_j f + c_j f\right) +  \sum_{j \in N \setminus S} c_j f + \sqrt{\sum_{i \in S} c_i} \\
    & = \sum_{i \in N} c_i f + \sum_{i \in S}\left( \frac{1}{\sqrt{2}} + (\frac{1}{\sqrt{2}} - 1)f \right) c_i+ \sqrt{\sum_{i \in S} c_i} \\
    & \overset{(i)}{=} 2f - \frac{\sum_{i \in S} c_i}{2}+ \sqrt{\sum_{i \in S} c_i},
\end{align*} 
where we use $\sum_{i\in N} c_i = 2$ and $f = \frac{3}{2}\sqrt{2} + 2$ in $(i)$. Using the result we discussed at the beginning of the proof, we have the function $2f - x/2 + \sqrt{x}$ attains its maximum at $x = 1$ so $2f - \frac{\sum_{i \in S} c_i}{2}+ \sqrt{\sum_{i \in S} c_i} \geq 2f + 1/2$ if and only if there exists an assortment $S \subseteq N$ such that $\sum_{i \in S} c_i = 1$. Therefore, we show that there exists $S\subseteq \{1, 2, \dots, n\}$ with $\Pi(S \cup N') \geq 2f + 1/2$ if and only if $\sum_{i \in S} c_i = 1$ and the \ref{eqn:cnl_assort} problem is NP-hard. 

\section{Proof of Results in Section \ref{sec:product-and-nest-graphs}}

\subsection{Proof of Theorem \ref{thm:using-fixed-point-to-solve-assortment-problem}}
\label{sec:using_fixed_point_proof}

Before proving the theorem, we first prove an auxiliary lemma where we bound the optimal objective value of the \ref{eqn:cnl_assort} problem.

\begin{lem}\label{lem:cnl_bound}
    Let $S^*$ be the optimal solution to the \ref{eqn:cnl_assort} problem and $i^* = \argmax_{i \in N } \Pi(\{i\})$, we have $\Pi(\{i^*\}) \le \Pi(S^*) \le n\,\Pi(\{i^*\}).$
\end{lem}
\noindent{\it Proof.} 
The inequality $\Pi(\{i^*\}) \le \Pi(S^*)$ follows trivially since $\{i^*\}$ is a feasible solution to the \ref{eqn:cnl_assort} problem. To see $\Pi(S^*) \le n\,\Pi(\{i^*\})$, for any assortment $S$, we upper bound $\Pi(S)$ by
\begin{align*}
    \Pi(S) & = \sum_{k \in K} \frac{V_k(S)^{\gamma_k}}{1 + \sum_{\ell \in K} V_\ell(S)^{\gamma_\ell}} \frac{\sum_{i \in S} v_{ik} f_i}{V_k(S)} 
    = \sum_{ i \in S} \sum_{k \in \pr{\nest(i)}} \frac{V_k(S)^{\gamma_k - 1}}{1 + \sum_{\ell \in K} V_\ell(S)^{\gamma_\ell}}  v_{ik} f_i \\
    & \overset{(i)}{\leq} \sum_{ i \in S} \sum_{k \in \pr{\nest(i)}} \frac{v_{ik}^{\gamma_k - 1}}{1 + \sum_{\ell \in K} v_{i\ell}^{\gamma_\ell}}  v_{ik} f_i = \sum_{ i \in S} \sum_{k \in K} \frac{v_{ik}^{\gamma_k }}{1 + \sum_{\ell \in K} v_{i\ell}^{\gamma_\ell}}  f_i 
    = \sum_{ i \in S} \Pi(\{i\}) \overset{(ii)}{\leq} |S| \cdot  \Pi(\{i^*\}),
\end{align*}
\noindent where $(i)$ follows by the observation that $V_k(S)^{\gamma_k - 1} \leq v_{ik}^{\gamma_k - 1}$ and $V_k(S)^{\gamma_k} \geq v_{ik}^{\gamma_k}$ because \mbox{$\gamma_k \in (0, 1]$} for all $k$ and $v_{ik} > 0$ for all $k \in \nest(i)$. Then, $(ii)$ follows by the definition of $i^*$. The inequality implies that we must have \mbox{$\Pi(S^*) \leq n \cdot \Pi(\{i^*\})$}. \qed

Here is the proof of  Theorem \ref{thm:using-fixed-point-to-solve-assortment-problem}.

\noindent {\bf \underline{Proof of  Theorem \ref{thm:using-fixed-point-to-solve-assortment-problem}}:} Let $S^*$ be an optimal solution to the \ref{eqn:cnl_assort} problem, and let \mbox{$Z^*=\Pi(S^*)$}. We first show that we can efficiently obtain a value $\zhat$ satisfying 
$
Z^*/(1+\epsilon)\le \zhat \le Z^*.
$
Let $i^* \in \arg\max_{i\in N}\Pi(\{i\})$. By Lemma~\ref{lem:cnl_bound}, we have 
$
\Pi(\{i^*\}) \le Z^* \le n\,\Pi(\{i^*\}).
$
By constructing a geometric grid over this interval with multiplicative spacing $(1+\epsilon)$, we obtain $O(\log(n)/\epsilon)$ candidate values, and at least one of them must satisfy 
$
Z^*/(1+\epsilon)\le \zhat \le Z^*.
$

Now fix such a value $\zhat$, and let $\Shat$ be a $(1-2\epsilon)$-approximate solution to $F(\zhat)$. We make one observation. Since 
$
\Pi(S^*) = 
\frac{\sum_{k\in K}V_k(S^*)^{\gamma_k}R_k(S^*)}
{1+\sum_{\ell\in K}V_\ell(S^*)^{\gamma_\ell}}
= Z^*,
$
a simple rearrangement yields
$
\sum_{k\in K}\frac{\sum_{i\in S^*}v_{ik}(f_i-Z^*)}{V_k(S^*)^{1-\gamma_k}}=Z^*.
$
We then have the following chain of inequalities
\begin{align*}
        \sum_{k \in K}  \frac{\sum_{i \in \Shat} v_{ik} (f_i - \zhat)}{V_k(\Shat)^{1-\gamma_k}} & \geq (1-2\epsilon) F(\zhat\,) 
        \overset{(i)}{\geq}  
        (1-2\epsilon)\sum_{k \in K}  \frac{\sum_{i \in S^*} v_{ik} (f_i - \zhat)}{V_k(S^*)^{1-\gamma_k}}  \\
         & \geq (1-2\epsilon)\sum_{k \in K}  \frac{\sum_{i \in S^*} v_{ik} (f_i - Z^*)}{V_k(S^*)^{1-\gamma_k}} \overset{(ii)}{=} (1-2\epsilon) Z^*,
    \end{align*}
where $(i)$ follows because $S^*$ is a feasible solution of the optimization defined over $F(\,\zhat\,)$ and $(ii)$ follows by the observation made at the beginning. Since $Z^*/(1+\epsilon) \leq \zhat$, the chain of the inequalities above gives 
\begin{align*}
    \sum_{k \in K}  \frac{\sum_{i \in \Shat} v_{ik} (f_i - \frac{Z^*}{1+\epsilon})}{V_k(\Shat)^{1-\gamma_k}} \geq \sum_{k \in K}  \frac{\sum_{i \in \Shat} v_{ik} (f_i - \zhat)}{V_k(\Shat)^{1-\gamma_k}} \geq (1-2\epsilon) Z^*.
\end{align*}
Focusing on the first and last terms and rearranging the terms gives us
\begin{align*}
    \sum_{k \in K}  \frac{\sum_{i \in \Shat} v_{ik} f_i}{V_k(\Shat)^{1-\gamma_k}} \geq (1-2\epsilon) Z^* + \frac{Z^*}{1+\epsilon} \sum_{k \in K} V_k(\Shat)^{\gamma_k}  \geq \big(1-2\epsilon) \big(1 + \sum_{k \in K} V_k(\Shat)^{\gamma_k}\big)Z^*.
\end{align*}
Consider the first and last terms and dividing both sides by $1 + \sum_{k \in K} V_k(\Shat)^{\gamma_k}$ gives us $\sum_{k \in K} \frac{V_k(\Shat)^{\gamma_k}}{1 + \sum_{k \in K} V_k(\Shat)^{\gamma_k}} \frac{\sum_{i \in \Shat} v_{ik} f_i}{V_k(\Shat)} \geq (1-2\epsilon) Z^*$. Therefore, we have $\Pi(\Shat) \geq (1-2\epsilon) Z^*$. Finally, we consider the running time. To retrieve the value of $\zhat$, we need to enumerate the value over a geometric grid with size $O(\log(n)/\epsilon)$. For each enumeration, we run the FPTAS for $F(z)$ \pr{with running time that is polynomial in $m$, $n$, and $1/\epsilon$.  
Then, the final running time of the algorithm is still 
polynomial in $m$, $n$, and $1/\epsilon$, giving the FPTAS for the \ref{eqn:cnl_assort} problem.}
%$O( ~m^2 \cdot (\frac{n}{\epsilon})^{O(\treewidth(G_\nest))} ~ )$. 
%$O( ~m^2 \cdot (\frac{n}{\epsilon})^{O(\treewidth(G_\nest))} ~ )$. 
\qed

\subsection{Proof of Theorem \ref{thm:subproblem_hardness}}\label{sec:subproblem_hardness}

In the proof, we describe a reduction that maps any instance of the maximum independent set problem defined over a graph $G = (V, E)$ where we let $V = \{u_1, u_2, \dots,u_n \}$ to an instance of the  \ref{eqn:sub_param_new} problem. For each vertex $u_i \in V$, we use $N(i)$ to \pr{denote the neighbors of $i$ with smaller indices than $i$}; that is, $N(i) = \{ j \in \{1,\ldots,n\}:(u_i, u_j) \in E \text{ and } j < i \}$. 

%be the index of $i$ with index smaller than $i$
    
Given an instance of the maximum independent set problem, we construct an instance of the \ref{eqn:sub_param_new} problem with $n$ products and $n$ nests so $N = \{1,2,\dots, n\}$ and $K = \{1, 2, \dots, n \}$. For a nest $k \in K$, we let $M_k = N(k) \cup \{k\}$. Let $i \in N$, we set the preference weight $v_{ik}$ as follows:
\begin{align*}
    v_{ik} = \begin{cases}
        n^{-6nk} & \text{ if } i = k \\
        n^{-6nk + 6k} & \text{ if } i \in N(k) \\
        0 & \text{ otherwise}.
    \end{cases}
\end{align*}
The revenue is $r_i = n^{6n(i-1/2)}$ for product $i \in N$ and the sensitivity parameter $\gamma_k = 1 - \frac{1}{2k}$ for nest~$k$. In the following, we let $R(S) = \sum_{k \in K}  \frac{ \sum_{i \in S} v_{ik}  r_i }{V_k(S)^{1-\gamma_k}}$ which corresponds to the objective function of the  \ref{eqn:sub_param_new} problem when an assortment $S$ is offered. 

We first show that for any independent set $U \subseteq V$, there exists a corresponding assortment $S_U$ with $R(S_U) \geq |U|$. To show this, we let $S_U = \{i : u_i \in U\}$. To evaluate $R(S_U)$, we start with an observation that since $U$ is an independent set, we must have $N(k) \cap S_U = \varnothing$ for $k \in S_U$. Therefore, the only product that is offered in nest $k$ is product $k$. We can then lower bound $R(S_U)$ by
\begin{align*}
    R(S_U) = \sum_{k \in K} \frac{ \sum_{i \in S_U} v_{ik}r_i   }{V_k(S_U)^{1-\gamma_k}} \geq  \sum_{k \in S_U } \frac{ n^{6n(k-1/2)} \cdot n^{-6nk}  }{(n^{-6nk})^{1/(2k)}} = |U|.
\end{align*}

Next, we show that for any assortment $S \subseteq N$, one can compute in polynomial time an independent set $U_S$ whose cardinality is at least $\rd{R(S)}$. Define $\bar{S} = \{i \in S : S \cap N(i) = \varnothing \}$. In what follows, we analyze the contribution to the objective function from three disjoint sets of nests: $\bar{S}$, $S \setminus \bar{S}$, and $K \setminus S$. First, let $k \in  \bar{S}$. By the definition of $\bar{S}$, we have $M_k \cap S = \{k\}$. Therefore the contribution of nest $k$ to the objective function is 
$\frac{ \sum_{i \in  S} v_{ik} r_i  }{ V_k(S)^{1-\gamma_k}} = \frac{ n^{6n(k-1/2)} \cdot n^{-6nk}  }{(n^{-6nk})^{1/(2k)}} = 1$.

We then consider $k \in S \setminus \bar{S}$, where we have $M_k \cap S = \{k\} \cup( N(k) \cap S)$. We can upper bound the contribution of nest $k$ to the objective function as follows
\begin{align*}
    \frac{ \sum_{i \in S}  v_{ik}r_i  }{V_k(S)^{1-\gamma_k}} & = \frac{v_{kk}r_k + \sum_{i \in N(k) \cap S} v_{ik}r_i}{(v_{kk} + \sum_{i \in N(k) \cap S} v_{ik})^{1-\gamma_k}} 
    = \frac{n^{-3n}+ n^{-6nk + 6k} \sum_{i \in N(k) \cap S} r_i}{(n^{-6nk} + n^{-6nk + 6k} |N(k) \cap S| )^{1/(2k)}} \\
    & \overset{(i)}{\leq} \frac{n^{-3n} + n^{-6nk + 6k} \cdot n \cdot n^{6n(k-3/2)}}{n^{-3n + 3}} 
    = \frac{n^{-3n} + n^{6k-9n+1}}{n^{-3n + 3}} 
    = \frac{1 + n^{6k-6n+1}}{n^{3}} 
    \overset{(ii)}{\leq} \frac{2}{n^2},
\end{align*}
where in $(i)$, we use the observation that $r_i \leq n^{6n(k-3/2)}$ for all $i \in N(k)$. Then, $(ii)$ follows by the fact that $k \leq n $. Finally, we let $k \in K\setminus S$, now we have $M_k \cap S = N(k) \cap S$. A similar derivation as previous case gives us:
\begin{align*}
    \frac{ \sum_{i \in S}   v_{ik}r_i  }{ V_k(S)^{1-\gamma_k}}  = \frac{\sum_{i \in N(k) \cap S} v_{ik}r_i}{(\sum_{i \in N(k) \cap S} v_{ik})^{1-\gamma_k}} 
    \leq \frac{ n^{-6nk + 6k} \cdot n \cdot n^{6n(k-3/2)}}{( n^{-6nk + 6k})^{1/(2k)}} \leq \frac{1}{n^2}.
\end{align*}
Consequently, the total contribution of the nest under $K \setminus \bar{S}$ is upper bounded by $2/n$.
\pr{By the preceding discussion, we have $|\bar{S}|  \leq R(S) \leq |\bar{S}| + 2/n$, which implies that $\rd{R(S)} = |\bar{S}|$.}
%This means that precisely $\rd{R(S)}$ contribute 1 to the objective function, and therefore $|\bar{S}| = \rd{R(S)}$. 
We then claim that the vertex set defined by $U_S = \{u_i : i \in \bar{S}\}$ forms an independent set in $G$. By contradiction, assume there exists $u_i, u_j \in U_S$ such that $(u_i, u_j) \in E$ and $i < j$. Then we must have $i \in N(j)$ and therefore $j \notin \bar{S}$, which raises a contradiction.

\pr{
Letting $I^* = \max\{ \, |U| \, \,:\, U \text{ is an independent set}\}$ denote the size of the largest independent set and  $F^* = \max_{S \subseteq N} R(S)$ denote the maximum revenue, the above analysis shows that \mbox{$F^* - \frac{2}{n} \leq I^* \leq F^*$}.  Moreover, if an assortment $\widehat{S}$ is an $\alpha$-approximation of $F^*$, then we can construct an independent set $\hat{U}$ such that
$$
    | \widehat{U} | ~\geq~ R \big( \, \widehat{S} \, \big) - \frac{2}{n} ~\geq~ \alpha F^*  - \frac{2}{n} ~\geq~ \alpha I^* - \frac{2}{n} ~\geq~ \left( \alpha  - \frac{2}{n} \right) \; I^*~,
$$
where the last inequality follows because $I^* \geq 1$. This shows that $\widehat{U}$ is a $(\alpha - \frac{2}{n})$-approximation to the maximum independent set.
%Altogether, this shows that an $\alpha$-approximation for solving $F(z)$ implies an $\alpha$-approximation for the maximum independent set problem. 
Because the maximum independent set is hard to approximate within a factor better than $O( 1/n^{1-\delta}) $ for any constant $\delta > 0$, the above reduction implies the same hardness of the \ref{eqn:sub_param_new} problem.} \qed

\section{Proofs of Results in  Section \ref{sec:fptas}} \label{sec:fptas_additional}

\subsection{Proof of Lemma \ref{lem:nest_graph_domain}}

We will prove the lemma by contradiction. Suppose, on the contrary, that there exists $k \in \nest(i)$ such that $k \notin \subtree{q} \cupdot \head{q}$. Because \mbox{$i \in \product(\subtree{q} )$}, it follows that there must exist $k' \in \subtree{q} $ such that $i \in M_{k'}$. We let $c$ be the node in $\descendant(q)$ that contains nest $k'$, and thus, $k' \in \vtcs{c}$. By construction, we have $i \in M_{k'} \cap M_{k} $ and therefore there exists an edge $(k, k')$ in the nest graph. By Property 2 of the tree decomposition, there exists $a \in \bag_\nest$ such that $k,k' \in \vtcs{a}$. Then, since $k \notin \subtree{q} $, node $a$ does not lie under the subtree under $q$. Notice that $a$ does not lie under the subtree of $q$ while $c$ lies under the subtree of $q$, so we must have nodes $q$ and $\parent(q)$ on the path between node $a$ and $c$. By Property~3 of the tree decomposition, it follows that $\vtcs{a} \cap \vtcs{c} \subseteq \vtcs{q}$ and $\vtcs{a} \cap \vtcs{c} \subseteq \vtcs{\parent(q)}$.  Because $k' \in \vtcs{a}  \cap \vtcs{c}$, it follows that
$ k' \in \vtcs{q}$ and $k' \in \vtcs{\parent(q)}$,  so $k' \in \head{q}$. However, this is a contradiction!  Therefore, $\nest(i) \subseteq   \subtree{q} \cupdot \head{q}$. \qed

\subsection{Proof of Lemma \ref{lem:nest_graph_prod_sep}}

Consider an arbitrary $q \in \bag_{\nest}$. For notational brevity, we define
$\Dcal_q = \subtree{q} \setminus \vtcs{q}$
and \mbox{$\Nbar_q = \big\{i \in \product(\tail{q}) : \nest(i) \subseteq \vtcs{q} \big\}$}. 
We  partition $\product(  \subtree{q}  )$ into two disjoint sets:
\begin{align*}
    \product(  \subtree{q}  ) = \Big( \product( \subtree{q} ) \setminus  \product( \Dcal_q) \Big) \bigcupdot \product( \Dcal_q).
\end{align*}
To prove the lemma, it suffices to establish the following two claims.

{\bf Claim 1:} $ \product( \subtree{q} ) \setminus  \product( \Dcal_q) = \Nbar_q$.

We begin with the first claim. To see that  $\Nbar_q \subseteq \product\big( \subtree{q}  \big) \setminus \product\big(  \Dcal_q \big)$, we consider an arbitrary $j \in \Nbar_q$. By definition, we have $j \in \product(\tail{q})$ such that $\nest(j) \subseteq \vtcs{q}$. It then follows that \mbox{$j \in \product\big( \subtree{q}  \big)$} because $\tail{q} \subseteq \subtree{q} $, and $j \notin \product\big(  \Dcal_q \big)$ because $\nest(j) \subseteq \vtcs{q}$. This gives us \mbox{$j \in \product\big( \subtree{q}  \big) \setminus \product\big(  \Dcal_q \big)$}, and thus, it follows that $\Nbar_q \subseteq \product\big( \subtree{q}  \big) \setminus \product\big(  \Dcal_q \big)$. 

To establish the other containment, consider an arbitrary $j \in \product\big( \subtree{q}  \big) \setminus \product\big(  \Dcal_q \big)$. 
Because \mbox{$\Dcal_q = \subtree{q} \setminus \vtcs{q}$},
$\subtree{q} =  \left( \bigcupdot_{a \in \children(q)} \subtree{a} \right) \bigcupdot \tail{q}$ by Lemma~\ref{lem:sep_vertices}, and \mbox{$\vtcs{q} = \tail{q} \cupdot \head{q}$}, it follows that we must have $j \in \product\big( \tail{q}  \big)$. Moreover,
it follows from Lemma~\ref{lem:nest_graph_domain} that $ \nest(j) \subseteq \subtree{q} \cupdot \head{q}$. Because $j \notin \product\big(  \Dcal_q \big)$, we must have \mbox{$\nest(j) \cap \Dcal_q = \varnothing$}. Therefore, it follows that  $\nest(j) \subseteq \subtree{q} \cup \head{q}  \setminus \Dcal_q = \vtcs{q}$.  Thus, we have that $j \in \product\big( \tail{q} \big)$ and $\nest(j) \subseteq \vtcs{q}$, which means that $j \in \Nbar_q$.  
This gives $\product\big( \subtree{q}  \big) \setminus \product\big(  \Dcal_q \big) \subseteq \Nbar_q$ and therefore $\Nbar_q = \product( \subtree{q} ) \setminus  \product( \Dcal_q)$.

{\bf Claim 2:} $ \product( \Dcal_q) = \bigcupdot_{a \in \children(q)} \product(\subtree{a})$.

We proceed to show the second claim. By Lemma \ref{lem:sep_vertices}, 
$\subtree{q} =  \left( \bigcupdot_{a \in \children(q)} \subtree{a} \right) \bigcupdot \tail{q}$, and it thus follows that
$\Dcal_q =  \bigcupdot_{a \in \children(q)} \subtree{a}$. By definition, we must have $\product( \Dcal_q) = \product(\bigcupdot_{a \in \children(q)} \subtree{a}) = \bigcup_{a \in \children(q)} \product(\subtree{a})$. To establish the second claim, it suffices to show that \mbox{$\product(\subtree{a}) \cap \product(\subtree{a'}) = \varnothing$} for $a, a' \in \children(q)$ such that $a \ne a'$. Suppose, on the contrary, that there exists $i \in \product(\subtree{a}) \cap \product(\subtree{a'}) $ where $a \ne a'$. Then, by Lemma \ref{lem:nest_graph_domain}, we have \mbox{$\nest(i) \subseteq \subtree{a} \cupdot \head{a}$} and $\nest(i) \subseteq \subtree{a'} \cupdot \head{a'}$. Notice that in Lemma \ref{lem:sep_vertices}, we show that $\subtree{a} \cap (\subtree{a'}\cup \head{a'}) = \varnothing $ so we must have $\nest(i) \subseteq \head{a} \cap \head{a'} $. This indicates that $i \notin  \product(\subtree{a}) $ and $i \notin \product(\subtree{a'})$. However, this is a contradiction! Thus,  \mbox{$\product(\subtree{a}) \cap \product(\subtree{a'}) = \varnothing$} for $a, a' \in \children(q)$ such that $a \ne a'$, and this completes Claim 2. \qed 

\subsection{Proof of Lemma \ref{lem:nest_approx}} \label{sec:nest_approx}

Let $(\, \Shat, \tvechat\,\,)$ be an optimal solution to the \ref{eqn:nest_reformulate} problem and let $S^*$ be the optimal solution to  the \ref{eqn:sub_param_new} problem. We let $t^*_k \in \grid_k(\epsilon)$ such that $\frac{t^*_k}{1+\epsilon} \leq V_k(S^*) \leq t^*_k$. Such $t^*_k$ must exist by the construction of $\grid_k$. 

To prove the claim, we make two observations. First, because $(\, \Shat, \tvechat\,\,)$ is a feasible solution for the \ref{eqn:nest_reformulate} problem, we have that 
$
\sum_{i \in \Shat }  \Big(\frac{4nv_{ik}}{\epsilon \, \that_k} - 1 \Big) \leq  \sum_{i \in \Shat } \Big \lfloor \frac{4nv_{ik}}{\epsilon \, \that_k} \Big \rfloor \leq  \Big \lceil \frac{4n}{\epsilon} \Big \rceil \leq \frac{4n}{\epsilon} + 1$, which implies that
$$
	\sum_{i \in \Shat} v_{ik} \leq \frac{ \epsilon \that_k}{4n} \left( \frac{4n}{\epsilon} + 1 + |\Shat| \right) \leq \that_k + \frac{n+1}{4n} \epsilon \that_k \leq (1 + \epsilon)\that_k~.
$$
Second, we claim that $(S^*, \tvec^*)$ is a feasible solution for the \ref{eqn:nest_reformulate} problem. To see this, let $k \in K$ and we have the following chain of inequalities
$$
\sum_{i \in S^* } \Big \lfloor \frac{4nv_{ik}}{\epsilon \, t^*_k} \Big \rfloor \leq \sum_{i \in S^* }  \frac{4nv_{ik}}{\epsilon \, t^*_k} \overset{(i)}{\leq} \frac{4n}{\epsilon} \leq \Big \lceil \frac{4n}{\epsilon} \Big \rceil ,
$$
where $(i)$ follows by $V_k(S^*) \leq t^*_k$. Using these two observations, we obtain the following chain of inequalities where we lower bound the objective function achieved under $\Shat$ 
\begin{align*}
    & \sum_{k \in K} \sum_{i \in \Shat} \frac{v_{ik} r_i}{V_k(\Shat)^{1-\gamma_k}}  \overset{(ii)}{\geq} \sum_{k \in K} \sum_{i \in \Shat} \frac{v_{ik} r_i}{\big((1+\epsilon)\that_k \big)^{1-\gamma_k}} 
    \overset{(iii)}{\geq} (1-\epsilon)\sum_{k \in K} \sum_{i \in \Shat} \frac{v_{ik} r_i}{\that_k^{\, 1-\gamma_k}} \\
    & \qquad \overset{(iv)}{\geq} (1-\epsilon)\sum_{k \in K} \sum_{i \in S^*} \frac{v_{ik} r_i}{(t^*_k)^{\, 1-\gamma_k}} 
    \overset{(v)}{\geq} (1-\epsilon)\sum_{k \in K} \sum_{i \in S^*} \frac{v_{ik} r_i}{((1+\epsilon)V_k(S^*))^{\, 1-\gamma_k}} \\
    & \qquad \qquad \overset{(vi)}{\geq} (1-\epsilon)^2\sum_{k \in K} \sum_{i \in S^*} \frac{v_{ik} r_i}{V_k(S^*)^{\, 1-\gamma_k}} \geq (1-2\epsilon) F^*,
\end{align*}
where $(ii)$ follows from our first observation, and $(iii)$ follows from $(1+\epsilon)^{1-\gamma_k} \leq 1+\epsilon$ when $\gamma_k \in (0, 1]$. Also, we have $ \frac{1}{1+\epsilon} \geq (1-\epsilon)$. Then, $(iv)$ follows because $(S^*, \tvec^*)$ is a feasible solution for the \mbox{\ref{eqn:nest_reformulate}} problem but $(\Shat, \tvechat\,\,) $ is the optimal solution. Finally, $(v)$ follows because $t^*_k/(1+\epsilon) \leq V_k(S^*)$ by construction and $(vi)$ uses $(1+\epsilon)^{1-\gamma_k} \leq (1+\epsilon)$ and $ \frac{1}{1+\epsilon} \geq (1-\epsilon)$ again. The last inequality is from $(1-\epsilon)^2 \geq (1-2\epsilon)$. \qed

\subsection{Proof of Theorem \ref{thm:fptas_dp_correctness}}
\label{sec:proof-dp-equation-nest-graph}

Consider an arbitrary $q \in \bag_\nest$ and $(\hvec,\gvec) \in \budget_q \times \prefW_q$. Define
\begin{align}
\Zhat_q ~=~ \displaystyle
\max_{ (S_q,\tvec),\,  ( \wvec_a \,:\,  a \in \children(q) ) }
&~~~\sum_{i \in S_q} \theta_i(\gvec + \tvec)
+
\sum_{a \in \children(q)} J^\nest_a\big( \wvec_a ,\, \gvec + \tvec \big)
\label{eqn:nest_graph_dp}  \tag{{\sf DP-Recursion}}\\
{\rm s.t.} ~~~~~~~&
~~~\sum_{i \in S_q} \rdBig{\frac{4nv_{ik}}{\epsilon\,g_k}} ~+~ \sum_{a \in \children(q)} w_{ak} ~\leq~ h_k
&&\!\! \forall k \in \head{q}, \notag
\\
&
~~~\sum_{i \in S_q} \rdBig{\frac{4nv_{ik}}{\epsilon\,t_k}} ~+~ \sum_{a \in \children(q)} w_{ak} ~\leq~ \ruBig{\frac{4n}{\epsilon}}
&&\!\! \forall k \in \tail{q}, \notag\\
&~~~~(S_q, \tvec) \in \reduceA_q,~~~  \wvec_a \in \budget_a~~\forall~a \in \children(q) \nonumber. 
\end{align}
We will show that $\Zhat_q = J^{\nest}_q(\hvec,\gvec)$.  
We proceed in showing two claims, where we first establish that $\Zhat_q \ge J^{\nest}_q(\hvec,\gvec)$, and we then show $\Zhat_q \le J^{\nest}_q(\hvec,\gvec)$.

{\underline{\it Claim 1: $\Zhat_q \ge J^{\nest}_q(\hvec,\gvec)$}}. We let $(\Shat, \tvechat \,\,)$ be the optimal solution for $J^{\nest}_q(\hvec,\gvec)$. We construct a feasible solution for the \ref{eqn:nest_graph_dp} problem. Let $\Stild_q = \Shat \cap \Nbar_q$ and $\ttild_k = \idc(k \in \tail{q})\, \that_k$ for all $k \in K$. For $a \in \children(q)$, we define
\begin{align*}
    \wtild_{ak} = \begin{cases}
        \sum_{i \in \Shat \cap N_a} \rdbig{\cfrac{4nv_{ik}}{\epsilon\,g_k}} & ~\text{if}~ k \in \head{q} \\
        \sum_{i \in \Shat \cap N_a} \rdbig{\cfrac{4nv_{ik}}{\epsilon\, \that_k}} & ~\text{if}~ k \in \tail{q} \\
        0 & ~\text{ otherwise}.
    \end{cases}
\end{align*}
We show the solution is indeed a feasible solution for the \ref{eqn:nest_graph_dp} problem. Both $\Stild_q$ and $\tvectild$ are clearly feasible for the \ref{eqn:nest_graph_dp} problem. We then show all the constraints are satisfied. We start by considering $k \in \head{q}$. Since $\Shat$ is a feasible solution for $J^{\nest}_q(\hvec,\gvec)$,  Lemma \ref{lem:nest_graph_prod_sep} gives us
$$\sum_{i \in \Shat } \rdbig{\frac{4nv_{ik}}{\epsilon\,g_k}} 
= 
\sum_{i \in \Shat \cap \Nbar_q } \rdbig{\frac{4nv_{ik}}{\epsilon\,g_k}} + \sum_{a \in \children(q)} \sum_{i \in \Shat \cap N_a } \rdbig{\frac{4nv_{ik}}{\epsilon\,g_k}} 
= 
\sum_{i \in \Stild_q } \rdbig{\frac{4nv_{ik}}{\epsilon\,g_k}} + \sum_{a \in \children(q)} \wtild_{ak} \leq h_k.$$
Similarly, we have $ \sum_{i \in \Stild_q } \rdbig{\frac{4nv_{ik}}{\epsilon\,\that_k}} + \sum_{a \in \children(q)} \wtild_{ak} \leq \rubig{\frac{4n}{\epsilon}}$ for $k \in \tail{q}$ so the solution we construct is  feasible for the \ref{eqn:nest_graph_dp} problem. Next, we show that we can lower bound $\Zhat_q$ by $J^\nest_q\big( \,\hvec ,\, \gvec\, \big)$. We make two observations. First, recall that $\Stild_q = \Shat \cap \Nbar_q$, and every product $i \in \Nbar_q$ satisfies $\nest(i) \subseteq \vtcs{q}$. Hence, in order to evaluate $\theta_i(\cdot)$ for such products, we only require the entries corresponding to nests in $\vtcs{q}$. Since $\ttild_k = \idc(k \in \tail{q})\,\that_k$ and $\gvec$ only has nonzero entries only on $\head{q}$, it follows that
$
\theta_i(\gvec + \tvechat\,) = \theta_i(\gvec + \tvectild\,)
$
for all $i \in \Stild_q$. 

Second, define a vector $\tvectild_a \in \grid$ by $\ttild_{ak} = \idc(k \in \subtree{a})\,\that_k$ for $k \in K$ and we claim that \mbox{$(\Shat \cap N_a,\tvectild_a)$} is feasible for $J^\nest_a\big(\wvectild_a ,\, \gvec + \tvectild \,\,\big)$. To see this, we examine the constraints by cases. First, suppose \mbox{$k \in \head{a} \cap \head{q}$}. By construction $g_k + \ttild_k = g_k$ and feasibility then follows directly from
$
\sum_{i \in \Shat \cap N_a}
\rdbig{\frac{4nv_{ik}}{\epsilon\,g_k}}=\wtild_{ak}.
$
Similarly, if $k \in \head{a} \cap \tail{q}$, then $g_k + \ttild_k  = \that_k$. Feasibility follows from 
$\sum_{i \in \Shat \cap N_a}
\rdbig{\frac{4nv_{ik}}{\epsilon\,\that_k}}
=\wtild_{ak}.$
Finally, if $k \in \subtree{a}$, we have $\ttild_{ak} = \that_k$. Since $ \tvechat$ is a feasible solution in $J^\nest_q(\hvec, \gvec)$, we have 
$
\sum_{i \in \Shat}
\rdbig{\frac{4nv_{ik}}{\epsilon\,\that_k}}
=
\sum_{i \in \Shat \cap N_a}
\rdbig{\frac{4nv_{ik}}{\epsilon\,\that_k}}
\leq
\rubig{\frac{4n}{\epsilon}}.
$
The first equality holds because $N_a = \product(\subtree{a})$ so we must have $v_{ik}=0$ whenever $k \in \subtree{a}$ and $ i \notin N_a$. Moreover, since
$
g_k + \ttild_k + \ttild_{ak}
=
g_k + \that_k
$ when $k \in \subtree{a} \cup \head{a}$,
it follows that
\[
J^\nest_a\big(\wvectild_a ,\, \gvec + \tvectild \,\,\big)
\;\ge\;
\sum_{i \in \Shat \cap N_a}
\theta_i(\gvec + \tvectild + \tvectild_a)
\;=\;
\sum_{i \in \Shat \cap N_a}
\theta_i(\gvec + \tvechat \,\,).
\]
Applying these two observations gives the following chain of inequalities
\begin{align*}
\Zhat_q \overset{(i)}{\geq} & \sum_{i \in \Stild_q} \theta_i(\gvec + \tvectild \,\,) + \sum_{a \in \children(q)} J^\nest_a\big( \wvectild_a ,\, \gvec + \tvectild \,\, \big) \\
 & \qquad  \overset{(ii)}{\geq} \sum_{i \in \Shat \cap \Nbar_q} \theta_i(\gvec + \tvechat \,\,) + \sum_{a \in \children(q)} \sum_{i \in \Shat \cap N_a} \theta_i(\gvec + \tvechat \,\, ) = J^\nest_q\big( \,\hvec ,\, \gvec\, \big),
\end{align*}
where the $(i)$ follows from the feasibility of $\Stild_q$, $\tvectild$ and $\wvectild_a$ and $(ii)$ follows from the two observations. 

{\underline{\it Claim 2: $\Zhat_q \le J^{\nest}_q(\hvec,\gvec)$}}. To show this claim, we let $\Stild_q$, $\tvectild$ and $\wvectild_a$ for $a \in \children(q)$ be the optimal solution for the \ref{eqn:nest_graph_dp} problem. We also let $(\Stild_a, \tvectild_a)$ denote the optimal solution for $J^\nest_a(\wvectild_a, \gvec +\tvectild \,\,)$. Notice that we can assume $\ttild_{ak} = 0$ for all $k \notin \subtree{a}$. In what follows, we construct a feasible solution $(\Shat, \tvechat\,\,)$ for $J^{\nest}_q(\hvec,\gvec)$ as follows
\begin{align*}
    \Shat = \Stild_q ~\bigcupdot~ \big( \bigcupdot_{a \in \children(q)} \Stild_a  \big), ~~ \that_k = \begin{cases}
        \ttild_k & \text{if} ~ k \in \tail{q} \\
        \ttild_{ak} & \text{if} ~ k \in \subtree{a} \\
        0 & \text{otherwise}
    \end{cases}.
\end{align*}
We begin by showing that $(\Shat,\tvechat\,\,)$ is feasible for $J^\nest_q\big( \,\hvec ,\, \gvec\, \big)$. We verify that all constraints associated with nests in $\head{q}$ and $\subtree{q}$ are satisfied.  
By Lemma~\ref{lem:sep_vertices}, the set $\subtree{q}$ can be further decomposed into
$
\subtree{q}
=
\tail{q}
\bigcupdot
\big(\bigcupdot_{a \in \children(q)} \subtree{a}\big).
$
We therefore check feasibility case by case.  
First, consider any $k \in \head{q}$. We have
\begin{align*}
    \sum_{i \in \Shat}
    \rdbig{\frac{4nv_{ik}}{\epsilon\,g_k}} = \sum_{i \in \Stild_q}
    \rdbig{\frac{4nv_{ik}}{\epsilon\,g_k}} + \sum_{ a \in \children(q)}\sum_{i \in \Stild_a}
    \rdbig{\frac{4nv_{ik}}{\epsilon\,g_k}} 
    \overset{(i)}{\leq} 
    \sum_{i \in \Stild_q}
    \rdbig{\frac{4nv_{ik}}{\epsilon\,g_k}} + \sum_{ a \in \children(q)}\wtild_{ak} 
    \overset{(ii)}{\leq}h_k,
\end{align*}
where $(i)$ follows because $\Stild_a$ is a feasible solution for $J^\nest_a(\wvectild_a, \gvec +\tvectild \,\,)$ and $(ii)$ follows from $\Stild_q$ and $\wvectild_a$ are feasible solutions for the \ref{eqn:nest_graph_dp} problem. We then consider $k \in \tail{q}$ and $\sum_{i \in \Shat} \rdbig{\frac{4nv_{ik}}{\epsilon\,\that_k}} \leq \rubig{\frac{4n}{\epsilon}}$ follows from the same derivation. Finally, when $k \in \subtree{a}$, we only need to consider the products under $N_a = \product(\subtree{a})$. Therefore, we have 
$\sum_{i \in \Shat} \rdbig{\frac{4nv_{ik}}{\epsilon\,\that_k}} = \sum_{i \in \Shat \cap N_a} \rdbig{\frac{4nv_{ik}}{\epsilon\,\that_k}}=  \sum_{i \in \Stild_a} \rdbig{\frac{4nv_{ik}}{\epsilon\,\ttild_{ak}}} \leq \rubig{\frac{4n}{\epsilon}}$. The last inequality follows from the feasibility of $(\Stild_a, \tvectild_a)$ in $J^\nest_a(\wvectild_a, \gvec + \tvectild \,\,)$.

We then make two observations. First, for all $i \in \Stild_q$, we have $\nest(i) \subseteq \vtcs{q}$, so
$\theta_i(\; \gvec + \tvectild \;) = \theta_i(\; \gvec + \tvechat \;)$, because for every
$k \in \nest(i)$ we have $g_k + \widehat{t}_k = g_k + \widetilde{t}_k$.
Second, for all $i \in N_a$, we have \mbox{$\nest(i) \subseteq \subtree{a} \cup \head{a}$},
and the same reasoning gives
$\theta_i(\gvec + \tvectild + \tvectild_a) = \theta_i(\gvec + \tvechat)$. We then have the following chain of inequalities
\begin{align*}
    \Zhat_q & = \sum_{i \in \Stild_q} \theta_i(\gvec + \tvectild\,\,) + \sum_{a \in \children(q)} J^\nest_a(\wvectild_a, \gvec + \tvectild \,\,) 
     = \sum_{i \in \Stild_q} \theta_i(\gvec + \tvectild\,\,) + \sum_{a \in \children(q)} \sum_{i \in \Stild_a} \theta_i(\gvec + \tvectild + \tvectild_a) \\
     & \qquad \overset{(i)}{=} \sum_{i \in \Stild_q} \theta_i(\gvec + \tvechat\,\,) + \sum_{a \in \children(q)} \sum_{i \in \Stild_a} \theta_i(\gvec + \tvechat\,\,) = \sum_{i \in \Shat} \theta_i(\gvec + \tvechat\,\,) \overset{(ii)}{\leq} J^\nest_q\big( \,\hvec ,\, \gvec\, \big),
\end{align*}
where $(i)$ follows by the two observations and $(ii)$ follows by the feasibility of $(\Shat, \tvechat\,\,)$. 

\section{Numerical Study}\label{sec:numerics}

In this section, we conduct computational experiments to evaluate the benefits of fitting a GNL model to synthetic datasets. Our objective is to demonstrate that when customers in reality follow a more complex choice process, fitting a GNL model can substantially outperform fitting  MNL, NL, and MMNL models in terms of both its ability to capture customer choice behavior and its ability to identify assortments that achieve higher revenue. \yw{We adopt a similar experimental setup as in \cite{feldman2017revenue} to generate the synthetic datasets.}

\subsection{Experimental Setup}

In our computational experiments, we generate the purchase history under a ground-truth choice model that does not belong to the GNL, NL, MNL, or MMNL models. We use the ranking-based model as the ground-truth choice model. In this ground-truth choice model, customers are grouped into $m$ distinct types. A customer of type $\ell$ considers a particular subset of products and ranks the products in this subset according to a strict preference order. When a customer of type $\ell$ arrives and is presented with an offer set, the customer purchases the highest-ranked product in the offer set among the products she considers. If none of her considered products are offered, the customer leaves without making a purchase. We use $\beta_{\ell}$ to denote the probability that an arriving customer is of type $\ell$. We specify the subset of products the customer considers and the corresponding preference ranking through a tuple
$
\sigma^\ell = \big( \sigma^{\ell}(1),\ldots,\sigma^{\ell}(k^\ell) \big),
$
where $\sigma^{\ell}(1)$ is the most preferred product and $\sigma^{\ell}(k^\ell)$ is the least preferred product. 

To construct the preference ranking $\sigma^\ell$, we proceed as follows. We assume that there exist $B$ basic rankings over the full set of products. These basic rankings can be interpreted as rankings that reflect various product attributes such as quality, price, functionality, or brand reputation. Each customer type $\ell$ is associated with one of these basic rankings. We introduce heterogeneity in the following way. First, customers do not evaluate all products. Customers of type $\ell$ only consider products whose rankings fall \pr{below} some threshold. 
%This restriction naturally induces a personalized subset of products that form the initial candidate set for that customer type. Second, even if a product falls within the interval, a customer of type $\ell$ may still exclude it from consideration with some probability. Lastly, among the products that remain in the consideration set, customers may occasionally deviate from the inherited basic ranking. That is, some adjacent products in the baseline ranking are randomly perturbed. These perturbations ensure that the induced ranking remains broadly consistent with the underlying ordering while allowing for some deviations.
\yw{This restriction defines an initial set of candidate products for each customer type. Second, customers of type~$\ell$ may independently exclude products from this initial consideration set. Lastly, we allow perturbations to the relative ordering of the remaining products by randomly swapping adjacent products. The resulting ranking preserves the order specified by the basic ranking while allowing for customer-specific deviations.}

% In our computational experiments, we assume that the product universe consists of 10 distinct products. We first generate 5 basic rankings over these products, so $B = 5$. We independently generate five basic rankings by uniformly sampling permutations of the ten products. To construct the ranking for a customer type~$\ell$, we uniformly select one of these basic rankings and denote it as $\sigma = \big(\sigma(1),\ldots,\sigma(10)\big)$. We then form a personalized interval by sampling two indices $L$ and $U$ independently and uniformly from $\{1,\ldots,10\}$ such that $L < U$. Customers of type~$\ell$ focus on an interval of products  $\big(\sigma(L),\ldots,\sigma(U)\big)$. To incorporate heterogeneity across customers, we then perturb the interval as follows. First, we independently remove each product in this interval with probability~$0.1$. Second, to introduce small deviations from the underlying basic ranking, we apply a local perturbation to the ordering. With probability~$0.5$, the ordering is left unchanged. Otherwise, we uniformly at random select one product from $\big( \sigma(L),\ldots,\sigma(U-1) \big)$ and swap it with its successor. 

\yw{In our computational experiments, the product universe contains 10 distinct products. We generate $B = 5$ basic rankings by independently sampling permutations of these products uniformly at random. For each customer type~$\ell$, we select one of the basic rankings with equal probability and write it as $\sigma = \big(\sigma(1),\ldots,\sigma(10)\big)$. We next draw two indices $L$ and $U$ independently and uniformly from $\{1,\ldots,10\}$ such that $L < U$. The resulting interval $\big(\sigma(L),\ldots,\sigma(U)\big)$ specifies the products on which customers of type~$\ell$ focus. Following our discussion in last paragraph, we introduce customer heterogeneity through two perturbations of this interval. First, each product is independently excluded with probability~$0.1$. Second, we allow a small change in the relative ordering of products. With probability~$0.5$, we retain the original ordering. Otherwise, we select a product uniformly at random from $\big(\sigma(L),\ldots,\sigma(U-1)\big)$ and exchange its position with that of its successor.}

To generate the arrival probabilities $(\beta_1,\ldots,\beta_m)$, we set $\beta_1 = \beta_2 = \ldots = \beta_m = 1/m$ so that customers of each type arrive with equal probability. In all of our computational experiments, we set the number of products to $n = 10$ and the number of customer types to $m = 100$. Once the ground-truth choice model is generated, we simulate the past purchase history under the assumption that customers make their choices according to this model. We represent the purchase history of $\tau$ transactions as $\{(S_t, j_t): t = 1,\ldots,\tau\}$, where $S_t$ denotes the subset of products offered to customer~$t$ and $j_t$ denotes the purchased product. If the customer does not make a purchase, we let $j_t = 0$. To construct the offer sets, we sample 100 distinct offer sets and assume that only these offer sets are available throughout the experiment. We generate these offer sets by including each product independently with probability~$0.5$. For each time $t$, we uniformly sample one of these 100 offer sets as $S_t$. We then sample the customer type $\ell_t$ according to $(\beta_1,\ldots,\beta_m)$. The purchased product $j_t$ is the first product in the tuple $\big(\sigma^{\ell_t}(1),\ldots,\sigma^{\ell_t}(k^{\ell_t}) \big)$ that belongs to $S_t$. If none of the considered products are included in $S_t$, the customer leaves without making a purchase. Finally, to enrich the dataset, we generate 10 distinct ground-truth models. For each ground-truth model, we vary the number of transactions $\tau$ in $\{5000, 10000, 15000\}$, corresponding to three different levels of data availability. In total, this procedure yields 30 distinct transaction datasets. For each transaction dataset, we additionally generate 5,000 transactions as the validation data from the same ground-truth model and the same collection of 100 offer sets.

\subsection{Predicting Customer Purchases}

We use the same model-fitting and validation procedures as in Section~\ref{sec:empirical} to fit the GNL, MNL, NL, and MMNL models. We first evaluate the predictive performance of the fitted choice models. Since we have access to the ground-truth choice model, we can directly compare the choice probabilities predicted by each fitted model with the true choice probabilities. Specifically, for each fitted model, we compute the predicted choice probability $\hat{\phi}_i(S)$ for every assortment $S$ and every $i \in S \cup \{0\}$. We then compare these predicted probabilities with the ground-truth choice probabilities, \pr{which is denoted by $\phi_i^{\GRN}(S)$}. Since the number of products is $n=10$, we can enumerate all possible assortments. Let $\mathcal S$ denote the collection of assortments used for evaluation. We measure the prediction error using the root mean square error, defined as
\[
\mathrm{RMSE}
=
100 \times \left(
\frac{
\sum_{S \in {\mathcal S}}
\sum_{i \in S \cup \{0\}}
\left(
\hat{\phi}_i(S) - \phi_i^{\GRN}(S)
\right)^2
}{
\sum_{S \in {\mathcal S}} |S \cup \{0\}|
}
\right)^{1/2}.
\]
A smaller RMSE indicates that the fitted model more accurately captures the ground-truth choice behavior. 

Table~\ref{tab:pred_performance} reports the RMSE of the fitted models under different levels of data availability. In the table, the first column corresponds to the $10$ ground-truth models that we generate. Columns $2$--$5$ correspond to the performance of the MNL, NL, MMNL, and GNL models when we use $\tau = 5000$ data points. The next four columns correspond to the case where we use $\tau=10000$ data points, and the last four columns correspond to the case where we use $\tau=15000$ data points.

Similar to our findings in Section~\ref{sec:empirical}, the results on synthetic data also demonstrate the superior performance of the GNL model. In particular, we find the GNL model consistently improves upon the MNL, NL, and MMNL models. To further elaborate on this finding, we first focus on the comparison between the GNL and NL models. We note that the NL model is a special case of the GNL model in which the nests do not overlap with one another. Our results show that the GNL model consistently achieves a lower RMSE than the NL model. Moreover, the performance gap between the GNL and NL models becomes more pronounced as the amount of observed data increases. This suggests that the GNL model benefits more substantially from additional data.

We next focus on the comparison between the GNL and MMNL models. Both models are designed to capture richer substitution patterns, and have exactly the same number of model parameters. On average, the GNL model achieves a lower RMSE than the MMNL model under all three levels of data availability. When $\tau=5000$, the average RMSE decreases from $3.24$ under the MMNL model to $3.01$ under the GNL model, corresponding to a percentage decrease of $7.1$\%. When $\tau=10000$, the average RMSE decreases from $2.64$ to $2.48$, corresponding to a percentage decrease of $6.1$\%. When $\tau=15000$, the average RMSE decreases from $2.52$ to $2.32$, corresponding to a percentage decrease of $7.9$\%. The advantage of the GNL model is also consistent across different ground-truth models. Across the $30$ problem instances, the GNL model achieves a lower RMSE than the MMNL model in $26$ instances. These results suggest that although the MMNL model is flexible and can capture customer heterogeneity through latent segments, the GNL model provides a better fit to the ground-truth choice models in our experiments.

\begin{table}[t]
\centering
\footnotesize
\setlength{\tabcolsep}{4pt}
\begin{tabular}{ccccccccccccc}
\toprule
GRN. & \multicolumn{4}{c}{$\tau=5000$}
& \multicolumn{4}{c}{$\tau=10000$}
& \multicolumn{4}{c}{$\tau=15000$} \\
\cmidrule(lr){2-5} \cmidrule(lr){6-9} \cmidrule(lr){10-13}
Instance
& MNL & NL & MMNL & GNL
& MNL & NL & MMNL & GNL
& MNL & NL & MMNL & GNL \\
\midrule
1  & 3.97 & 3.29 & 3.39 & 2.98 & 3.91 & 3.31 & 2.64 & 2.48 & 3.85 & 3.16 & 2.57 & 2.46 \\
2  & 3.79 & 3.44 & 2.96 & 2.85 & 3.82 & 3.19 & 2.32 & 2.42 & 3.80 & 3.17 & 2.20 & 2.22 \\
3  & 4.10 & 3.42 & 3.19 & 2.98 & 4.00 & 3.27 & 2.70 & 2.53 & 4.02 & 3.25 & 2.74 & 2.66 \\
4  & 4.01 & 3.43 & 2.84 & 3.10 & 4.00 & 3.28 & 2.60 & 2.55 & 3.99 & 3.28 & 2.70 & 2.37 \\
5  & 3.62 & 2.98 & 3.33 & 3.00 & 3.56 & 2.87 & 2.71 & 2.50 & 3.56 & 2.87 & 2.33 & 2.23 \\
6  & 3.63 & 2.96 & 3.39 & 2.92 & 3.61 & 3.08 & 2.43 & 2.44 & 3.62 & 2.82 & 2.43 & 2.14 \\
7  & 4.20 & 3.42 & 3.57 & 3.28 & 3.98 & 3.23 & 2.89 & 2.57 & 3.93 & 3.18 & 2.72 & 2.34 \\
8  & 4.09 & 3.33 & 3.39 & 3.21 & 4.04 & 3.42 & 3.00 & 2.54 & 3.98 & 3.37 & 2.42 & 2.16 \\
9  & 3.92 & 3.16 & 3.41 & 2.93 & 3.82 & 3.00 & 2.43 & 2.30 & 3.85 & 3.15 & 2.61 & 2.25 \\
10 & 4.03 & 3.52 & 2.96 & 2.83 & 3.87 & 3.21 & 2.73 & 2.48 & 3.85 & 3.24 & 2.48 & 2.37 \\
\midrule
\textbf{Ave.} & 3.94 & 3.29 & 3.24 & 3.01 & 3.86 & 3.19 & 2.64 & 2.48 & 3.85 & 3.15 & 2.52 & 2.32 \\
\bottomrule
\end{tabular}
\caption{Comparison of model performance under different levels of data availability.}
\label{tab:pred_performance}
\end{table}

\subsection{Finding Good Assortments}

In the previous section, we focused on how well each model predicts customer choices. In this section, we are interested in the downstream revenue implications of using the MNL, NL, MMNL, and GNL models to make assortment decisions. To evaluate this, we compute the true expected revenue attained by the assortment recommended by each model fitted from the data. Recall that $f_i$ denotes the revenue of product $i$, and $\hat{\phi}_i(S)$ denotes the choice probability of product $i$ from assortment $S$ under a fitted model. For each fitted model, we first compute the revenue-maximizing assortment:
$
\hat{S}
=
\argmax_{S \subseteq N}
\sum_{i \in S} f_i \hat{\phi}_i(S).
$
Since we only have $n=10$ products, we can enumerate all possible assortments to find $\hat{S}$. We then evaluate the true expected revenue of this assortment under the ground-truth choice model. Recall that we use $\phi_i^{\GRN}(S)$ to denote the choice probability of product $i$ under the ground-truth model. The true expected revenue achieved by $\hat{S}$ is given by
$
\sum_{i \in \hat{S}} f_i \phi_i^{\GRN}(\hat{S}).
$
Therefore, the fitted model is used only to identify the assortment, while the final revenue is always evaluated under the ground-truth model. This allows us to compare different fitted models based on the true revenue performance of the assortments that would be chosen. In our experiment, we randomly generate the revenue $f_i$ from a uniform distribution over $[1,10]$. Additionally, to reduce the influence of randomness in the product revenues, we randomly generate $100$ revenue instances and report the average true expected revenue over these instances. We follow the same procedure for the MNL, NL, MMNL, and GNL models.

\begin{table}[t]
\centering
\footnotesize
\setlength{\tabcolsep}{4pt}
\begin{tabular}{ccccccccccccc}
\toprule
GRN & \multicolumn{4}{c}{$\tau=5000$}
& \multicolumn{4}{c}{$\tau=10000$}
& \multicolumn{4}{c}{$\tau=15000$} \\
\cmidrule(lr){2-5} \cmidrule(lr){6-9} \cmidrule(lr){10-13}
Instance
& MNL & NL & MMNL & GNL
& MNL & NL & MMNL & GNL
& MNL & NL & MMNL & GNL \\
\midrule
1  & 6.616 & 6.818 & 6.775 & 6.794 & 6.632 & 6.785 & 6.833 & 6.859 & 6.618 & 6.820 & 6.809 & 6.854 \\
2  & 6.449 & 6.612 & 6.721 & 6.715 & 6.436 & 6.670 & 6.743 & 6.723 & 6.439 & 6.679 & 6.766 & 6.770 \\
3  & 6.478 & 6.708 & 6.706 & 6.746 & 6.512 & 6.717 & 6.763 & 6.775 & 6.504 & 6.750 & 6.732 & 6.725 \\
4  & 6.551 & 6.724 & 6.729 & 6.749 & 6.547 & 6.792 & 6.801 & 6.789 & 6.546 & 6.765 & 6.798 & 6.819 \\
5  & 6.436 & 6.630 & 6.482 & 6.557 & 6.448 & 6.628 & 6.618 & 6.656 & 6.449 & 6.628 & 6.647 & 6.664 \\
6  & 6.483 & 6.685 & 6.540 & 6.617 & 6.481 & 6.626 & 6.701 & 6.690 & 6.471 & 6.667 & 6.657 & 6.700 \\
7  & 6.445 & 6.736 & 6.645 & 6.677 & 6.483 & 6.715 & 6.756 & 6.772 & 6.479 & 6.736 & 6.718 & 6.761 \\
8  & 6.431 & 6.644 & 6.657 & 6.717 & 6.430 & 6.566 & 6.685 & 6.758 & 6.430 & 6.570 & 6.752 & 6.794 \\
9  & 6.332 & 6.582 & 6.549 & 6.612 & 6.354 & 6.615 & 6.635 & 6.663 & 6.326 & 6.566 & 6.580 & 6.639 \\
10 & 6.606 & 6.699 & 6.857 & 6.865 & 6.634 & 6.778 & 6.863 & 6.870 & 6.634 & 6.756 & 6.893 & 6.890 \\
\midrule
Ave. & 6.483 & 6.684 & 6.666 & 6.705 & 6.496 & 6.689 & 6.740 & 6.755 & 6.490 & 6.694 & 6.735 & 6.762 \\
\bottomrule
\end{tabular}
\caption{Comparison of revenue performance under different levels of data availability.}
\label{tab:revenue_performance}
\end{table}

The numerical results are summarized in Table~\ref{tab:revenue_performance}. The first column corresponds to the $10$ ground-truth models that we generate. The next four columns report the revenue performance of the MNL, NL, MMNL, and GNL models when $\tau=5000$. Similarly, the following two groups of four columns report the revenue performance of these models when $\tau=10000$ and $\tau=15000$. The final row reports the average revenue performance across the $10$ ground-truth models. The results in Table~\ref{tab:revenue_performance} show that the GNL model achieves the strongest overall revenue performance across all three levels of data availability. In particular, the GNL model attains the highest average expected revenue for all three levels of data availability, with average revenues of $6.705$, $6.755$, and $6.762$ for $\tau=5000$, $\tau=10000$, and $\tau=15000$ respectively. This corresponds to improvements of $0.58\%$, $0.23\%$, and $0.39\%$ over the MMNL model. For some individual instances, the improvement exceeds $1\%$. We note that, in the context of online platforms, even a $1\%$ improvement in revenue can be economically significant. Lastly, we note that the improvement of the GNL model over MMNL model is consistent across the different instances, with GNL achieving a higher revenue in 24 out of the 30 instances.

\section{Fitting the Benchmark Models} \label{sec:fitting}

We begin by discussing how we fit the NL model. Since the NL model is a special case of the GNL model in which the nests do not overlap with one another, our estimation technique for the GNL model naturally extends to the estimation of the NL model. We make a key observation. If $M_k$ is given for all $k \in K$, estimating the NL model reduces to estimating the continuous parameters $\{\mu_{ik}: k\in K, i\in M_k\}$ and $\{\gamma_k:k\in K\}$. These parameters can be estimated using the same procedure developed for the GNL model. Therefore, the remaining challenge is to determine how the products should be partitioned across the nests. To this end, we start from an initial feasible partition of the products and then apply a greedy heuristic to iteratively improve the partition. At each iteration, we consider two types of local modifications. First, we consider switching the nest assignments of any pair of products that currently belong to different nests. Second, we consider moving a single product from its current nest to another nest. For each candidate modification, we fix the resulting nest structure and re-estimate the model parameters using the estimation procedure developed for the GNL model. We then compute the corresponding training log-likelihood and select the local modification that yields the largest improvement. We repeat this procedure until none of the local modifications leads to an improvement in the log-likelihood. 

We \yw{next} discuss our procedures to estimate the MNL and MMNL models. Under the MNL model, we assume that a customer associates a random utility with each offered product and the no-purchase option. Specifically, the random utility that customer $t$ associates with alternative $i \in S_t \cup \{0\}$ is given by $U_i = \mu_i + \epsilon_i$, where $\mu_i$ is the deterministic component of the utility and $\epsilon_i$ is an i.i.d.\ Gumbel random variable. Without loss of generality, we normalize $\mu_0 = 0$. The probability of observing choice $i_t$ from assortment $S_t$ is
$
\phi_{i_t}^{\MNL}(S_t)
=
\frac{e^{\mu_{i_t}}}{1+\sum_{i \in S_t}e^{\mu_i}}.
$
Given the historical purchase data $\{(S_t,i_t): t=1,\ldots,\tau\}$, the log-likelihood under the MNL model is
$$
\mathcal{L}^{\MNL}
=
\sum_{t=1}^\tau
\log \phi_{i_t}^{\MNL}(S_t) = \sum_{t = 1 \,:\, i_t \neq 0}^\tau \mu_{i_t} - \sum_{t = 1}^\tau \log( 1+\sum_{i \in S_t}e^{\mu_i} ).
$$
We estimate the parameters $\{\mu_i: i \in N\}$ by maximizing this log-likelihood. We note that the log-likelihood under the MNL model is concave in the utility parameters, so the maximum-likelihood estimation problem can be solved to optimality. For consistency with the implementation of the GNL model, we use \textsf{fmincon} in \textsf{MATLAB} to solve this maximum-likelihood estimation problem.

We finally describe the estimation procedure for the MMNL model. Under the MMNL model, customers are drawn from a mixture of $R$ latent segments. Each segment follows an MNL model with its own utility parameters. Let $\beta_r$ denote the probability that an arriving customer belongs to segment $r$, where $\beta_r \geq 0$ and $\sum_{r=1}^R \beta_r = 1$. We also let $\mu_{ir}$ denote the deterministic utility of product $i$ for segment $r$, and normalize the utility of the no-purchase option to be $\mu_{0r}=0$ for all $r \in \{ 1,\ldots,R \}$. The probability of observing choice $i_t$ under the MMNL model is
\[
\phi_{i_t}^{\MMNL}(S_t)
=
\sum_{r=1}^R
\beta_r
\frac{e^{\mu_{i_t r}}}{1+\sum_{j \in S_t} e^{\mu_{jr}}}.
\]
We estimate $\{\mu_{ir}: i \in N, r=1,\ldots,R\}$ and $\{\beta_r: r=1,\ldots,R\}$ by maximizing the log-likelihood subject to $\beta_r \geq 0$ for all \yw{$r = 1, \dots, R$} and $\sum_{r=1}^R \beta_r = 1$. Unlike the MNL model, the maximum-likelihood estimation problem for the MMNL model is generally nonconvex. We therefore use \textsf{fmincon} in \textsf{MATLAB} to solve the maximum-likelihood estimation problem. 
%To find the optimal number of mixture components, we follow a validation procedure similar to the one used for the GNL model. Specifically, we estimate the MMNL model with $R \in \{1,2,3,4,5,6,7,8\}$ using the training dataset, and then evaluate the log-likelihood of each fitted model on the validation set. We choose the number of mixture that yields the highest validation likelihood. 

\section{\pr{An} Example of the Dynamic Program over the Product Graph}
\label{section:example-dp-product-graph}

In this section, we consider an instance of \ref{eqn:sub_param_new} problem discussed in Section \ref{sec:product-and-nest-graphs} and illustrate how the dynamic program is carried out over the product graph discussed in Section \ref{sec:product-graph-bounded-tree-width}. We consider an instance with ten products, so the product universe is given by $N=\{1,2,\ldots,10\}$. There are seven nests, so $K=\{1,2,\ldots,7\}$. Recall that for each $k\in K$, we use $M_k$ to denote the set of products contained in nest $k$. We define the nests as follows
\begin{align*}
& M_1=\{1,2,3\},\quad
M_2=\{3,4,5\},\quad
M_3=\{2,4,6\},\quad \\
& M_4=\{3,5\},\quad
M_5=\{2,6\},\quad
M_6=\{1,7\},\quad M_7=\{1,7,8,9,10\}.
\end{align*}
For each \(k\in K\), we let \(\gamma_k=1/2\) and for simplicity, we set $v_{ik}=1$ if $i\in M_k$ and $v_{ik}=0$ otherwise. Finally, the product revenues are given by $r_1 = r_4 = r_8 = 3$, $r_2 = r_5 = r_9 = 2$ and the rest of the products have a revenue of 1. Under this nest structure, products from the set $\{1,\ldots,7\}$ belong to at least two nests and are therefore overlapping products, so \(\widehat N=\{1,\ldots,7\}\). Products $8,9,10$ belong only to nest $7$ and are nest-specific products. 

\begin{figure}
    \centering
    \includegraphics[
        width=0.85\linewidth,
        trim=0 130 0 110,
        clip
    ]{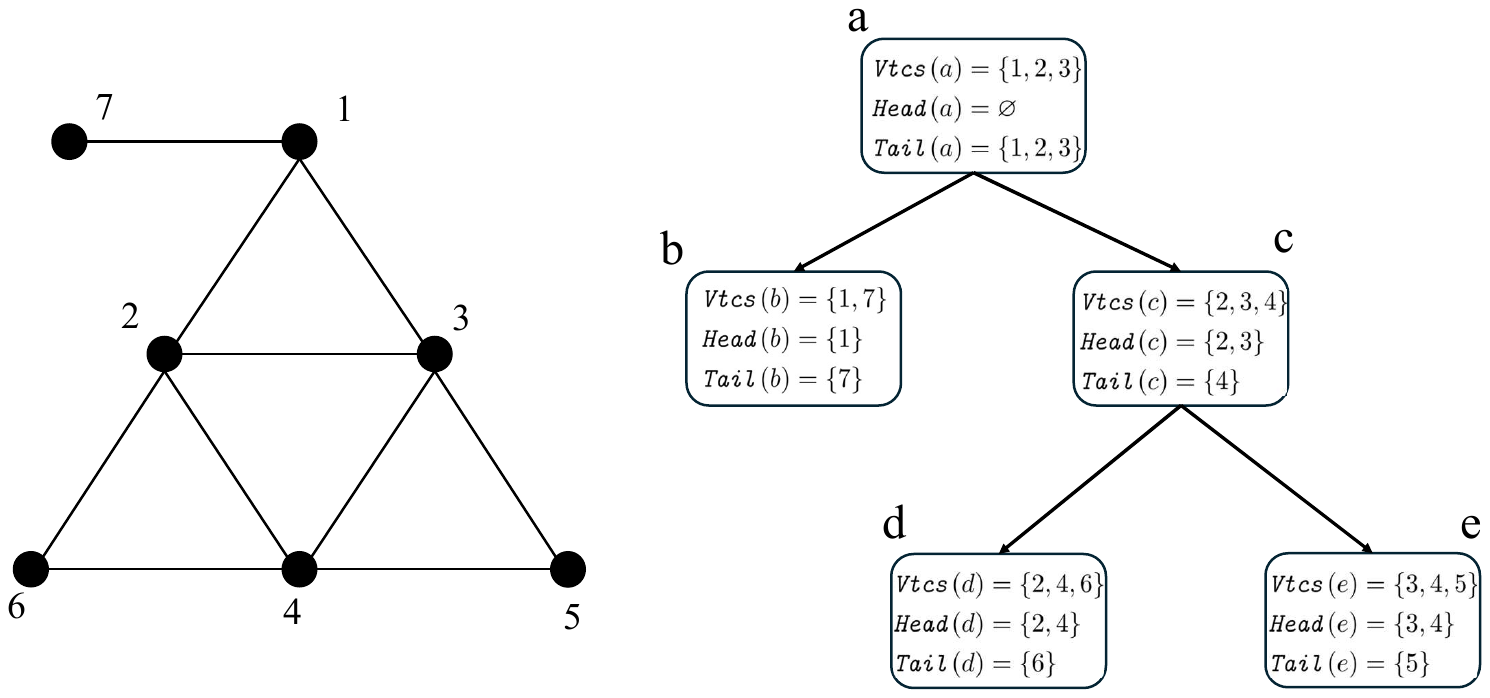}
    \caption{Left: The product graph for the illustrative example. 
    Right: A tree decomposition of the product graph. 
    Each node of the tree decomposition shows the corresponding bag of products 
    $\vtcs{p}$, together with its head $\head{p}$ and tail $\tail{p}$.}
    \label{fig:dp_example}
\end{figure}

Figure \ref{fig:dp_example} left gives the product graph. Recall that the product graph contains only the overlapping products, so its vertex set is $\widehat N=\{1,\ldots,7\}$, while the nest-specific products $8,9,10$ do not appear in the graph. Two vertices are connected by an edge if they belong to the same nest. The right side of Figure~\ref{fig:dp_example} shows a tree decomposition of this product graph. The tree decomposition contains five nodes labeled by $a,b,c,d,e$. We root the tree decomposition at node $a$. The corresponding head and tail are shown within each node in the figure. This tree decomposition has width $2$. 

We now perform the dynamic program discussed in Section~\ref{sec:product-graph-bounded-tree-width}. We begin with nodes $d$ and $e$. These two nodes are symmetric and therefore involve essentially the same computation. We first focus on node $d$. We note that this is a leaf node with $\head{d} = \{2,4\}$ and $\tail{d} = \{6\}$. In our dynamic program at node $d$, we consider the nests that contain products in $\tail{d}$, i.e., $\nest(\tail{d})$, which in this case is $\{3,5\}$. Neither of these two nests contains any nest-specific products, and we have $\Mhat_3 \subseteq \vtcs{d}$ and $\Mhat_5 \subseteq \vtcs{d}$. Therefore, the revenue contributions from both nests are evaluated at node $d$. By Theorem \ref{thm:exact_algo}, the value function at node $d$ is thus given by
$$
J_d^\product(S)
=
\max_{T_d \subseteq \tail{d}}
\left\{
\psi_3(S \cupdot T_d)
+
\psi_5(S \cupdot T_d)
\right\}
=
\max_{T_d \subseteq \{6\}}
\left\{
\psi_3(S \cupdot T_d)
+
\psi_5(S \cupdot T_d)
\right\}.
$$
Since $\head{d} = \{2,4\}$, there are four possible states:
$
\left\{
\varnothing,
\{2\},
\{4\},
\{2,4\}
\right\}.
$
For each state, we consider two possible decisions: $T_d = \varnothing$ and $T_d = \{6\}$. Finally, by how we define the preference weight, we have
$
\psi_k(A) = \frac{\sum_{i \in A \cap M_k} r_i}{\sqrt{|A \cap M_k|}}$. We can then easily evaluate the value function for $S \subseteq \{2,4\}$. 
We can then follow the same procedure for node $e$. For any $S \subseteq \{3,4\}$, the value function at node~$e$ is
$
J_e^\product(S)
=
\max_{T_e \subseteq \{5\}}
\left\{
\psi_2(S \cupdot T_e)
+
\psi_4(S \cupdot T_e)
\right\}.
$
The resulting value functions for nodes $d$ and $e$ are summarized in Table \ref{tab:dp_example_dec}.

\begin{table}[ht]
\centering
\begin{tabular}{c|c}
\multicolumn{2}{c}{\textbf{Node $d$}} \\ \hline
\rule{0pt}{2.7ex} State $S$ & $J_d^\product(S)$ \\ \hline
$\varnothing$ & $2.00$ \\
$\{2\}$ & $4.24$ \\
$\{4\}$ & $3.83$ \\
$\{2,4\}$ & $5.59$ \\ \hline
\end{tabular}
~~~~
\begin{tabular}{c|c}
\multicolumn{2}{c}{\textbf{Node $e$}} \\ \hline
\rule{0pt}{2.7ex} State $S$ & $J_e^\product(S)$ \\ \hline
$\varnothing$ & $4.00$ \\
$\{3\}$ & $4.24$ \\
$\{4\}$ & $5.54$ \\
$\{3,4\}$ & $5.59$ \\ \hline
\end{tabular}
~~
\begin{tabular}{c|c}
\multicolumn{2}{c}{\textbf{Node $c$}} \\ \hline
\rule{0pt}{2.7ex} State $S$ & $J_c^\product(S)$ \\ \hline
$\varnothing$ & $9.36$ \\
$\{2\}$ & $11.12$ \\
$\{3\}$ & $9.41$ \\
$\{2,3\}$ & $11.17$ \\ \hline
\end{tabular}
\caption{Dynamic programming values at nodes $d$, $e$, and $c$ for the illustrative example.}
\label{tab:dp_example_dec}
\end{table}

We then consider node $c$. From Figure~\ref{fig:dp_example}, we have
$\head{c}=\{2,3\}$ and $\tail{c}=\{4\}$. The children of node $c$ are node $d$ and $e$. We also note that there does not exist nest $k \in \nest(\tail{c})$ such that $\Mhat_k \subseteq \vtcs{c}$, so there is no ``local contribution" to the objective function at node $c$. For each $S \subseteq \head{c}$, we only need to account for the contributions from the children nodes, which gives us
\[
J_c^\product(S)
=
\max_{T_c\subseteq\{4\}}
\left\{
J_d^\product\big((S\cupdot T_c)\cap\head{d}\big)
+
J_e^\product\big((S\cupdot T_c)\cap\head{e}\big)
\right\}.
\]
For each fixed assortment $S\subseteq\{2,3\}$, the value function at node $c$ can be computed directly using the values already obtained at nodes $d$ and $e$. We report the resulting values of $J_c^\product(S)$ in Table~\ref{tab:dp_example_dec}.

We next consider node $b$. From Figure~\ref{fig:dp_example}, we have
$\head{b}=\{1\}$ and $\tail{b}=\{7\}$. The two nests associated with $\tail{b}$ are
$M_6=\{1,7\}$ and $M_7=\{1,7,8,9,10\}$. Since the products 8, 9, 10 are nest-specific item, so we have $\Mhat_6=\Mhat_7=\{1,7\}\subseteq\vtcs{b}$. Therefore, the contributions from both nests are evaluated at node $b$. Notice that nest $M_7$ contains the nest-specific products $\{8,9,10\}$,
with revenues $r_8=3$, $r_9=2$, and $r_{10}=1$. By
Theorem~\ref{thm:nest-specific-products}, it suffices to consider the
revenue-ordered subsets
$
\Ocal_7
=
\left\{
\varnothing,\{8\},\{8,9\},\{8,9,10\}
\right\}.
$
Thus, for each fixed assortment $S\subseteq\{1\}$, the value function at
node $b$ is given by
\[
J_b^\product(S)
=
\max_{T_b\subseteq\{7\}}
\left\{
\psi_6(S\cupdot T_b)
+
\max_{A_7\in\Ocal_7}
\psi_7(S\cupdot T_b\cupdot A_7)
\right\}.
\]
Evaluating the above expression gives
$J_b^\product(\varnothing)=4.50$ and
$J_b^\product(\{1\})=7.62$.

Finally, we consider the root node $a$, where we have
$\head{a}=\varnothing$ and $\tail{a}=\{1,2,3\}$. The children of
node $a$ are nodes $b$ and $c$. The set of nests that contain products in $\tail{a}$ is $\{1,2,3,4,5,6,7\}$, but nest 1 is the only nest where $M_1 \subseteq \vtcs{a}$. Since
$\head{a}=\varnothing$, there is only one state for the value function,
and the dynamic programming equation is given by
\[
J_a^\product(\varnothing)
=
\max_{T_a\subseteq\{1,2,3\}}
\left\{
\psi_1(T_a)
+
J_b^\product(T_a\cap\head{b})
+
J_c^\product(T_a\cap\head{c})
\right\}.
\]
Recall that $\head{b}=\{1\}$ and $\head{c}=\{2,3\}$. Using the value functions already
computed at nodes $b$ and $c$, we can evaluate the above expression for
all eight possible subsets of $\{1,2,3\}$. The largest value is attained
at $T_a=\{1,2\}$, which gives
$
J_a^\product(\varnothing)=22.28.
$
Since node $a$ is the root of the tree decomposition, we have
$J_a^\product(\varnothing)=F^*$, which is the optimal value of the  \ref{eqn:sub_param_new} problem for this instance.

\end{APPENDICES}

\end{document}